\documentclass[%
 reprint,
superscriptaddress,
amsmath,amssymb,
aps,
]{revtex4-2}

\usepackage{graphicx}
\usepackage{dcolumn}
\usepackage{bm}
\usepackage{hyperref}
\usepackage{verbatim} 
\hypersetup{
colorlinks,
citecolor=blue,
filecolor=blue, 
linkcolor=blue, 
urlcolor=blue
}

\usepackage{color}
\usepackage{xcolor}
\usepackage{ulem}
\usepackage{physics}

\newcommand{\yambo}{\textsc{yambo}}
\newcommand{\qe}{\textsc{Quantum Espresso}}

\newcommand{\q}{{\mathbf q}}
\newcommand{\kk}{{\mathbf k}}

\renewcommand{\Im}{\mathrm{Im}} 
\renewcommand{\Re}{\mathrm{Re}}

\newcommand{\editor}[2]{%
  \expandafter\newcommand\csname #1note\endcsname[1]{%
    \textcolor{#2}{(\textbf{#1:} \textit{##1})}}%
  \expandafter\newcommand\csname #1\endcsname[1]{%
    \textcolor{#2}{##1}}%
  \expandafter\newcommand\csname #1cancel\endcsname[1]{%
    \textcolor{#2}{\sout{##1}}}%
\expandafter\newcommand\csname #1can\endcsname[1]{%
    \textcolor{#2}{\sout{##1}}}%
  \expandafter\newcommand\csname #1change\endcsname[2]{%
    \textcolor{#2}{\sout{##1} ##2}}%
      \expandafter\newcommand\csname #1ch\endcsname[2]{%
    \textcolor{#2}{\sout{##1} ##2}}%
  \newenvironment{#1text}{\color{#2}}{\color{black}}
}
\definecolor{Blu}{rgb}{0.00,0.00,1.00}
\definecolor{Red}{rgb}{1.00,0.00,0.00}
\definecolor{Cyan}{rgb}{0.00,0.50,0.50}
\definecolor{Green}{rgb}{0.00,0.70,0.00}

\editor{DAL}{Green}
\editor{CC}{Cyan}
\editor{kb}{orange}
\editor{KB}{orange}
\editor{R}{Red}

\renewcommand{\emph}{\textit}
\newcommand{\suppinfo}{Ref.~\cite{supp-info}}

\begin{document}

\title{Bulk plasmons in elemental metals} 

\author{Dario A. Leon}
\email{dario.alejandro.leon.valido@nmbu.no}
\affiliation{
 Department of Mechanical Engineering and Technology Management, \\ Norwegian University of Life Sciences, NO-1432 Ås, Norway
}%

\author{Claudia Cardoso}
\affiliation{
 S3 Centre, Istituto Nanoscienze, CNR, 41125 Modena, Italy
}

\author{Kristian Berland}
\affiliation{
 Department of Mechanical Engineering and Technology Management, \\ Norwegian University of Life Sciences, NO-1432 Ås, Norway
}

\date{\today}

\begin{abstract}
The spectral properties, momentum dispersion, and broadening of bulk plasmonic excitations of 26 elemental metals are studied from first principles calculations in the random-phase approximation. 
Spectral band structures are constructed from the resulting momentum- and frequency-dependent inverse dielectric function.  
We develop an effective analytical representation of the main collective excitations in the  dielectric response, 
extending our earlier model based on multipole-Pad\'e approximants (MPAs) to incorporate both momentum and frequency dependence [MPA($\q$)]. 
With this representation, 
we identify plasmonic quasiparticle dispersions exhibiting complex features, including non-parabolic energy and intensity dispersions, discontinuities due to anisotropy, and overlapping effects that lead to band crossings and anti-crossings. 
Comparing with available experimental data, mainly in the optical limit, we  
find good agreement with the computed spectra. 
The results for elemental metals and their effective MPA($\q$) representation establish a reference point that can guide both fundamental studies and practical applications in plasmonics and spectroscopy. 
\end{abstract}

\maketitle

\section{Introduction}
Elemental metals, composed of a single type of metallic atom, are foundational to both modern technology and condensed matter physics~\cite{Ashcroft-Mermin1976book,Kittel2005book}. Examples encompass alkali metals (e.g., Na, K), alkaline earth metals (e.g., Mg, Ca), and noble (e.g., Al, Cu, Ag, Au) or other transition metals (e.g., Cr, Ti, Fe, Ni). These materials are characterized by high electrical and thermal conductivity, a strong but ductile metallic bonding, and a conduction band partially filled with delocalized electrons. 
Delocalized electrons respond collectively to external fields and are responsible for the characteristic optical and electronic properties that distinguish metals from insulators and semiconductors.
Thus, understanding the electronic properties of elemental metals, including their collective excitations, is of significant theoretical and practical importance.
Metals with a relatively simple electronic band structure, such as Al and alkali metals, host delocalized conduction electrons that behave nearly as a free-electron gas~\cite{Melnyk1970PRB,book_Palik1985,Maksimov1988JPEMP,martin2016book}. This property, along with a substantial amount of available experimental data, makes them ideal systems for studying fundamental electronic excitations and their underlying many-body interactions.

Collective oscillations of the electron density give rise to plasmonic quasiparticles, or plasmons. Their origin lies in the long-range Coulomb restoring force acting on charge-density fluctuations, which produces well-defined modes in the free-electron gas, while in real materials, the plasmonic dispersions are modulated by their underlying electronic band structure and dielectric screening~\cite{Pines-Nozieres2005book,Giuliani-Vignale2005book,Mahan2013book}. Plasmons can be excited by electromagnetic perturbations or incident charged particles, and appear as bulk or surface modes~\cite{Raether2006book,Egerton2011book}.
Plasmons are central in several technological applications, including nanophotonics~\cite{Barnes2003Nature, Karabchevsky2020Nanopho}, surface-enhanced spectroscopy~\cite{Moskovits1985RMP}, energy harvesting~\cite{Atwater2010NatMat,Cushing2016JPCL}, and emerging quantum plasmonic 
technologies~\cite{Jacob2012MRSplas,Gjerding2017natcomm,Babicheva2023Nanomat}.
Plasmonic properties can also be engineered by modifying the dielectric environment or the electronic structure, for example through alloying~
\cite{Moller1980PRL, Orhan2019JPCM}.

Bulk plasmons correspond to longitudinal charge-density oscillations and are therefore optically inactive. They can be probed experimentally by methods such as electron energy-loss spectroscopy (EELS)~\cite{Raether2006book,Egerton2011book}, where they appear as well-defined peaks in the measured spectra. 
Other techniques such as inelastic x-ray scattering and reflection electron energy loss spectroscopy (REELS) can also 
provide complementary insights into plasmonic properties~\cite{Schulke1984PRL,Schulke1986PRB,Schulke1989PRB,Schulke2007book,Werner2009JPCRD,Garcia2010RMP,Babar2015ApplOpt}. In simple metals such as Al or Na, the plasmon peak is sharp and symmetric, akin to that of a free-electron gas~\cite{Maksimov1988JPEMP,Raether2006book}. 
In contrast, in noble metals, such as Ag and Au, the occupied $d$-bands introduce inter-band transitions that broaden and shift the plasmon resonance, and give rise to much richer spectra~\cite{Johnson1972PRB,book_Palik1985}.
Such experimental signatures are broadly employed for material characterization, from determining free-carrier concentrations, crystalline phases, and thicknesses to exploring nanoplasmonic effects and probing the electronic band structure~\cite{Garcia2010RMP,Egerton2011book,Leon2024PRB}.	

Although the dielectric properties of elemental metals in the optical limit have been extensively studied and tabulated~\cite{Johnson1972PRB,book_Palik1985,Maksimov1988JPEMP,Giuliani-Vignale2005book,Babar2015ApplOpt,Polyanskiy2024SciData}, properties at finite momentum are not as widely available. 
For gapless metallic systems, strong electronic screening suppresses bound electron–hole (excitonic) effects, making the random-phase approximation (RPA) a well-established and widely used framework for describing bulk plasmon excitations~\cite{Giuliani-Vignale2005book,martin2016book}. 
In this work, we report the frequency and momentum dependent inverse dielectric functions of 26 elemental metals computed from first principles within the RPA level of the many-body theory, using Kohn–Sham Bloch (KS) states obtained from density functional theory (DFT). We also provide an effective analytical representation of the main collective excitations in the inverse dielectric function, 
by generalizing our previous model based on multipole-Pad\'e approximants (MPA)~\cite{Leon2021PRB,Leon2023PRB,Leon2025PRB} to include momentum alongside frequency dependence. The generalized MPA model in two variables, MPA($\q$), is used to describe the main spectral properties of such excitations. 
It also shows potential to develop efficient methods to further reduce the
computational cost of first-principles calculations such as $GW$ and BSE~\cite{martin2016book,Leon2024PRB}, with respect to the original MPA method~\cite{Leon2021PRB}.

The paper is organized as follows: Sec.~\ref{sec:methods} details the theoretical framework (Sec.~\ref{sec:theo}), the MPA and MPA($\q$) representations (Sec.~\ref{sec:mpa}), the set of studied materials (Sec.~\ref{sec:mat}), and the computational details of the first-principle calculations (Sec.~\ref{sec:com}). Sec.~\ref{sec:results} presents the results for the optical limit of intra-band contributions (Sec.~\ref{sec:intra-inter}), the comparison between RPA and the experimental spectra  (Sec.~\ref{sec:optical_limit}), and the deviations from the free-electron gas model (Sec.~\ref{sec:PPA_like}), while results at finite momentum are used for the construction of spectral band structures  (Sec.~\ref{sec:rpa_spectral}), and the generalized MPA($\q$) model (Sec.~\ref{sec:mpa_spectral}). Finally, Sec.~\ref{sec:con} holds the conclusions.

\section{Methods}
\label{sec:methods}

\subsection{Theoretical framework}
\label{sec:theo}
The complex dielectric function $\varepsilon$ governs the response of a material to external electromagnetic fields. It depends on momentum $\q$ and frequency $\omega$. In the optical, or long-wavelength limit, $\q \to 0$, the dielectric function of the free-electron gas reduces to the Drude form, characterized by a single plasmon pole~\cite{Allen1977PRB,book_Palik1985,Smith1986PRB,Maksimov1988JPEMP,Lee1994PRB}. Within the RPA approximation, Lindhard theory extends the classical Drude model to include the $\q$ dependence~\cite{Mahan2013book,Giuliani-Vignale2005book}, providing a first-order quantum-mechanical description of the free-electron response. 
The dielectric response of real materials, often deviates from that of the homogeneous free-electron behavior, and is often much richer. The RPA dielectric function can be computed from first principles using KS-DFT states to evaluate the independent-particle microscopic polarizability, as follows
\begin{multline}
    \chi_{0\mathbf{G G'}}(\mathbf{q},\omega) = \sum_{n,m} \int_{\mathrm{BZ}} \frac{d \mathbf{k}}{(2\pi)^3} \rho^*_{n m \mathbf{k}}(\mathbf{q},\mathbf{G}) \rho_{n m \mathbf{k}}(\mathbf{q},\mathbf{G'}) \times \\
     f_{n \mathbf{k-q}}(1-f_{m \mathbf{k}})
    \left [ \frac{2 \Omega^{\mathrm{KS}}_{n m \mathbf{k} \mathbf{q}}}{\omega^2- (\Omega^{\mathrm{KS}}_{n m \mathbf{k} \mathbf{q}})^2} \right ],
    \label{eq:Xipa}
\end{multline}
where the $n$ and $m$ indices run over the bands, $\rho_{n m \mathbf{k}}(\mathbf{q},\mathbf{G}) \equiv \langle n \mathbf{k} | e^{i (\mathbf{q} + \mathbf{G}) \cdot \mathbf{r}} | m \mathbf{k-q} \rangle$ are transition matrix elements, the $f$ factors are the occupations of the KS states, $\Omega^{\mathrm{KS}}_{n m \mathbf{k} \mathbf{q}} = (\epsilon_{m \mathbf{k}}-\epsilon_{n \mathbf{k-q}}) -i\delta$ are KS single-particle transitions, and the limit $\delta \to 0^+$ is implicit and ensures the correct time ordering~\cite{martin2016book}.

The interacting microscopic polarizability and the inverse dielectric function are obtained by the Dyson equation:
\begin{align}
  \chi &= \chi_0 + \chi_0 v \chi \label{eq:dyson}, \\
  \varepsilon^{-1} &= 1 + v \chi,
  \label{eq:inv_epsilon}
\end{align}
where $v$ is the Coulomb potential. 
Even though the macroscopic polarizability, $\chi(\mathbf{q},\omega)$, is given by the $\mathbf{G}=\mathbf{G}'=0$ component of the microscopic $\chi$, it is affected by the coupling of $\mathbf{G},\mathbf{G}'\neq 0$ components through Eq.~\eqref{eq:dyson}. These contributions are commonly referred to as local-field effects, as they account for microscopic variations of the induced fields within the crystal. Their inclusion accounts for the difference between the independent-particle approximation (IPA) and RPA.

In metallic systems, $\chi_0$ 
encompasses both intra- and inter-band transitions, which in Eq.~\eqref{eq:Xipa} correspond to $n=m$ and $n\neq m$ respectively, and thus $\chi_0 = \chi_0^{\mathrm{intra}} +\chi_0^{\mathrm{inter}}$. 
In the interacting polarizability $\chi$, such contributions mix according to the Dyson equation of Eq.~\eqref{eq:dyson} and cannot be simply splitted. 
In order to define intra- and inter-band frequencies from the interacting $\chi$, we can then take the case of semiconductors, that have only inter-band contributions, to define the following terms 
\begin{align}
\chi^{\mathrm{intra}} &= \chi - \chi^{\mathrm{inter}}, \\
\chi^{\mathrm{inter}} &=  \chi_0^{\mathrm{inter}} + \chi_0^{\mathrm{inter}} v \chi^{\mathrm{inter}}.
\end{align}
This separation carries over to the dynamic part of the inverse dielectric function, $Y \equiv v \chi = \varepsilon^{-1}-1$, resulting in~\cite{Leon2023PRB}
\begin{equation}
    Y(\q, \omega) =
    Y_{\mathrm{intra}}(\q, \omega) + Y_\mathrm{inter}(\q, \omega),
    \label{eq:y_intra+inter}
\end{equation}
where $Y_\mathrm{intra} \equiv v\chi^\mathrm{intra}$ and $Y_\mathrm{inter} \equiv v\chi^\mathrm{inter}$. 
Since $\chi_0^\mathrm{intra}$ vanishes in the $\q \to 0$ limit, so does $\chi^\mathrm{intra}$. However, 
$Y_\mathrm{intra}$ remains finite due to the divergence of the Coulomb potential $v$.
Therefore, in metals, the plasmonic quasiparticles observed in $Y(\q, \omega)$ at any value of $\q$ emerge from a mixture of intra- and inter-band contributions.

An important property of the imaginary part of $Y$, or the loss function $\Im [-\varepsilon^{-1}]$, is the f-sum rule, which is conserved in the inversion of the Dyson equation (see Supplemental materials of Ref.~\cite{Leon2023PRB}) and thus can be used to define corresponding intra-band, inter-band, and plasma frequencies: 
\begin{align}
    \label{eq:intra_freq} 
    \omega_{\mathrm{intra}}^2(\q) &\equiv \frac{2}{\pi} \int_0^{\infty} d \omega \,\omega  |\Im [Y_{\mathrm{intra}}(\q, \omega)]| , 
    \\\label{eq:inter_freq} 
    \omega_{\mathrm{inter}}^2(\q) &\equiv \frac{2}{\pi} \int_0^{\infty} d \omega\, \omega  |\Im [Y_{\mathrm{inter}}(\q, \omega)]| ,
    \\\label{eq:plasma_freq} 
    \omega_{\mathrm{pl}}^2 &\equiv \frac{2}{\pi} \int_0^{\infty} d \omega  \,\omega |\Im [Y(\q, \omega)]| .
\end{align}
From Eq.~\eqref{eq:y_intra+inter} it follows that $   \omega_{\mathrm{pl}}^2 = \omega_\mathrm{intra}^2 (\q) + \omega_\mathrm{inter}^2 (\q)$.  
Whereas $\omega_{\mathrm{pl}}$ is $\q$-independent, the respective intra- and inter-band contributions are $\q$ dependent, but with a dependence constrained by this condition.

The interacting $\chi$, and hence also $Y$, has a pole structure similar to that of $\chi_0$ in Eq.~\eqref{eq:Xipa}. However, the large number of single-particle transitions $\Omega^{\mathrm{KS}}_{n m \mathbf{k} \mathbf{q}}$ are mixed in the Dyson equation in Eq.~\eqref{eq:dyson}, giving rise to plasmonic quasiparticles and thus resulting in much simpler envelope functions~\cite{Leon2023PRB}. 
For many systems, 
$Y$ in the optical limit $\q \to 0$ resembles the response of the free-electron gas~\cite{Hedin1965PR,Fetter-Walecka1971book,Giuliani-Vignale2005book}, and  can therefore be approximated by a single plasmon-pole model (PPA):
\begin{equation}
    Y^{\mathrm{PPA}} (\q \to 0, \omega) = \frac{\Omega_0^2}{\omega^2-\Omega_0^2}, 
    \label{eq:PPA_q0}
\end{equation}
where the pole is located at the plasmon energy, $\Re[\Omega_0] = \omega_{\mathrm{pl}}$. In the free-electron gas the plasmon energy is related to the electronic density, $\rho_e$, as (in Gaussian units)
\begin{equation}
    \label{eq:plasma_energy}
    \omega_{\mathrm{pl} }= \sqrt{4\pi \rho_e}. 
\end{equation}

The extension of the PPA model to finite $\q$ is given by~\cite{Hybertsen1986PRB,vonderLinden1988PRB,Zhang1989PRB,Godby1989PRL,Engel1993PRB,Stankovski2011PRB,Miglio2012EPJB,Larson2013PRB}
\begin{equation}
    Y^{\mathrm{PPA}} (\q, \omega) = \frac{2 R(\q) \Omega(\q)}{\omega^2-[\Omega(\q)]^2} ,
    \label{eq:y_ppa}
\end{equation}
where the pole has a quadratic dispersion according to the Drude/Lindhard model, 
$\Omega(q) \approx \Omega_0 (1 + \Omega'' q^2/2)$, with an effective electron mass factor $\Omega''$.  The dispersion of the spectral weight is linked to the position of the pole by the f-sum rule of Eq.~\eqref{eq:plasma_freq}:
\begin{equation}
   \Omega_0^2 = 2 R(\q) \Omega(\q),
    \label{eq:f-sumrule_PPA}
\end{equation}
which implies that $R(\q)$ decreases with the inverse of $\Omega(\q)$, as $\propto 1/q^{2}$.

\subsection{Simple analytical representation: The multipole approach (MPA)}
\label{sec:mpa}
The RPA approximation is at the core of {\it ab initio} many-body theories beyond DFT, such as
$GW$; however, expressing the operators in terms of single-particle excitations, as in the Lehmann representation of Eq.~\eqref{eq:Xipa}, is both computationally impractical and hinders the physical interpretation of collective features.
Within first principles approaches, analytical models simplifying the frequency dependence of the dielectric function, starting from the (single-pole) PPA, have been used to reduce the computational cost.   
However, the use of PPA is subject to accuracy limitations~\cite{Leon2021PRB,Leon2023PRB}. 
More complex models, such as Pad\'e approximants and many-pole schemes, have often been used to fit and interpret individual experimental spectra~\cite{Lee1996PRB,Jin1999PRB,Soininen2005PS,Kas2007PRB,Kas2009JPCS,Schulke2007book,Bourke2012JPCA}.  
Within {\it ab initio} $GW$ calculations, effective MPA representations have been shown to provide an efficient description of the dielectric response, achieving an accuracy comparable to that of full-frequency approaches~\cite{Leon2021PRB,Leon2023PRB,Leon2025PRB}. Such models generalize the PPA inverse dielectric response of Eq.~\eqref{eq:y_ppa} as
\begin{equation}
    Y^{\mathrm{MPA}}(\q, \omega) = \sum_p^{n_{Y}} \frac{2 R_p (\q) \Omega_p(\q)}{\omega^2 - [\Omega_p(\q)]^2},
    \label{eq:mpa_model}
\end{equation}
where the number of poles $n_{Y}$ is typically around 10.

In practice, poles and residues are obtained, for each $\q$, by interpolation with a coarse sampling in the complex frequency plane~\cite{Leon2021PRB,Leon2023PRB}.  
Despite considering each $\q$ separately, the procedure provides a good description of the $\q$ dispersion of the main poles~\cite{Leon2023PRB}. 
MPA has also been combined with numerical methods to accelerate convergence with respect to the $\q$-mesh, in the so-called 
constant approximation (CA) of Ref.~\cite{Leon2023PRB} and the Monte Carlo $W$ averaging (W-av) of Refs.~\cite{Guandalini2024PRB,Sesti2025Arxiv}.

In this work, we further advance this type of 
representation by proposing a generalized MPA model with an explicit $\q$ dependence, denoted MPA($\q$). We consider a polynomial dispersion of the poles and residues, following the Taylor expansion around $\q = 0$ of the independent-particle response and the free-electron gas. 
In practice, we truncate the expansion at the third order and consider specific $\q$-directions:
\begin{equation}
\begin{split}
    R_p(q) &\approx R_p (1 + R'_p q + R''_p q^2/2 + R'''_p q^3/6) , \\
    \Omega_p(q) &\approx \Omega_p (1 + \Omega'_p q + \Omega''_p q^2/2 + \Omega'''_p q^3/6),
    \label{eq:mpa_taylor}
\end{split}
\end{equation}
where $R_p$ and $\Omega_p$ are the residues and pole positions in the optical limit $q \to 0$ along the selected direction, and the respective pairs $R'_p$ and $\Omega'_p$, $R''_p$ and $\Omega''_p$, and $R'''_p$ and $\Omega'''_p$ are the corresponding linear, quadratic, and cubic coefficients of the polynomial expansions.
Although other $q$ dependencies can be explored, we find a third-order polynomial to be sufficiently flexible to obtain accurate representations with relatively few poles, 
between 2 and 16, for all the systems studied here. Therefore, the generalized MPA($\q$) model provides a fairly compact representation of the momentum and frequency dependent $Y(\q,\omega)$, with the total number of parameters given by $8 \times n_{Y}$. 
All the parameters are determined 
in a non-linear fit with physical constraints on the position of the poles, satisfying the time-ordering in a finite energy interval (see Refs.~\cite{Leon2021PRB,Leon2023PRB}): 
\begin{equation}
\begin{split}
      & 0 < \Re [\Omega_p (\q)] < \omega_{\mathrm{max}} \\
      & - \Re [\Omega_p (\q)] < \Im [\Omega_p (\q)] < 0.
\end{split}  
\label{eq:pole_constraints}
\end{equation}

Another relevant constraint is the f-sum rule, which for MPA($\q$) takes the following form
\begin{equation}
    \omega_{\mathrm{pl}}^2 = \sum_p^{n_Y} 2 \Re [R_p (\q) \Omega_p(\q)].
    \label{eq:f-sumrule_MPA}
\end{equation}
However, this is strictly respected only when the integration interval over energy in Eq.~\eqref{eq:plasma_freq} is sufficiently large. Therefore, we do not explicitly impose it in any of the present calculations, since they are done for limited energy ranges.

\subsection{Reference data for the set of elemental metals}
\label{sec:mat}
The dielectric properties of metals in the optical limit are widely studied and reported, e.g., in several handbooks and websites~\cite{eels_atlas, book_Palik1985,Polyanskiy2024SciData}. However, the frequency dependent dielectric function is not always available and different sources can be inconsistent  with respect to precision, energy range, etc. Experimental techniques such as EELS only measure the imaginary part of the response and need careful post-processing to retrieve the inverse dielectric function, which is nontrivial at finite momentum~\cite{Leon2024PRB}. 

Here we consider reference experimental data from several sources. Data corresponding to the optical limit $\q \to 0$ were taken from the REELS measurements of Ref.~\cite{Werner2009JPCRD} (set of 17 elemental metals: Ag, Au, Cu, Ni, Pb, Pt, Pd, Fe, Mo, Ta, V, W, Co, Ti, Zn, Bi, and Te), the EELS atlas~\cite{eels_atlas} (for Be, Al, V, Cr, Cu, Zn, Ag, Sn, Te, W, Os, Pt, Au, Tl, Pb, Bi), and  Ref.~\cite{Mathewson1971PhySc,Nilsson1977PRB} for Ca. In the case of EELS, we use a standard procedure, based on a fit to a power law model~\cite{Stoger2008Micron}, to remove the background intensity from the zero-loss peak~\cite{Egerton2011book}. 
Reference data at finite momentum are available for Li~\cite{Schulke1984PRL,Schulke1986PRB,Hiraoka2023PRB}, Be~\cite{Schulke1989PRB,Seidu2018PRB},  Na~\cite{vomFelde1989PRB,Cazzaniga2011PRB}, Mg~\cite{Kloos1973PLA}, Al~\cite{Kloos1973PLA,Perti1976SSC,Batson1976PRL,Batson1983PRB,Sprosser-Prou1989PRB,Abril1998PRA,Cazzaniga2011PRB}, and K~\cite{vomFelde1989PRB}, and for Cu, Ag, and Au~\cite{Alkauskas2013PRB}.

A complete list of the 26 elemental metals studied here is shown in Table~\ref{tab:metals}. The table includes the atomic number, the electronic configuration of valence electrons, the symmetry group, the experimental lattice parameters~\cite{periodic_table}, and the volume of the unit cell. The computed intra- and inter-band frequencies are also provided.

\subsection{Computational details}
\label{sec:com}
The crystalline structure of elemental metals is well characterized experimentally and theoretically~\cite{Sun2023PNAS}. 
In this work, we use the experimental geometries and lattice parameters listed in Table~\ref{tab:metals}.
DFT calculations were performed using the {\qe}~\cite{QE1, QE2} plane wave package with the Perdew-Burke-Ernzerhof variant of the GGA functional~\cite{Perdew1996PRL}. 
We adopted the norm-conserving optimized Vanderbilt pseudopotentials of Ref.~\cite{Hamann2013PRB}, with a kinetic plane wave energy cutoff of 70 Ry. 
In the DFT calculations, the Brillouin zone was sampled with a $36\times 36 \times 36$ Monkhorst-Pack grid for cubic, and $36\times 36 \times 24$ for hexagonal materials. 
Metallic occupations were treated using a smearing technique, employing ordinary Gaussian spreading with a smearing width of 0.01 Ry. 
Spin polarization was included in the DFT calculations of ferromagnetic elemental metals (Fe, Co, Ni).
The calculations of dielectric spectra within RPA were performed with the {\yambo} code~\cite{Marini2009CPC,Sangalli2019JPCM}. Eq.~\eqref{eq:Xipa} is evaluated with a damping parameter of $\delta=0.1$~eV and a kinetic plane wave energy cutoff of 5 Ry.

\section{Results}
\label{sec:results}

\begin{table*}[hbt!]
  \begin{ruledtabular}
    \begin{tabular}{cccccccccc} 
      \\[-8pt]
      System & $Z$ & val. electrons & space group (num.) & lattice par. (\AA) & unit vol. (\AA$^3$)  & $\omega_\mathrm{intra}$ (eV) & $\omega_\mathrm{inter}$ (eV) & $\omega_\mathrm{intra}^2/\omega_\mathrm{pl}^2$ (\%) & $Z_{\mathrm{eff}}$\\
      \hline\\[-8pt]     
      Li & 3  & 2s$^1$ & Im\_3m (229) & 3.51 & 21.6218 & 6.56 & 4.66 & 66.5 & 0.72 \\
      Be & 4  & 2s$^2$ & P6$_3$/mmc (194) & 2.2858, \; 3.5843 & 8.1093 & 5.35 & 17.83 & 8.2 & 2.20 \\
      Na & 11 & 3s$^1$ & Im\_3m (229) & 4.2906 & 39.4934 & 5.62 & 1.42 & 94.0 & 0.90 \\
      Mg & 12 & 3s$^2$ & P6$_3$/mmc (194) & 3.2094,\; 5.2108 & 23.2409 & 6.97 & 7.39 & 47.1 & 1.76 \\
      Al & 13 & 3s$^2$ 3p$^1$ & Fm\_3m (225) & 4.0495 & 16.6014 & 11.84 & 9.22 & 62.2 & 2.63 \\
      K  & 19 & 4s$^1$ & Im\_3m (229) & 5.328 & 75.6245 & 4.15 & 10.58 & 13.4 & 0.76 \\
      Ca & 20 & 4s$^2$ & Fm\_3m (225) & 5.5884 & 43.6317 & 4.20 & 15.08 & 7.2 & 2.10 \\
      Ti & 22 & 3d$^2$ 4s$^2$ & P6$_3$/mmc (194) & 2.9508,\; 4.6855 & 17.6659 & 3.74 & 26.48 & 2.0 & 3.77 \\
      V  & 23 & 3d$^3$ 4s$^2$ & Im\_3m (229) & 3.03 & 13.9091 & 7.96 & 29.61 & 6.7 & 4.22 \\
      Cr & 24 & 3d$^5$ 4s$^1$ & Im\_3m (229) & 2.91 & 12.3211 & 6.80 & 32.72 & 4.1 & 4.85 \\
      Fe & 26 & 3d$^6$ 4s$^2$ & Im\_3m (229) & 2.8665 & 11.7768 & 6.40 & 32.15 & 3.8 & 5.48 \\
      Co & 27 & 3d$^7$ 4s$^2$ & P6$_3$/mmc (194) & 2.5071,\; 4.0695 & 11.0761 & 6.00 & 31.90 & 3.4 & 5.43 \\
      Ni & 28 & 3d$^8$ 4s$^2$ & Fm\_3m (225) & 3.524 & 10.9408 & 6.56 & 34.09 & 3.6 & 7.94 \\
      Cu & 29 & 3d$^{10}$ 4s$^1$ & Fm\_3m (225) & 3.6149 & 11.8094 & 8.61 & 31.65 & 6.9 & 5.76 \\
      Zn & 30 & 3d$^{10}$ 4s$^2$ & P6$_3$/mmc (194) & 2.6649,\; 4.9468 & 15.2120& 9.14 & 29.96 & 8.5 & 4.00 \\
      Mo & 42 & 4d$^{5}$ 5s$^1$ & Im\_3m (229) & 3.147 & 15.5833 & 8.51 & 32.41 & 6.5 & 6.68 \\
      Pd & 46 & 4d$^{10}$ & Fm\_3m (225) & 3.8907 & 14.7239 & 6.72 & 34.90 & 3.6 & 10.17 \\
      Ag & 47 & 4d$^{10}$ 5s$^1$ & Fm\_3m (225) & 4.0853 & 17.0456 & 8.84 & 34.48 & 6.2 & 7.57 \\
      Sn & 50 & 4d$^{10}$ 5s$^2$ 5p$^2$ & I41/amd (141) & 5.8318,\; 3.1819 & 27.0540& 8.85 & 29.45 & 8.3 & 3.35 \\
      Ta & 73 & 4f$^{14}$ 5d$^2$ 6s$^2$ & Im\_3m (229) & 3.3013 & 17.9897 & 8.21 & 47.77 & 2.9 & 5.82 \\
      W  & 74 & 4f$^{14}$ 5d$^3$ 6s$^2$ & Im\_3m (229) & 3.1652 & 15.8553 & 6.91 & 50.89 & 1.81 & 6.85 \\
      Os & 76 & 4f$^{14}$ 5d$^6$ 6s$^2$ & P6$_3$/mmc (194) & 2.7344,\;  4.3173 & 13.9777& 8.32 & 38.81 & 4.4 & 10.94 \\
      Pt & 78 & 4f$^{14}$ 5d$^9$ 6s$^1$ & Fm\_3m (225) & 3.9242 & 15.1075 & 8.38 & 38.66 & 4.5 & 12.05 \\
      Au & 79 & 4f$^{14}$ 5d$^{10}$ 6s$^1$ & Fm\_3m (225) & 4.0782 & 16.9569 & 8.69 & 36.78 & 5.3 & 7.35 \\
      Tl & 81 & 4f$^{14}$ 5d$^{10}$ 6s$^2$ 6p$^1$ & P6$_3$/mmc (194) & 3.4566,\; 5.5248 & 28.5835& 6.72 & 32.65 & 4.1 & 2.17 \\
      Pb & 82 & 4f$^{14}$ 5d$^{10}$ 6s$^2$ 6p$^2$ & Fm\_3m (225) & 4.9508 & 30.3365 & 8.94 & 32.43 & 7.1 & 3.32 \\
      
    \end{tabular}
  \end{ruledtabular}
\caption{List of properties of the 26 elemental metals studied in this work. The experimental lattice parameters and the corresponding unit cell volume are taken from the compilation made in Ref.~\cite{periodic_table}. 
}
    \label{tab:metals}
\end{table*}

\subsection{Intra-band vs inter-band contributions}
\label{sec:intra-inter}
\begin{figure}[hbt!]
    \centering
    \includegraphics[width=0.49\textwidth]{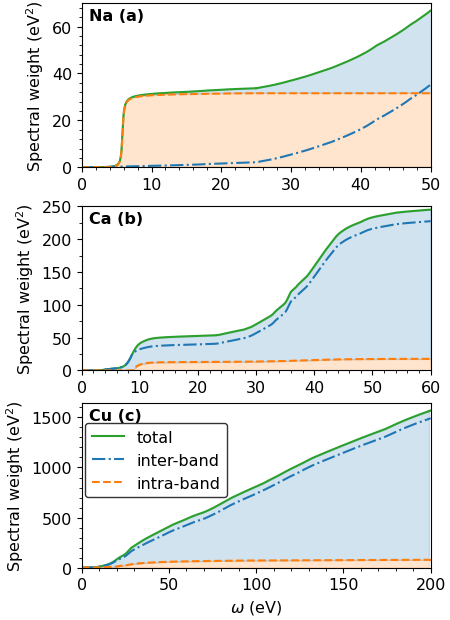}
    \caption{Spectral contributions of the intra-band (dashed orange line), the inter-band (dash-dotted blue line), and the plasma (solid green line) frequency as a function of energy, evaluated as running integrals of Eqs.~\eqref{eq:intra_freq}, \eqref{eq:inter_freq}, and \eqref{eq:plasma_freq}, respectively. The total area under the plasma curve is filled with each corresponding contribution.}
    \label{fig:sum-rule}
\end{figure}

Fig.~\ref{fig:sum-rule} shows the inter- and intra-band contributions to the plasma frequency 
as a function of the energy for Na, Ca, and Cu, computed at at the optical limit with Eqs.~\eqref{eq:intra_freq}, \eqref{eq:inter_freq} and \eqref{eq:plasma_freq}. 
Inter-band contributions from semi-core and core states appear at large energies, e.g., above $25$~eV for the case of Na, and were not considered when evaluating the inter-band and plasma frequencies.

Table~\ref{tab:metals} reports the respective intra- and inter-band frequencies, and the percentage of intra-band contributions to the plasma frequency for the 26 studied metals. 
While the intra-band frequencies of all the metals range from 3.74~eV for Ti to $11.84$~eV for Al, the inter-band frequencies have a far wider range, from only $1.42$~eV for Na to $50.89$~eV for W. This huge span is due to inter-band contributions of d- and f-orbitals, thus intra-band contributions are relatively more relevant for metals with only $s$ and $p$ electrons.
The plasmon of Na is dominated by intra-band contributions (94.0~\%), followed by Li (66.5~\%), and Al (62.2~\%). 
Next in the list are Mg, with less than half (47.1~\%), and K (13.4~\%). The percentage for the rest of the metals is much lower.

\subsection{Spectra in the optical limit} 
\label{sec:optical_limit}
 \begin{figure}[hbt!]
    \centering
    \includegraphics[width=0.49\textwidth]{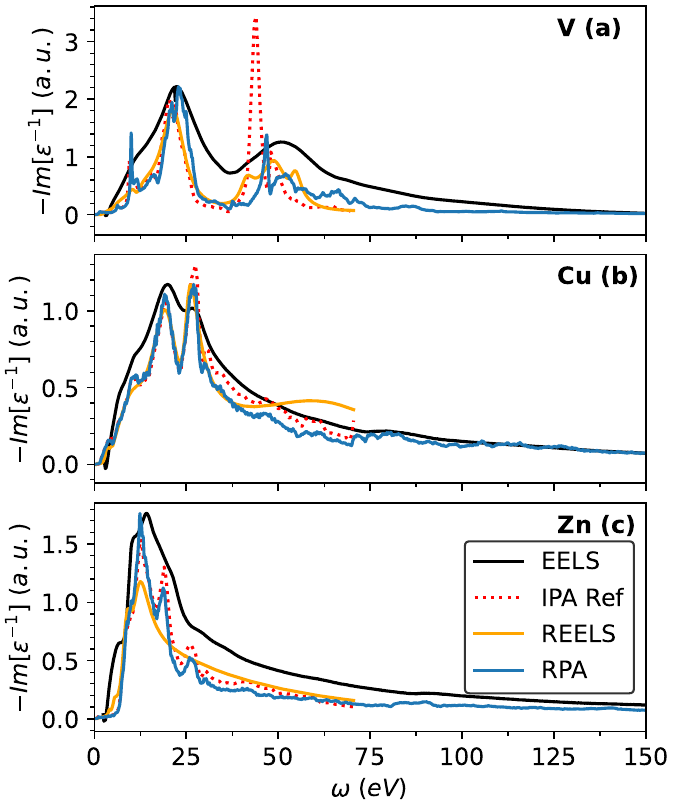}
    \caption{Comparison of the computed RPA loss function of V (a), Cu (b), and Zn (c) in the optical limit, with experimental EELS data from Ref.~\cite{eels_atlas}, REELS from Ref.~\cite{Werner2009JPCRD}, and theoretical IPA calculations from Ref.~\cite{Werner2009JPCRD}.}
    \label{fig:loss_q0}
\end{figure}

In Fig.~\ref{fig:loss_q0} we compare the computed RPA loss function of V, Cu, and Zn in the optical limit, with EELS measurements from the EELS Atlas~\cite{eels_atlas} and REELS data and IPA of Ref.~\cite{Werner2009JPCRD}. These are examples of materials with complex screening effects resulting in several peaks in their response function, that require a description beyond a single pole PPA model~\cite{Nilsson1977PRB,book_Palik1985,Leon2023PRB}.

As mentioned in Sec.~\ref{sec:mat}, the raw EELS data provided in Ref.~\cite{eels_atlas} includes a background intensity from the zero-loss peak, was removed using a standard power law model~\cite{Stoger2008Micron,Egerton2011book}, while the remaining intensity was normalized according to the maximum RPA intensity. 
These measurements have a lower resolution and a larger instrumental broadening than those of Ref.~\cite{Werner2009JPCRD}; however, for most cases, the energy range is larger, providing information on the tail of the spectra. The experimental loss function of Ref.~\cite{Werner2009JPCRD} was obtained by performing Kramers-Kronig transformations~\cite{Werner2009JPCRD,Egerton2011book} on the raw REELS data. The results in the tail region are therefore affected by the finite energy interval in the frequency integral of such transformations, and in several cases, like Cu, are less accurate than the direct EELS measurements.

In all cases, both our RPA calculations and the IPA results of 
Ref.~\cite{Werner2009JPCRD} are able to accurately describe the main features of the experimental data. The inclusion of local field effects with RPA improves the description of the intensity with respect to IPA. Refining the DFT starting point, e.g., with DFT+U or hybrid functionals, may improve the description of the finer features in the loss function of materials with $d$-states and strong spin-orbit interactions~\cite{Zhukov2002PRB,Rangel2012PRB,Prandini2019npjComputMater,Leon2024PRB}.
Such refinements are beyond the scope of this work. 

 \begin{figure}[hbt!]
    \centering
    \includegraphics[width=0.49\textwidth]{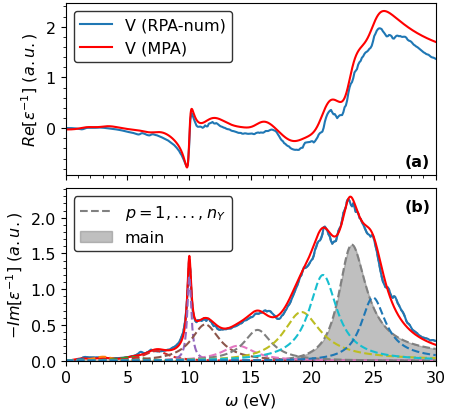}
    \caption{Real (a) and imaginary (b) parts of the inverse dielectric function of V computed with RPA in the optical limit. The numerical data is compared with its corresponding MPA model with $n_Y=13$ poles. Dashed lines in the bottom panel represent the individual contributions of each pole. The most prominent pole is highlighted with a gray filling.}
    \label{fig:mpa_q0}
\end{figure}

Fig.~\ref{fig:mpa_q0} compares the numerical
RPA results for the real and imaginary parts of $\varepsilon^{-1} (\q \to 0, \omega)$ of V in a smaller energy range, up to $30$~eV, with the fitted MPA model of Eq.~\eqref{eq:mpa_model} with $n_Y = 13$. 
Even with such a small number of poles, the MPA model accurately reproduces the main features of the inverse dielectric response. 
The bottom panel shows dashed curves representing the individual contributions of each pole. Many of them carry a significant fraction of the total spectral weight and present a large broadening. The overlap of the poles illustrates the complex nature of plasmonic excitations in elemental metals like V.

\subsection{Deviation from the free-electron gas and PPA
}
\label{sec:PPA_like}
\begin{figure}[hbt!]
    \centering
    \includegraphics[width=0.49\textwidth]{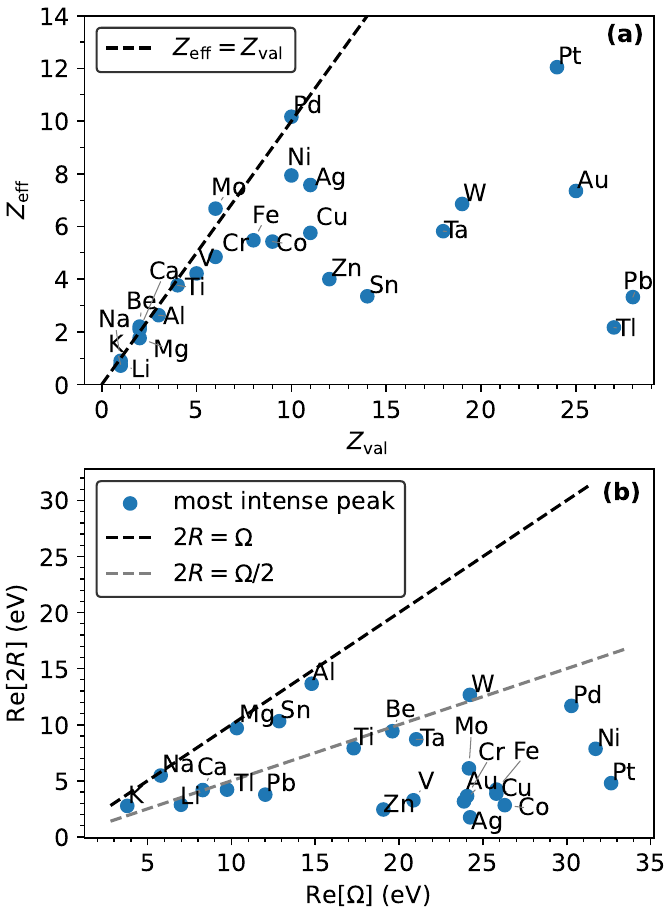}
    \caption{ (a) Relation between the effective number of electrons, $Z_{\mathrm{eff}}$, and the number of valence electrons, $Z_{\mathrm{val}}$, listed in Table~\ref{tab:metals} for the set of studied metals. $Z_{\mathrm{eff}}$ is computed from the energy position of the main MPA pole, with Eq.~\eqref{eq:Zeff}. The black dashed line represents the identity $Z_{\mathrm{eff}} = Z_{\mathrm{val}}$. 
    (b) Relation between the spectral weight, $\Re[2 R]$, and the energy position, $\Re[\Omega]$, of the main MPA pole of all the studied metals. The black dashed line corresponding to $\Re[2 R]=\Re[\Omega]$, represents the ideal PPA model where all the spectral weight is concentrated in a single pole. The presence of other poles with significant spectral weights lowers the position with respect to this PPA line. The gray dashed line represents a pole with half the spectral weight.}
    \label{fig:main-peak}
\end{figure}

The free-electron gas model has often been used to describe simple metals like Al and Na~\cite{Ashcroft-Mermin1976book,Kittel2005book,Wooten1972book}, considering an electronic density given by the number of their valence electrons. Such an approach can be used for other metals, but it is not always clear which electronic density to consider. With MPA in place, we can define an effective number of electrons from the expression of the classical plasmon energy in Eq.~\eqref{eq:plasma_energy}:
\begin{equation}
    Z_{\mathrm{eff}} \equiv \frac{\Re[\Omega]^2 V}{4 \pi},
    \label{eq:Zeff}
\end{equation}
where $\Omega$ is the main pole of the MPA representation and $V$ is the volume of the unit cell.

Fig.~\ref{fig:main-peak}(a) shows the relation between the computed $Z_{\mathrm{eff}}$ and the number of valence electrons,  $Z_{\mathrm{val}}$, for the set of 26 studied metals. The values are listed in the last column of Table~\ref{tab:metals}.
For some of the cases $Z_{\mathrm{eff}}$ is close to the number of electrons in the outer orbitals. For example, the value of $0.9$, $2.1$, and $10.17$ are very close to $1$, $2$, and $10$, corresponding to the number of electrons in the 3s$^1$, 4s$^2$, and 4d$^{10}$ orbitals of Na, Ca, and Pd, respectively. However, in several cases like Co, Ag, and Ta, the mixed nature of the different orbital contributions to the plasmon energy and thus $Z_{\mathrm{eff}}$, prevents us from establishing any simple association.

On the {\it ab initio} side, the (single-pole) PPA model, has been extensively used in $GW$ and similar approaches beyond DFT, even for systems with several prominent poles in their response functions. The MPA representation captures such a multipole behavior, and the spectral weight of its dominant pole provides a quantitative measure of the unsuitability of a (single-pole) PPA description. 

Fig.~\ref{fig:main-peak} shows the spectral weight and the energy position of the main MPA pole in the response function of the 26 studied metals. According to Eq.~\eqref{eq:PPA_q0}, in PPA the spectral weight should equal the plasmon energy, which is 
represented by the black dashed line. As a guide to eye, the gray dashed line indicates half of the total spectral, $\Re[2 R]=\Re[\Omega]/2$.
The results show that Na, Mg, Al, K, and Sn are the systems well described by a (single-pole) PPA model, and the plasmon of Al has the largest spectral weight of all the metals. 
For Li, Be, Ca, W, Os, Ta, Tl, and Pb the main pole carries nearly half the total spectral weight, whereas for the remaining metals it contributes even less, with Ag exhibiting the smallest spectral weight and representing one of the systems furthest from a (single-pole) PPA picture.

\subsection{Spectral $Y (\q, \omega)$ band structures}
\label{sec:rpa_spectral}
\begin{figure*}[hbt!]
    \centering
    \includegraphics[width=0.99\textwidth]{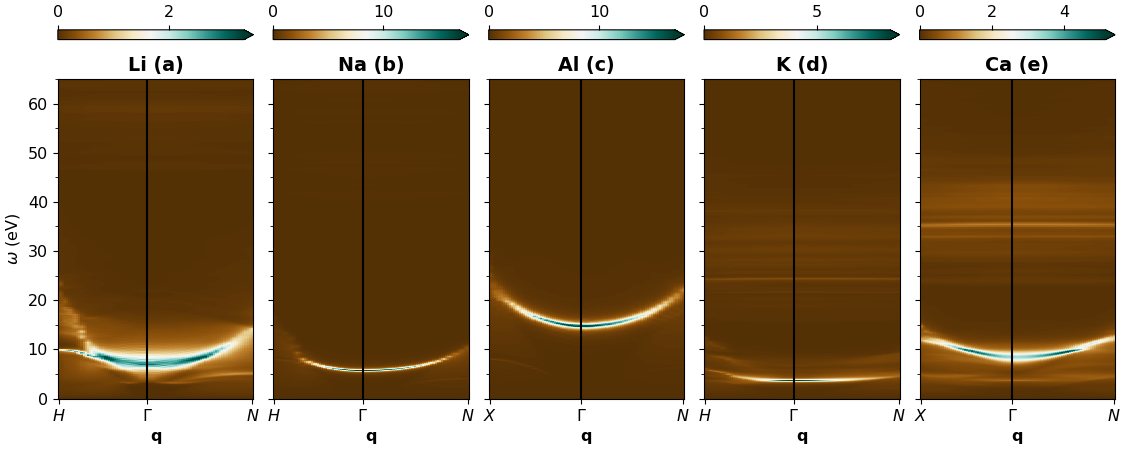}
    \caption{Spectral band structures corresponding to the RPA loss function, $|\Im [Y (\omega, \q)]|$, of cubic elemental metals from rows II-IV of the periodic table. The plots are made for two high-symmetry $\q$-lines of bulk Li (a), Na (b), Al (c), K (d), and Ca (e).}
    \label{fig:qeel_II-IV}
\end{figure*}

Fig.~\ref{fig:qeel_II-IV} shows the dispersion of the main excitations in $Y (\q, \omega)$ 
along two high-symmetry $\q$-lines of bulk Li (a), Na (b), Al (c), K (d), and Ca (e). The selected $\q$-paths correspond to $H \Gamma N$ and $X \Gamma N$ for the bcc and fcc Bravais lattices, respectively. Such simple elemental metals show a parabolic-like dispersion of the main plasmon, with an effective mass that increases with the plasmon energy. Secondary excitations are more visible for heavier elements, showing a much flatter dispersion than that of the main plasmon.

 \begin{figure*}[hbt!]
    \centering
    \includegraphics[width=0.99\textwidth]{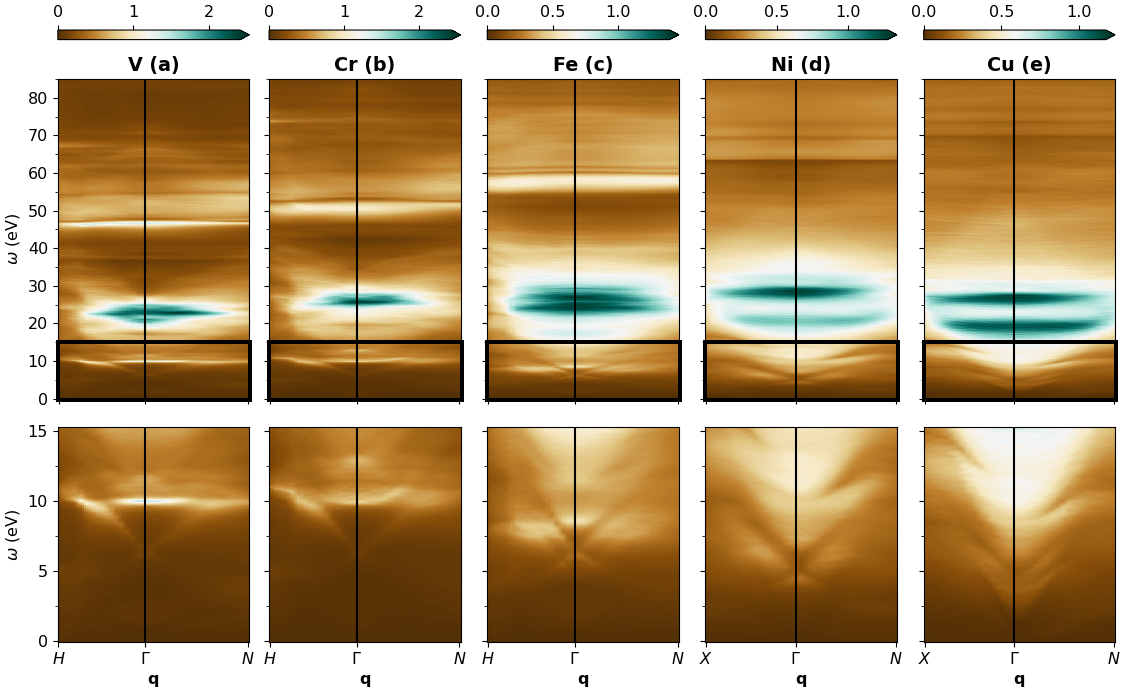}
    \caption{Spectral band structures corresponding to the RPA loss function, $|\Im [Y (\omega, \q)]|$, of cubic elemental metals from rows IV of the periodic table. The plots are made for two high-symmetry $\q$-lines of bulk V (a), Cr (b), Fe (c), Ni (d), and Cu (e). A zoomed region is included in the lower panels.}
    \label{fig:qeel_IV_zoom}
\end{figure*}

\begin{figure*}[hbt!]
    \centering
    \includegraphics[width=0.99\textwidth]{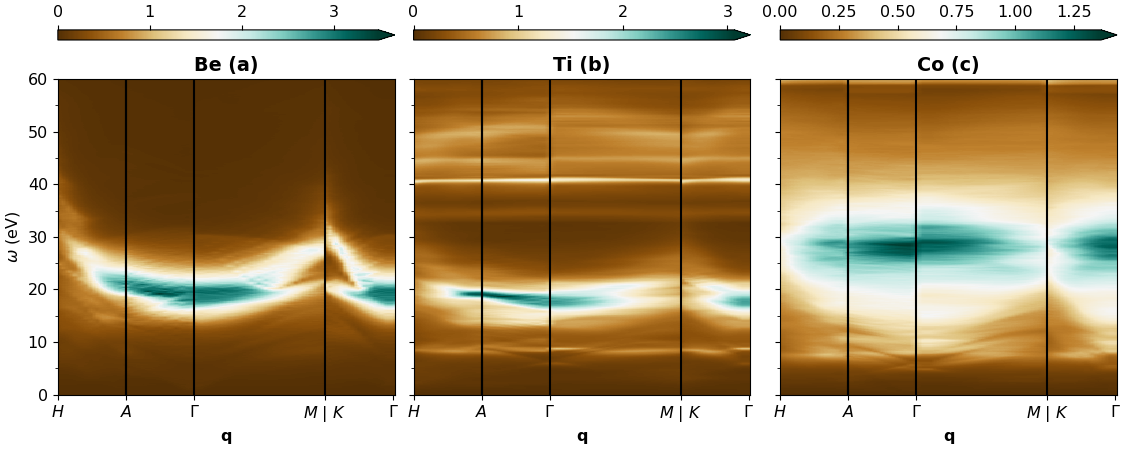}
    \caption{Spectral band structures corresponding to the RPA loss function, $|\Im [Y (\omega, \q)]|$, of hexagonal elemental metals from rows II-IV of the periodic table. The plots are made for four high-symmetry $\q$-lines of bulk Be (a), Ti (b), and Co (c).}
    \label{fig:qeel_H1v2}
\end{figure*}

Fig.~\ref{fig:qeel_IV_zoom} shows analogous plots for V (a), Cr (b), Fe (c), Ni (d), and Cu (e), while Fig.~\ref{fig:qeel_H1v2} shows plots along the $H A \Gamma M|K \Gamma$ path for metals with hexagonal symmetry: Be (a), Ti (b), and Co (c).
The complexity of the excitation landscape of these metals increases very quickly with the atomic number, particularly when valence $d$-electrons are present, contrasting with the light cubic elements in Fig.~\ref{fig:qeel_II-IV} and the light hexagonal ones (Be and Mg). 
In these cases, the main plasmon is split into a multiple-peak structure with mostly flat dispersions, in contrast to what would be expected from the free-electron gas model with corresponding electronic densities.
In addition, there are many more secondary excitations with a significant spectral weight. 
The dispersion of such excitations is also more complex, and gives rise to diverse features in the spectral band structure. 

Complex features in the spectra include anisotropic plasmonic dispersions and discontinuities that arise due to
anisotropies in the underlying electronic band structure~\cite{Perti1976SSC,Sturm1978ZPBCM,Sprosser-Prou1989PRB,Budagosky2019PRB}. Examples of such discontinuities are discernible 
in Fig.~\ref{fig:qeel_H1v2}, such as
for Be around $30$~eV  
at the $\Gamma$ point along the $A \Gamma M$ path, 
and correspondingly at $45$-$55$~eV for Ti, and $20$-$35$~eV for Co.  
For the case of Co, the discontinuity extends to the region of the plasmon, similar to other hexagonal systems with $d$-electrons showing strong anisotropy signatures, such as ZnO~\cite{Leon2024PRB}. 
Other spectral features, such as non-parabolic energy and intensity dispersions~\cite{Kaltenborn2013PRB,Budagosky2019PRB}, including negative dispersions and indirect excitations, i.e., with a finite intensity at finite $\q$, that vanishes at $\q = \Gamma$, can be seen in many of the studied metals. There is also quasi-particle overlapping leading to band crossings and anti-crossings, as shown in Fig.~\ref{fig:qeel_IV_zoom} in the zoomed region around $10$~eV. 

\subsection{Spectral properties within MPA($\q$)}
\label{sec:mpa_spectral}
\begin{figure}[hbt!]
    \centering
    \includegraphics[width=0.49\textwidth]{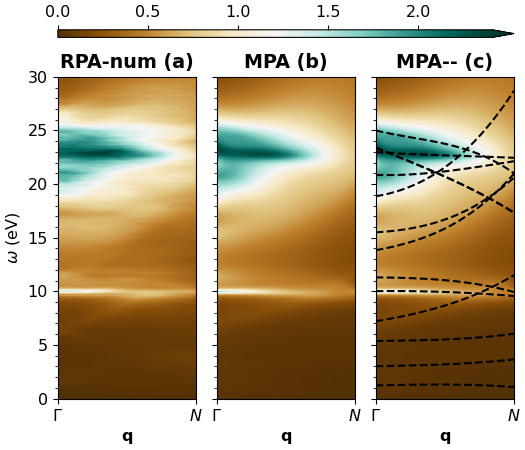}
    \caption{Comparison of the spectral $Y (\q, \omega)$ band structure of V in the $\Gamma N$ direction, generated from the numerical RPA data with its corresponding MPA($\q$) model with $n_Y=13$ poles.}
    \label{fig:qeel_ff_vs_mpa}
\end{figure}
Spectral $Y (\q, \omega)$ band structures for each of the studied 26 metals are provided in \suppinfo{}. 
Many of them exhibit complex features, analogous to those in Fig.~\ref{fig:qeel_H1v2}. 
Nonetheless, at least for the most prominent peaks, the dispersion with $\q$ is smooth enough to be easily followed. This property suggests that $Y (\q, \omega)$ can be modeled with a small set of plasmonic excitations with 
simple $\q$ dependence.
To test this hypothesis, we have fitted an MPA($\q$) model to the numerical $Y(\q, \omega)$ data of each of the 26 metals studied, using a number of poles that varies from $n_Y = 2$ to $n_Y = 15$, depending on the complexity of the spectra. The fit is performed along a selected high symmetry $\q$-line, $\Gamma N$ for cubic and $\Gamma M$ for hexagonal systems. Analogous fits can be made along the other $\q$-paths.  
In all the cases, the resulting $Y^{\mathrm{MPA}}(\q, \omega)$ reproduces very accurately the numerical data, as shown in \suppinfo{} and in the following for the specific examples of V, Ca, and Ni.  

\begin{figure*}[hbt!]
    \centering
    \includegraphics[width=0.99\textwidth]{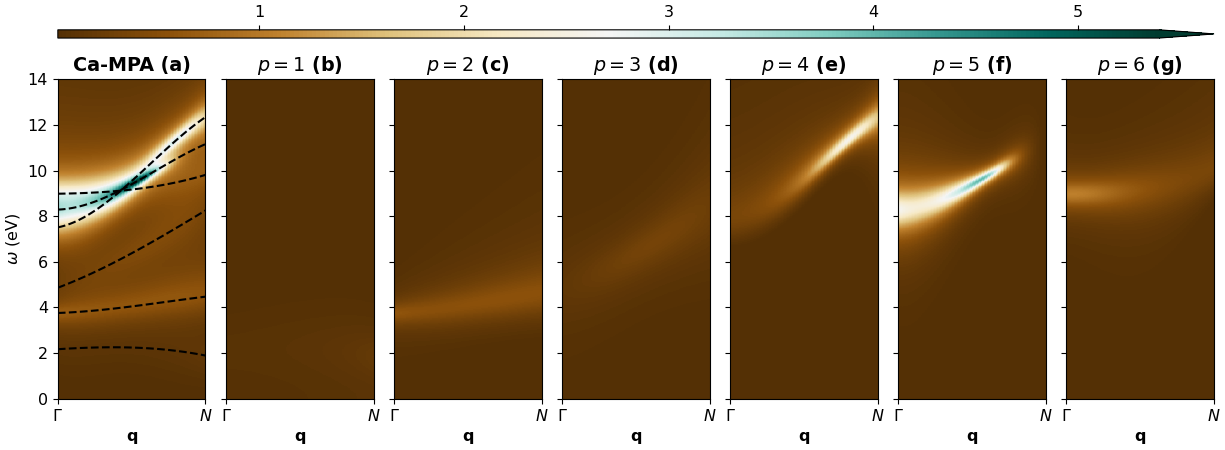}
    \caption{Spectral $Y (\q, \omega)$ band structure of Ca reconstructed with MPA($\q$) in Eqs.~\eqref{eq:mpa_model} and \eqref{eq:mpa_taylor} with a number of $n_Y = 6$ poles. The total spectral function along the $\Gamma N$ $\q$-path is shown in (a), while the individual contribution of each pole is plotted in (b-g) according to their energy position at $\Gamma$. The dashed lines in (a) indicate the energy dispersion of $\Re [\Omega_p]$.}
    \label{fig:qeel_mpa_Ca}
\end{figure*}

\begin{figure}[hbt!]
    \centering
    \includegraphics[width=0.49\textwidth]{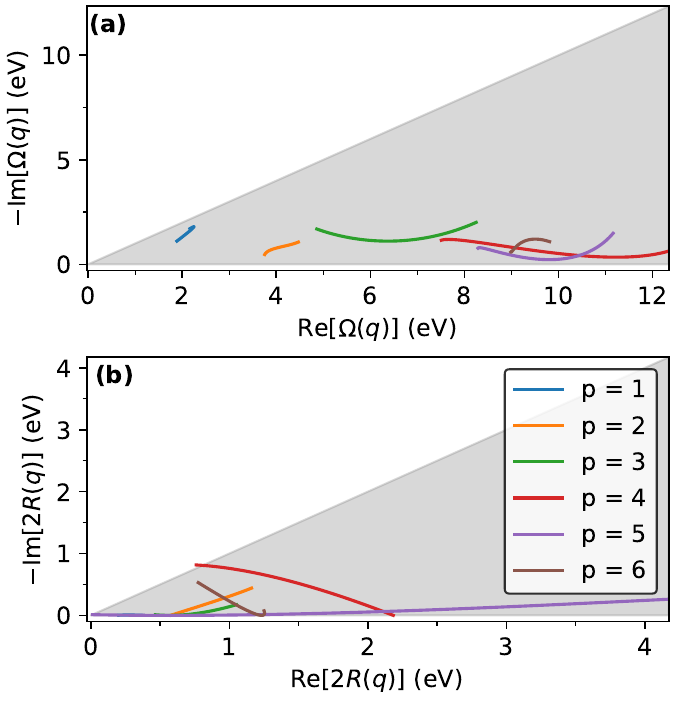}
    \caption{Relation between the real and the imaginary parts of the poles $\Omega_p(q)$ (a) and the residues $2 R_p(q)$ (b) of the MPA($\q$) representation of Ca, in the complex frequency plane.}
    \label{fig:constrainER}
\end{figure}

Figure~\ref{fig:qeel_ff_vs_mpa} illustrates the case of V. The numerical $Y(\q, \omega)$ spectral band structure,  shown in Fig.~\ref{fig:qeel_ff_vs_mpa}(a), is accurately reproduced with a fitted MPA($\q$) model with $n_Y=13$ poles, the same number used in Fig.~\ref{fig:mpa_q0}, as shown here in Fig.~\ref{fig:qeel_ff_vs_mpa}(b). To facilitate the interpretation,   Fig.~\ref{fig:qeel_ff_vs_mpa}(c) plots the same MPA($\q$) model with dashed black lines indicating the energy position of the poles, $\Re [\Omega_p (\q)]$.

The analytical MPA($\q$) representation also provide means to disentangle overlapping poles, which is highly nontrivial.  
Figure~\ref{fig:qeel_mpa_Ca}(a) shows the total spectral band structure of Ca, reconstructed with MPA($\q$) with 6 poles. 
Figures~\ref{fig:qeel_mpa_Ca}(b)-(g) shows the individual contribution of each pole to the sum in Eq.~\eqref{eq:mpa_model}, labeled according to their energy position at $\q=0$. 
The MPA($\q$) poles of Ca have sufficiently small broadening and smoothly varying intensity to allow identifying them as well-defined spectral bands along the whole $\q$ range. 
The spectrum shows two main poles in line with Ref.~\cite{Nilsson1977PRB}. 
For most values of $\q$, the most intense pole is $p = 5$, while $p = 4$ dominates around the $\q = N$ high-symmetry point. Both poles present a non-monotonic spectral weight with a maximum at finite $\q$. The three upper poles ($p=4,5,6$) cross at a $\q$-point slightly before half the $\Gamma N$ distance. The maximum total intensity is located around this region, as a consequence of a constructive pole superposition. 

In Fig.~\ref{fig:constrainER} we show the position of the poles $\Omega_p (q)$ (a) and the residues $2 R_p (q)$ (b) in the complex frequency plane. The poles satisfy the constraint of Eq.~\eqref{eq:pole_constraints}, which is indicated by the gray area in Fig.~\ref{fig:constrainER}(a), as imposed in the fitting. Even if no constraints are imposed on the residues, they locate in the gray area of Fig.~\ref{fig:constrainER}(b), representing a similar condition,  
$\Re [R_p (q)] > - \Im [R_p (q)] > 0$, 
except for $p=4$, which is slightly outside in the limit $q \to 0$ [See the parameters of the MPA($\q$) representation in Table~\ref{tab:poles_Ca_Ni}]. This result indicates the constructive character of the pole superposition of Ca.

\begin{figure*}[hbt!]
    \centering
    \includegraphics[width=0.99\textwidth]{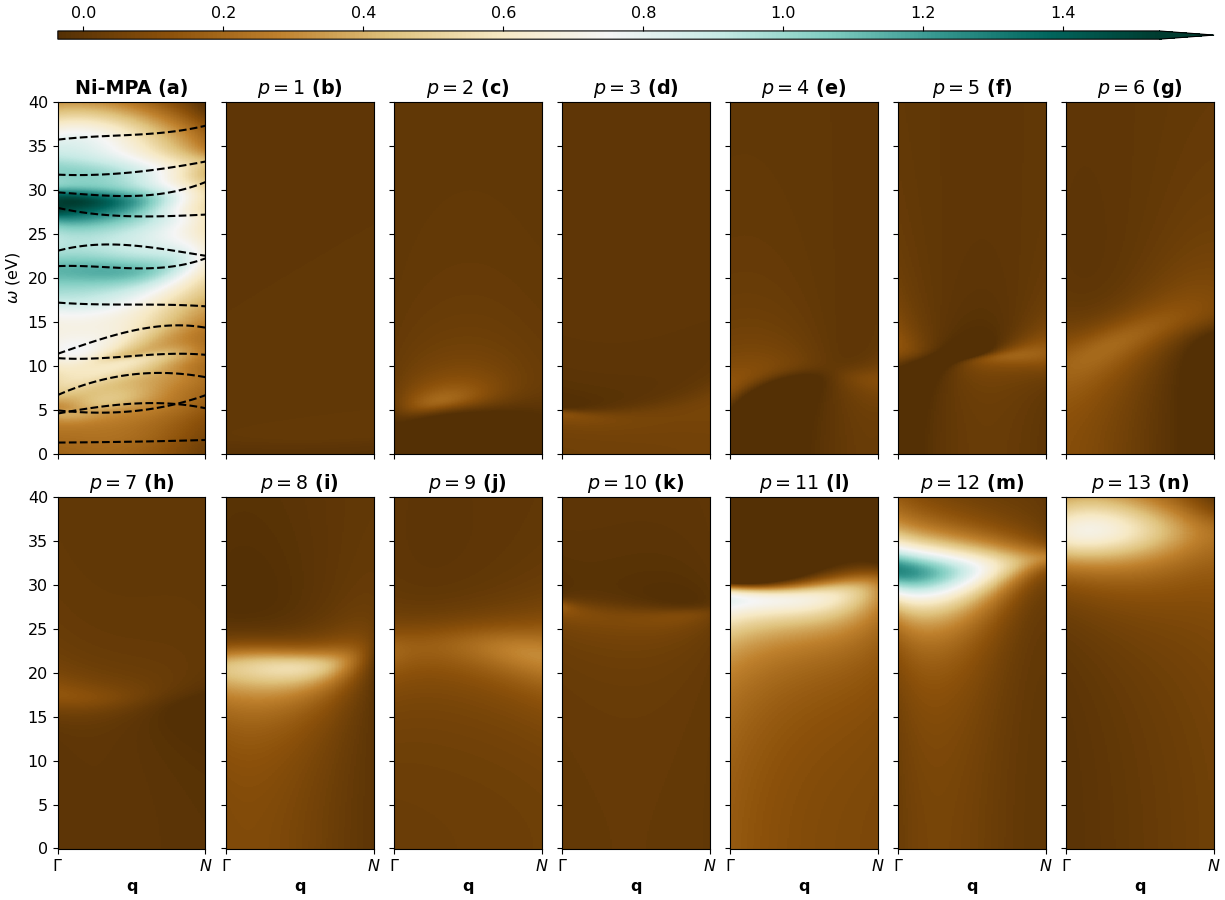}
    \caption{Spectral $Y (\q, \omega)$ band structure of Ni reconstructed with MPA($\q$) in Eqs.~\eqref{eq:mpa_model} and \eqref{eq:mpa_taylor} with a number of $n_Y = 13$ poles. The total spectral function along the $\Gamma N$ $\q$-path is shown in (a), while the individual contribution of each pole is plotted in (b-n) according to their energy position at $\Gamma$. The dashed lines in (a) indicate the energy dispersion of $\Re [\Omega_p]$.}
    \label{fig:qeel_mpa_Ni}
\end{figure*}

\begin{table*}[hbt!]
  \begin{ruledtabular}
    \begin{tabular}{cccccccc} 
      \\[-8pt]
       $\Omega_p~$(eV) & $\Omega'_{p}$ & $\Omega''_{p}$ & $\Omega'''_{p}$  & $R_p~$(eV) & $R'_{p}$ & $R''_{p}$ & $R'''_{p}$ \\
      \hline\\[-8pt]     
      \multicolumn{8}{c}{{Ca} ($n_Y=6$)} \\[4pt]
      $2.16 - 1.71 i$ & $0.48 - 0.14 i$ & $-0.50 + 0.23 i$ & $-0.40 + 0.12 i$ & $0.10 - 0.00 i$ & $0.96 - 0.10 i$ & $0.58 + 0.03 i$ & $0.24 + 0.12 i$ \\
$3.75 - 0.47 i$ & $0.19 - 0.51 i$ & $0.40 + 0.55 i$ & $-0.18 - 0.31 i$ & $0.29 - 0.00 i$ & $1.87 - 1.33 i$ & $0.65 - 0.38 i$ & $-0.50 + 0.20 i$ \\
$4.86 - 1.69 i$ & $0.65 + 1.04 i$ & $0.62 - 0.21 i$ & $0.02 - 0.52 i$ & $0.23 - 0.00 i$ & $1.69 + 0.38 i$ & $0.77 - 0.57 i$ & $0.07 - 0.46 i$ \\
$7.51 - 1.15 i$ & $0.37 - 0.20 i$ & $1.71 + 1.96 i$ & $-0.85 - 1.44 i$ & $0.38 - 0.41 i$ & $0.86 + 1.19 i$ & $0.12 + 1.13 i$ & $-0.31 + 0.55 i$ \\
$8.29 - 0.78 i$ & $0.07 - 0.17 i$ & $1.12 + 1.75 i$ & $-0.49 - 1.69 i$ & $2.09 - 0.13 i$ & $-1.70 + 0.08 i$ & $-0.63 + 0.04 i$ & $0.33 - 0.12 i$ \\
$8.98 - 0.59 i$ & $0.02 + 0.06 i$ & $0.13 - 0.76 i$ & $0.04 + 0.60 i$ & $0.63 - 0.04 i$ & $0.06 + 0.67 i$ & $-0.40 - 0.72 i$ & $-0.37 - 0.72 i$ \\[2pt]
\multicolumn{8}{c}{Ni ($n_Y=13$)} \\[4pt]
$1.25 - 1.25 i$ & $0.17 + 0.03 i$ & $0.21 - 0.02 i$ & $0.09 - 0.01 i$ & $0.00 - 0.02 i$ & $1.39 + 1.40 i$ & $0.50 - 0.25 i$ & $-0.07 - 0.54 i$ \\
$4.55 - 2.09 i$ & $0.39 + 2.04 i$ & $1.21 - 3.38 i$ & $-1.09 + 0.81 i$ & $0.03 - 0.16 i$ & $18.07 + 14.32 i$ & $-27.23 - 29.25 i$ & $11.13 + 14.58 i$ \\
$4.90 - 0.62 i$ & $-0.63 - 0.55 i$ & $1.12 - 0.39 i$ & $0.31 + 0.14 i$ & $0.03 + 0.10 i$ & $-0.21 - 0.77 i$ & $1.00 + 1.50 i$ & $0.65 - 1.61 i$ \\
$6.66 - 2.47 i$ & $1.73 + 1.36 i$ & $-1.54 - 0.52 i$ & $0.28 - 0.48 i$ & $0.28 - 0.42 i$ & $-1.71 - 6.13 i$ & $-4.76 + 13.35 i$ & $4.63 - 6.51 i$ \\
$10.85 - 2.35 i$ & $-0.51 + 1.33 i$ & $1.25 - 1.49 i$ & $-0.69 + 0.22 i$ & $0.20 - 0.50 i$ & $-10.73 - 4.15 i$ & $13.80 + 11.53 i$ & $-4.22 - 6.32 i$ \\
$11.35 - 4.07 i$ & $0.66 + 1.06 i$ & $0.30 - 0.81 i$ & $-0.56 + 0.08 i$ & $0.56 + 0.48 i$ & $-0.11 + 0.45 i$ & $-3.47 - 5.61 i$ & $0.62 + 3.78 i$ \\
$17.17 - 2.08 i$ & $-0.16 + 0.10 i$ & $0.30 - 0.07 i$ & $-0.18 - 0.08 i$ & $0.27 - 0.13 i$ & $-2.46 + 1.61 i$ & $0.48 - 5.19 i$ & $-0.43 + 2.42 i$ \\
$21.32 - 2.64 i$ & $0.07 - 0.04 i$ & $-0.51 + 0.43 i$ & $0.53 - 0.40 i$ & $1.00 + 0.80 i$ & $1.80 + 0.33 i$ & $-4.36 - 3.54 i$ & $0.62 + 2.95 i$ \\
$23.06 - 3.11 i$ & $0.40 - 0.05 i$ & $-0.74 + 0.24 i$ & $0.30 - 0.23 i$ & $0.56 + 0.26 i$ & $-2.15 + 1.85 i$ & $6.04 - 4.15 i$ & $-2.55 + 1.78 i$ \\
$27.94 - 0.83 i$ & $-0.26 - 0.46 i$ & $0.29 + 0.92 i$ & $-0.09 - 0.47 i$ & $0.11 + 0.13 i$ & $0.85 + 0.94 i$ & $-0.50 + 0.95 i$ & $-1.52 - 1.69 i$ \\
$29.70 - 1.87 i$ & $-0.08 + 0.05 i$ & $-0.05 - 0.21 i$ & $0.21 + 0.10 i$ & $0.52 + 2.27 i$ & $-2.34 - 0.51 i$ & $2.58 - 2.94 i$ & $-1.21 + 3.27 i$ \\
$31.72 - 3.09 i$ & $-0.04 - 0.08 i$ & $0.16 + 0.15 i$ & $-0.04 + 0.05 i$ & $3.93 + 0.02 i$ & $0.33 + 2.39 i$ & $-5.19 - 5.05 i$ & $3.06 + 2.67 i$ \\
$35.70 - 3.95 i$ & $0.11 + 0.06 i$ & $-0.16 - 0.22 i$ & $0.14 + 0.20 i$ & $2.54 - 0.62 i$ & $1.45 + 0.91 i$ & $-5.54 - 0.51 i$ & $2.42 + 0.27 i$ \\
    \end{tabular}
  \end{ruledtabular}
\caption{Parameters of the MPA($\q$) representation of Eq.~\eqref{eq:mpa_taylor} fitted to the RPA data of Ca and Ni. These data generate the spectral $Y(\q, \omega)$ bands structures of Figs.~\ref{fig:qeel_mpa_Ca} and \ref{fig:qeel_mpa_Ni}. Analogous MPA($\q$) representations for the 26 studied metals are provided in \suppinfo.
}
    \label{tab:poles_Ca_Ni}
\end{table*}

In Fig.~\ref{fig:qeel_mpa_Ni} we show the MPA($\q$) spectral band structure of Ni, in this case modeled with 13 poles. 
All the parameters of the MPA($\q$) representation can be found in Table~\ref{tab:poles_Ca_Ni}.
As for Ca, the most intense pole ($p = 12$) is well defined, with the intensity spreading over a larger energy range. However, many poles, especially in the low-energy region ($p=2$-$6$)  around $10$~eV, shown in the zoomed panel of Fig.~\ref{fig:qeel_IV_zoom}(d), exhibit residues that do not always comply with the relation $\Re [R_p (q)] > - \Im [R_p (q)] > 0$. This could, in principle, be imposed in the fitting, as constraints on the residues $R_p$, similar to the ones imposed on the poles $\Omega_p$ through Eq.~\eqref{eq:pole_constraints}. However, such an approach sacrifices the accuracy of the MPA($\q$) fit with respect to the numerical data, representing a considerable increase of the fitting error for Ni and many other metals. 
The fit with unconstrained residues allows for incoherent pole superposition and, therefore, underlines the complexity of plasmonic quasi-particle formation in non-homogeneous electronic materials like Ni.

\section{Conclusions}
\label{sec:con}
A comprehensive study of the frequency- and momentum-dependent inverse dielectric function of 26 elemental metals, computed from first principles at the RPA level of 
theory have been presented. 
The RPA results are found to accurately
reproduce available experimental measurements in the optical limit and were used to evaluate plasma, intra-, and inter-band frequencies, construct spectral band structures of the inverse dielectric function, and develop effective analytical representations of plasmonic excitations, using frequency- and momentum- dependent multipole-Pad\'e approximants, MPA($\q$). 

Plasmonic excitations in inhomogeneous metals were found to exhibit multipole structures, in contrast to the simple free-electron gas assumptions, which arise due to the mixed orbital character of the valence states. 
The energy position of the main MPA pole in the optical limit, $\q \to 0$, is used to compute an effective number of electrons that can be used within the free-electron gas model. In turn, the spectral weight of such pole gives a measure of the unsuitability of the commonly used (single-pole) PPA description. The analysis across all the 26 elemental metals studied, provides a general picture of the limited applicability of PPA. 

Moreover, the plasmonic dispersions are much flatter than what would be expected for the corresponding free-electron gas model with an equivalent electronic density or plasma energy. This is particularly the case when the spectral weight is distributed among several poles. 
Using the MPA($\q$) scheme introduced here, which is not restricted to free-electron or single-pole descriptions, we have accurately fitted effective MPA($\q$) models to the RPA numerical data with a relatively small number of poles, ranging from 2 to 16 for all the studied metals. The obtained MPA($\q$) representations are able to describe the 
complex features present in the spectral band structures, including non-parabolic energy and intensity dispersions and constructive and destructive quasiparticle overlapping, leading to band crossings and anti-crossings. 
The reported MPA($\q$) models can be used as a first-principle alternative to over-simplified analytical models such as the Drude/Lindhard dielectric function, or even as fitting schemes to experimental data. Although the present work focuses on elemental metals at the RPA level, the MPA($\q$) framework is general in the sense that it is constructed directly from the frequency- and momentum-dependent dielectric response. As such, it can in principle be extended to other classes of materials and levels of theory. The MPA($\q$) models constructed from RPA data, for example, could be used as a key building block to reduce the computational cost of $GW$/BSE calculations, and as a starting point towards $GW$/BSE performed on a $\kk$/$\q$-path, thereby significantly reducing the overall computational cost of such calculations.

\section*{Acknowledgments}
%
Work in Norway is supported by the Research Council of Norway in the MORTY project (664350).
The computations of this work were carried out on UNINETT Sigma2 high performance computing resources (Grant No. NN9650K), and through Project No. EHPC-EXT-2022E01-022 Joint Undertaking, which granted access to Leonardo-Booster@Cineca, Bologna, Italy, provided by EuroHPC. 

\section*{Data Availability}
The data generated in this article are openly available on Ref.~\cite{MPAq_data}.

%
%
\bibliography{biblio}

\end{document}


\title{Bulk plasmons in elemental metals:  Supplemental materials}

\author{Dario A. Leon}
\email{dario.alejandro.leon.valido@nmbu.no}
\affiliation{
 Department of Mechanical Engineering and Technology Management, \\ Norwegian University of Life Sciences, NO-1432 Ås, Norway
}%

\author{Claudia Cardoso}
\affiliation{
 S3 Centre, Istituto Nanoscienze, CNR, 41125 Modena, Italy
}

\author{Kristian Berland}
\affiliation{
 Department of Mechanical Engineering and Technology Management, \\ Norwegian University of Life Sciences, NO-1432 Ås, Norway
}

\date{\today}

\maketitle

In the following sections, we provide the inverse dielectric function computed in the RPA approximation for each of the 26 metals studied in this work. The RPA spectra in the optical limit $\q \to 0$ are compared with available optical measurements (see e.g. the compilation made in Ref.~\cite{Polyanskiy2024SciData}), EELS~\cite{eels_atlas}, and REELS and IPA calculations~\cite{Werner2009JPCRD}. Spectral band structures are constructed from the numerical RPA data within a frequency interval of $100$~eV, and in a $\q$-path corresponding to $H \Gamma N$, $X \Gamma M$, and $H A \Gamma M|K \Gamma$, for bcc, fcc, and hcp Bravais lattices, respectively. The low-energy part of the spectra along a single high symmetry $\q$-line, exhibiting the most relevant features, has been fitted to MPA($\q$) models with a number of poles between 2 and 16, depending on the complexity of the spectra. The reconstructed spectral band structures with MPA($\q$) are shown, while all the parameters of the model are reported in a dedicated Table.
\newpage

\subsection{Li} \FloatBarrier \begin{figure*}[hbt!]
    \centering
    \includegraphics[width=0.266\textwidth]{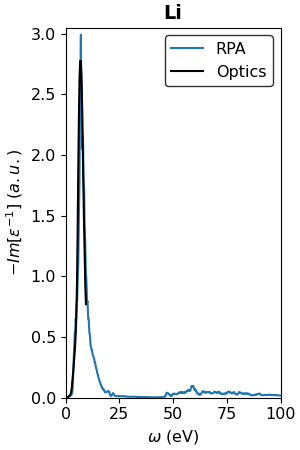}
    \includegraphics[width=0.2\textwidth]{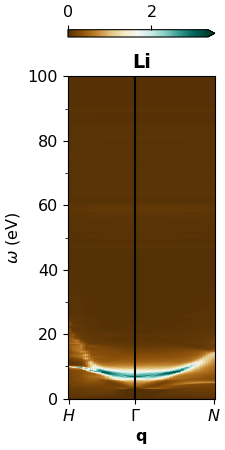}
    \includegraphics[width=0.465\textwidth]{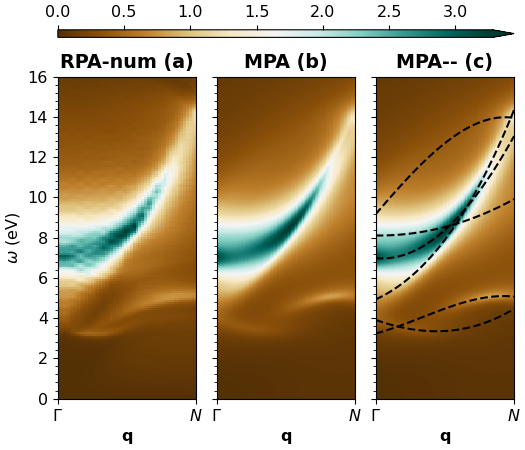}
    \caption{(left panel) Comparison of the computed RPA loss function of Li in the optical limit with optical measurements. (middle panel) Spectral $Y (\omega, \q)$ band structure of Li computed from the numerical RPA results along the $H \Gamma N$ $\q$-path. (right panels) Spectral $Y (\q, \omega)$ band structure of Li in the $\Gamma N$ direction reconstructed with MPA($\q$) with a number of $n_Y = 6$ poles.}
    \label{fig:Li}
\end{figure*} 

\begin{table*}[hbt!]
  \begin{ruledtabular}
    \begin{tabular}{cccccccc} 
      \\[-8pt]
       $\Omega_p~$(eV) & $\Omega'_{p}$ & $\Omega''_{p}$ & $\Omega'''_{p}$  & $R_p~$(eV) & $R'_{p}$ & $R''_{p}$ & $R'''_{p}$ \\
      \hline\\[-8pt]     
      $3.24 - 2.02 i$ & $-0.12 + 2.00 i$ & $4.87 + 2.43 i$ & $-17.38 - 33.33 i$ & $0.10 - 0.00 i$ & $0.72 - 0.35 i$ & $2.87 - 3.83 i$ & $-4.50 - 10.23 i$ \\
      $3.91 - 0.55 i$ & $-1.32 + 0.47 i$ & $5.33 - 3.48 i$ & $9.08 - 9.65 i$ & $0.10 - 0.02 i$ & $-1.68 + 0.11 i$ & $2.81 - 3.05 i$ & $13.94 - 11.92 i$ \\
      $4.93 - 1.26 i$ & $0.98 + 1.16 i$ & $8.00 - 0.54 i$ & $12.82 + 4.48 i$ & $0.38 - 0.37 i$ & $0.05 - 0.07 i$ & $-6.16 + 0.97 i$ & $-6.41 + 1.71 i$ \\
      $6.99 - 0.65 i$ & $-0.15 - 0.50 i$ & $7.53 + 6.80 i$ & $0.28 - 31.54 i$ & $1.43 - 0.10 i$ & $-0.52 + 0.46 i$ & $-1.60 - 2.43 i$ & $11.96 + 6.21 i$ \\
      $8.11 - 1.06 i$ & $-0.04 + 0.15 i$ & $2.42 - 4.98 i$ & $-2.15 + 20.38 i$ & $1.13 - 0.00 i$ & $0.93 + 0.01 i$ & $-2.61 - 0.05 i$ & $-14.20 + 0.13 i$ \\
      $9.17 - 1.51 i$ & $1.82 - 0.41 i$ & $-1.18 + 3.59 i$ & $-13.16 - 2.04 i$ & $0.42 - 0.00 i$ & $-1.15 + 0.02 i$ & $-0.63 - 0.18 i$ & $5.39 + 0.59 i$ \\
    \end{tabular}
  \end{ruledtabular}
\caption{List of poles and residues of the MPA($\q$) representation of Li.
}
    \label{tab:poles_Li}
\end{table*} \newpage

\subsection{Be} \FloatBarrier \begin{figure*}[hbt!]
    \centering
    \includegraphics[width=0.250\textwidth]{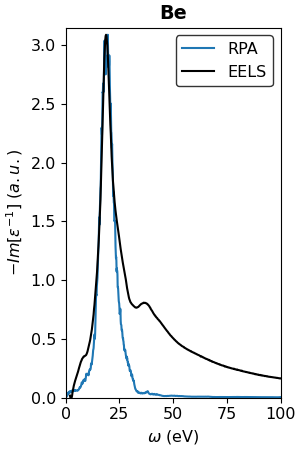}
    \includegraphics[width=0.333\textwidth]{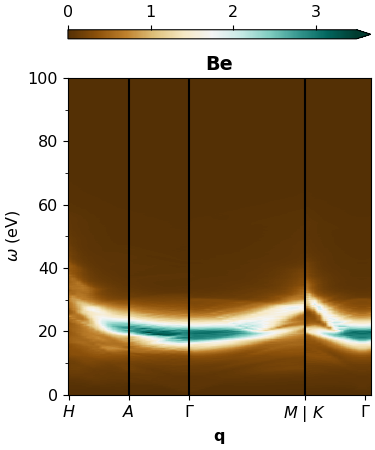}
    \includegraphics[width=0.406\textwidth]{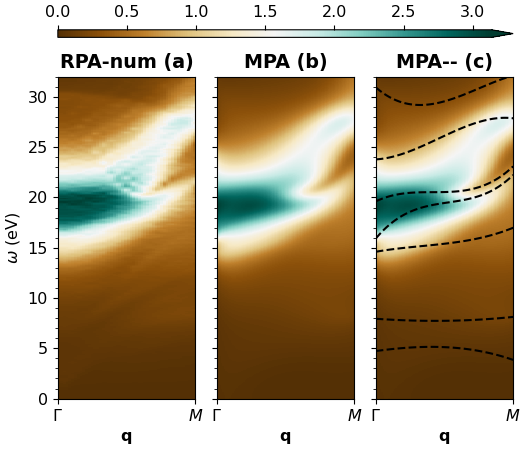}
    \caption{(left panel) Comparison of the computed RPA loss function of Be in the optical limit with EELS measurements. (middle panel) Spectral $Y (\omega, \q)$ band structure of Be computed from the numerical RPA results along the $H A \Gamma M|K \Gamma$ $\q$-path. (right panels) Spectral $Y (\q, \omega)$ band structure of Be in the $\Gamma M$ direction reconstructed with MPA($\q$) with a number of $n_Y = 7$ poles.}
    \label{fig:Be}
\end{figure*} 

\begin{table*}[hbt!]
  \begin{ruledtabular}
    \begin{tabular}{cccccccc} 
      \\[-8pt]
       $\Omega_p~$(eV) & $\Omega'_{p}$ & $\Omega''_{p}$ & $\Omega'''_{p}$  & $R_p~$(eV) & $R'_{p}$ & $R''_{p}$ & $R'''_{p}$ \\
      \hline\\[-8pt]     
      $4.72 - 5.04 i$ & $-1.51 + 2.16 i$ & $1.80 - 2.34 i$ & $-0.72 + 0.23 i$ & $0.29 + 0.20 i$ & $-3.78 - 3.22 i$ & $-0.06 + 2.77 i$ & $0.73 + 0.16 i$ \\
$7.94 - 7.29 i$ & $-0.93 + 0.76 i$ & $-0.14 + 0.42 i$ & $0.31 - 0.31 i$ & $-0.15 - 0.79 i$ & $-1.95 - 1.98 i$ & $-1.47 + 6.06 i$ & $1.45 - 3.91 i$ \\
$14.60 - 1.43 i$ & $0.36 - 0.32 i$ & $-0.59 + 1.51 i$ & $0.59 - 1.52 i$ & $1.22 + 0.29 i$ & $-5.88 - 6.52 i$ & $2.36 + 16.71 i$ & $3.25 - 9.84 i$ \\
$15.90 - 1.40 i$ & $2.13 - 0.64 i$ & $-3.59 + 1.66 i$ & $2.27 - 1.10 i$ & $0.05 - 2.59 i$ & $-5.25 + 17.05 i$ & $2.32 - 14.98 i$ & $0.83 + 0.24 i$ \\
$19.60 - 2.33 i$ & $0.67 + 0.04 i$ & $-1.69 + 0.71 i$ & $1.38 - 0.76 i$ & $4.71 + 1.42 i$ & $-4.10 + 0.02 i$ & $-6.07 - 4.95 i$ & $7.52 + 5.73 i$ \\
$23.79 - 2.70 i$ & $0.19 - 1.10 i$ & $0.75 + 2.59 i$ & $-0.60 - 1.42 i$ & $2.10 + 1.76 i$ & $-5.48 + 5.42 i$ & $15.05 - 3.38 i$ & $-8.77 - 3.44 i$ \\
$30.96 - 4.43 i$ & $-0.69 - 0.64 i$ & $1.42 + 0.76 i$ & $-0.66 - 0.16 i$ & $3.37 + 1.04 i$ & $-1.40 + 0.15 i$ & $1.47 + 3.13 i$ & $1.50 - 4.88 i$ \\
$32.82 - 6.15 i$ & $0.79 - 0.60 i$ & $-0.61 + 0.80 i$ & $-0.04 - 0.26 i$ & $-4.71 + 2.46 i$ & $-3.64 + 3.13 i$ & $-1.73 + 2.79 i$ & $4.92 - 3.00 i$ \\
$33.04 - 4.81 i$ & $0.51 - 0.43 i$ & $-0.23 + 0.56 i$ & $0.07 - 0.13 i$ & $1.58 - 4.26 i$ & $-4.11 + 2.70 i$ & $-1.35 + 1.08 i$ & $2.90 - 2.92 i$ \\
    \end{tabular}
  \end{ruledtabular}
\caption{List of poles and residues of the MPA($\q$) representation of Be.
}
    \label{tab:poles_Be}
\end{table*} \newpage

\subsection{Na} \FloatBarrier \begin{figure*}[hbt!]
    \centering
    \includegraphics[width=0.266\textwidth]{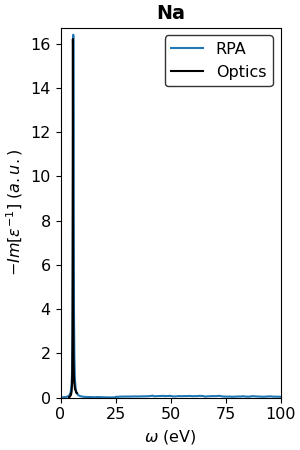}
    \includegraphics[width=0.2\textwidth]{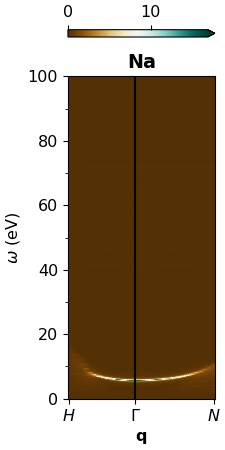}
    \includegraphics[width=0.465\textwidth]{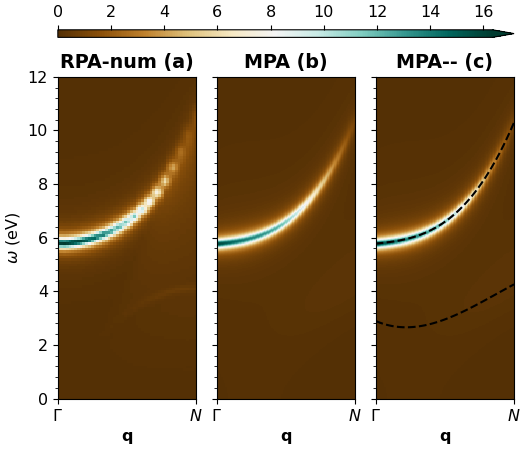}
    \caption{(left panel) Comparison of the computed RPA loss function of Na in the optical limit with optical measurements. (middle panel) Spectral $Y (\omega, \q)$ band structure of Na computed from the numerical RPA results along the $H \Gamma N$ $\q$-path. (right panels) Spectral $Y (\q, \omega)$ band structure of Na in the $\Gamma N$ direction reconstructed with MPA($\q$) with a number of $n_Y = 2$ poles.}
    \label{fig:Na}
\end{figure*} 

\begin{table*}[hbt!]
  \begin{ruledtabular}
    \begin{tabular}{cccccccc} 
      \\[-8pt]
       $\Omega_p~$(eV) & $\Omega'_{p}$ & $\Omega''_{p}$ & $\Omega'''_{p}$  & $R_p~$(eV) & $R'_{p}$ & $R''_{p}$ & $R'''_{p}$ \\
      \hline\\[-8pt]     
      $2.89 - 2.97 i$ & $-1.89 + 0.32 i$ & $2.00 + 2.00 i$ & $-0.12 - 1.39 i$ & $0.10 - 0.00 i$ & $1.65 + 0.41 i$ & $2.00 - 2.00 i$ & $2.00 - 2.00 i$ \\
$5.78 - 0.15 i$ & $0.17 - 0.07 i$ & $0.18 + 0.26 i$ & $1.22 - 0.23 i$ & $2.73 - 0.04 i$ & $0.27 - 0.07 i$ & $-2.00 + 0.31 i$ & $0.06 - 0.25 i$ \\
    \end{tabular}
  \end{ruledtabular}
\caption{List of poles and residues of the MPA($\q$) representation of Na.
}
    \label{tab:poles_Na}
\end{table*} \newpage

\subsection{Mg} \FloatBarrier \begin{figure*}[hbt!]
    \centering
    \includegraphics[width=0.250\textwidth]{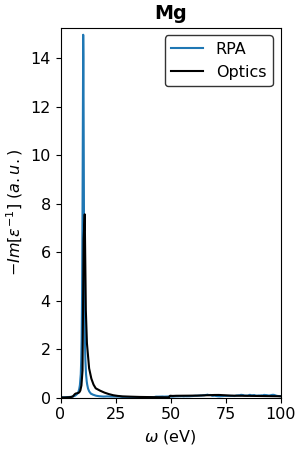}
    \includegraphics[width=0.333\textwidth]{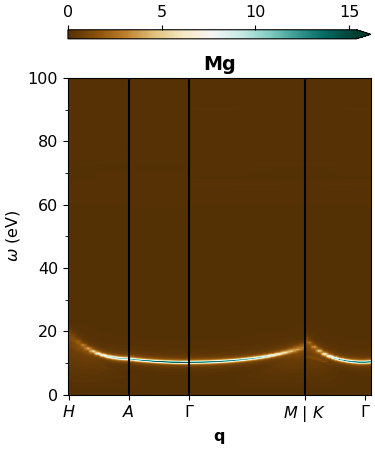}
    \includegraphics[width=0.406\textwidth]{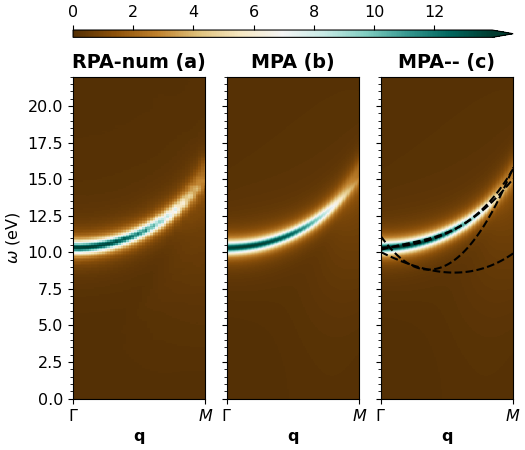}
    \caption{(left panel) Comparison of the computed RPA loss function of Mg in the optical limit with optical measurements. (middle panel) Spectral $Y (\omega, \q)$ band structure of Mg computed from the numerical RPA results along the $H A \Gamma M|K \Gamma$ $\q$-path. (right panels) Spectral $Y (\q, \omega)$ band structure of Mg in the $\Gamma M$ direction reconstructed with MPA($\q$) with a number of $n_Y = 4$ poles.}
    \label{fig:Mg}
\end{figure*} 

\begin{table*}[hbt!]
  \begin{ruledtabular}
    \begin{tabular}{cccccccc} 
      \\[-8pt]
       $\Omega_p~$(eV) & $\Omega'_{p}$ & $\Omega''_{p}$ & $\Omega'''_{p}$  & $R_p~$(eV) & $R'_{p}$ & $R''_{p}$ & $R'''_{p}$ \\
      \hline\\[-8pt]     
      $10.02 - 9.81 i$ & $-0.67 - 0.21 i$ & $-0.12 + 0.58 i$ & $0.42 - 0.01 i$ & $-0.28 - 0.34 i$ & $-2.93 + 9.59 i$ & $1.33 + 1.70 i$ & $0.50 + 0.18 i$ \\
$10.17 - 0.77 i$ & $0.49 + 0.55 i$ & $-0.77 - 0.89 i$ & $1.36 + 0.37 i$ & $-0.71 + 0.36 i$ & $-8.25 + 0.41 i$ & $1.29 - 2.61 i$ & $-1.71 + 0.12 i$ \\
$10.32 - 0.38 i$ & $0.07 + 0.06 i$ & $0.42 - 0.10 i$ & $0.43 + 0.08 i$ & $5.56 - 0.40 i$ & $-1.12 + 0.34 i$ & $-0.72 + 0.02 i$ & $-0.05 - 0.22 i$ \\
$11.11 - 7.95 i$ & $-2.86 + 0.80 i$ & $3.38 - 0.27 i$ & $-0.53 + 0.67 i$ & $0.24 - 0.12 i$ & $-8.59 + 0.50 i$ & $-1.28 + 1.68 i$ & $-0.60 + 1.99 i$ \\
    \end{tabular}
  \end{ruledtabular}
\caption{List of poles and residues of the MPA($\q$) representation of Mg.
}
    \label{tab:poles_Mg}
\end{table*} \newpage

\subsection{Al} \FloatBarrier \begin{figure*}[hbt!]
    \centering
    \includegraphics[width=0.266\textwidth]{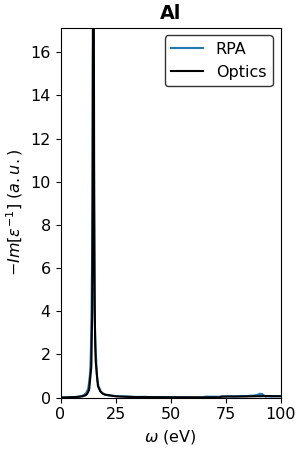}
    \includegraphics[width=0.2\textwidth]{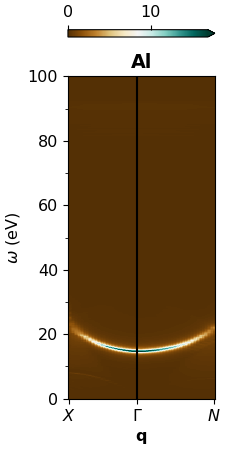}
    \includegraphics[width=0.465\textwidth]{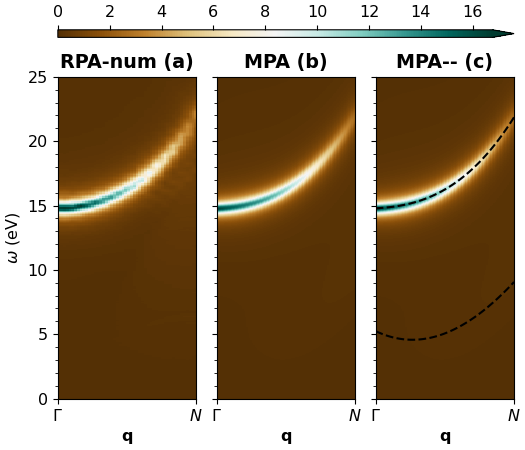}
    \caption{(left panel) Comparison of the computed RPA loss function of Al in the optical limit with optical measurements. (middle panel) Spectral $Y (\omega, \q)$ band structure of Al computed from the numerical RPA results along the $X \Gamma N$ $\q$-path. (right panels) Spectral $Y (\q, \omega)$ band structure of Al in the $\Gamma N$ direction reconstructed with MPA($\q$) with a number of $n_Y = 2$ poles.}
    \label{fig:Al}
\end{figure*} 

\begin{table*}[hbt!]
  \begin{ruledtabular}
    \begin{tabular}{cccccccc} 
      \\[-8pt]
       $\Omega_p~$(eV) & $\Omega'_{p}$ & $\Omega''_{p}$ & $\Omega'''_{p}$  & $R_p~$(eV) & $R'_{p}$ & $R''_{p}$ & $R'''_{p}$ \\
      \hline\\[-8pt]     
      $5.24 - 4.99 i$ & $-1.41 - 0.59 i$ & $2.00 + 2.00 i$ & $0.06 - 0.57 i$ & $0.17 - 0.00 i$ & $1.19 - 0.88 i$ & $2.00 - 2.00 i$ & $2.00 - 2.00 i$ \\
$14.79 - 0.38 i$ & $0.07 - 0.06 i$ & $0.42 + 0.15 i$ & $0.47 - 0.15 i$ & $6.83 - 0.16 i$ & $0.33 - 0.09 i$ & $-1.60 + 0.31 i$ & $0.14 - 0.20 i$ \\
    \end{tabular}
  \end{ruledtabular}
\caption{List of poles and residues of the MPA($\q$) representation of Al.
}
    \label{tab:poles_Al}
\end{table*} \newpage

\subsection{K} \FloatBarrier \begin{figure*}[hbt!]
    \centering
    \includegraphics[width=0.266\textwidth]{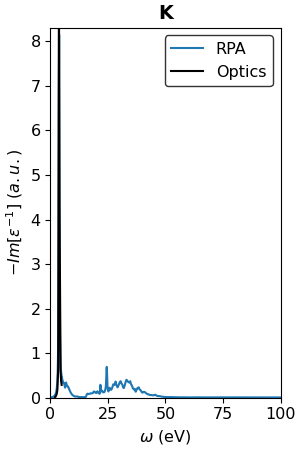}
    \includegraphics[width=0.2\textwidth]{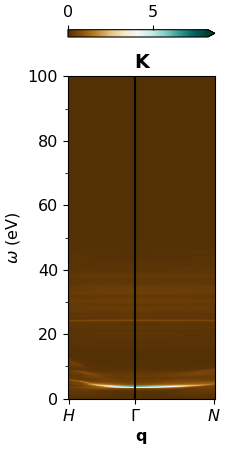}
    \includegraphics[width=0.465\textwidth]{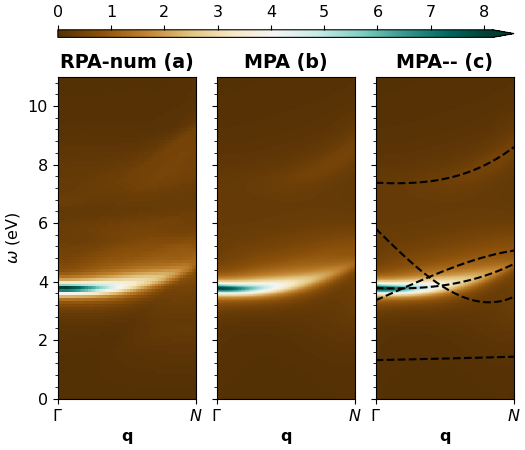}
    \caption{(left panel) Comparison of the computed RPA loss function of K in the optical limit with optical measurements. (middle panel) Spectral $Y (\omega, \q)$ band structure of K computed from the numerical RPA results along the $H \Gamma N$ $\q$-path. (right panels) Spectral $Y (\q, \omega)$ band structure of K in the $\Gamma N$ direction reconstructed with MPA($\q$) with a number of $n_Y = 5$ poles.}
    \label{fig:K}
\end{figure*} 

\begin{table*}[hbt!]
  \begin{ruledtabular}
    \begin{tabular}{cccccccc} 
      \\[-8pt]
       $\Omega_p~$(eV) & $\Omega'_{p}$ & $\Omega''_{p}$ & $\Omega'''_{p}$  & $R_p~$(eV) & $R'_{p}$ & $R''_{p}$ & $R'''_{p}$ \\
      \hline\\[-8pt]     
      $1.31 - 1.33 i$ & $-0.00 + 0.15 i$ & $-0.08 + 0.11 i$ & $0.03 - 0.03 i$ & $0.10 - 0.00 i$ & $-1.70 + 0.03 i$ & $-1.43 - 0.02 i$ & $1.13 - 0.02 i$ \\
$3.36 - 0.71 i$ & $1.15 + 0.35 i$ & $0.19 - 0.06 i$ & $-0.38 + 0.03 i$ & $0.10 - 0.11 i$ & $0.82 + 0.58 i$ & $0.59 + 0.02 i$ & $0.22 + 0.03 i$ \\
$3.78 - 0.16 i$ & $-0.11 - 0.01 i$ & $0.33 - 0.43 i$ & $0.22 + 0.41 i$ & $1.38 - 0.03 i$ & $-0.53 - 0.69 i$ & $-1.03 + 0.18 i$ & $-0.11 + 0.47 i$ \\
$5.82 - 1.63 i$ & $-1.38 - 0.85 i$ & $-0.00 + 0.27 i$ & $0.62 + 0.43 i$ & $0.28 - 0.03 i$ & $1.34 + 0.82 i$ & $0.94 - 0.28 i$ & $0.37 - 0.36 i$ \\
$7.38 - 0.90 i$ & $-0.07 + 0.08 i$ & $0.10 + 0.25 i$ & $0.28 - 0.25 i$ & $0.13 - 0.00 i$ & $0.89 + 0.08 i$ & $1.16 - 0.09 i$ & $0.83 + 0.05 i$ \\
    \end{tabular}
  \end{ruledtabular}
\caption{List of poles and residues of the MPA($\q$) representation of K.
}
    \label{tab:poles_K}
\end{table*} \newpage

\subsection{Ca} \FloatBarrier \begin{figure*}[hbt!]
    \centering
    \includegraphics[width=0.266\textwidth]{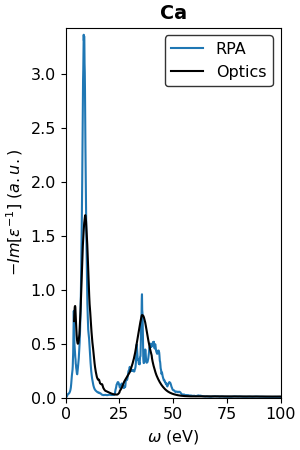}
    \includegraphics[width=0.2\textwidth]{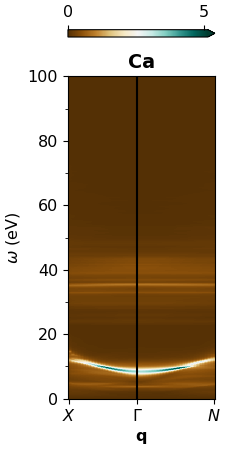}
    \includegraphics[width=0.465\textwidth]{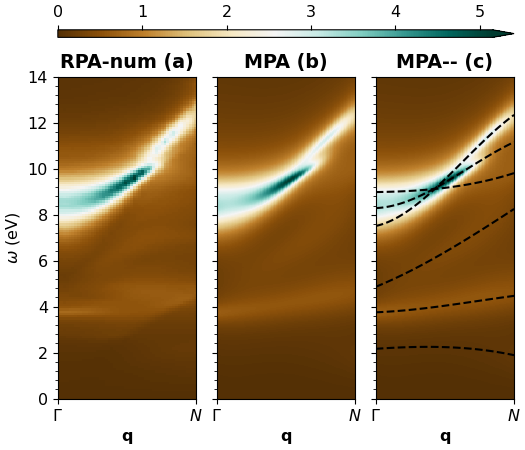}
    \caption{(left panel) Comparison of the computed RPA loss function of Ca in the optical limit with optical measurements. (middle panel) Spectral $Y (\omega, \q)$ band structure of Ca computed from the numerical RPA results along the $X \Gamma N$ $\q$-path. (right panels) Spectral $Y (\q, \omega)$ band structure of Ca in the $\Gamma N$ direction reconstructed with MPA($\q$) with a number of $n_Y = 6$ poles.}
    \label{fig:Ca}
\end{figure*} 

\begin{table*}[hbt!]
  \begin{ruledtabular}
    \begin{tabular}{cccccccc} 
      \\[-8pt]
       $\Omega_p~$(eV) & $\Omega'_{p}$ & $\Omega''_{p}$ & $\Omega'''_{p}$  & $R_p~$(eV) & $R'_{p}$ & $R''_{p}$ & $R'''_{p}$ \\
      \hline\\[-8pt]     
      $2.16 - 1.71 i$ & $0.48 - 0.14 i$ & $-0.50 + 0.23 i$ & $-0.40 + 0.12 i$ & $0.10 - 0.00 i$ & $0.96 - 0.10 i$ & $0.58 + 0.03 i$ & $0.24 + 0.12 i$ \\
$3.75 - 0.47 i$ & $0.19 - 0.51 i$ & $0.40 + 0.55 i$ & $-0.18 - 0.31 i$ & $0.29 - 0.00 i$ & $1.87 - 1.33 i$ & $0.65 - 0.38 i$ & $-0.50 + 0.20 i$ \\
$4.86 - 1.69 i$ & $0.65 + 1.04 i$ & $0.62 - 0.21 i$ & $0.02 - 0.52 i$ & $0.23 - 0.00 i$ & $1.69 + 0.38 i$ & $0.77 - 0.57 i$ & $0.07 - 0.46 i$ \\
$7.51 - 1.15 i$ & $0.37 - 0.20 i$ & $1.71 + 1.96 i$ & $-0.85 - 1.44 i$ & $0.38 - 0.41 i$ & $0.86 + 1.19 i$ & $0.12 + 1.13 i$ & $-0.31 + 0.55 i$ \\
$8.29 - 0.78 i$ & $0.07 - 0.17 i$ & $1.12 + 1.75 i$ & $-0.49 - 1.69 i$ & $2.09 - 0.13 i$ & $-1.70 + 0.08 i$ & $-0.63 + 0.04 i$ & $0.33 - 0.12 i$ \\
$8.98 - 0.59 i$ & $0.02 + 0.06 i$ & $0.13 - 0.76 i$ & $0.04 + 0.60 i$ & $0.63 - 0.04 i$ & $0.06 + 0.67 i$ & $-0.40 - 0.72 i$ & $-0.37 - 0.72 i$ \\
    \end{tabular}
  \end{ruledtabular}
\caption{List of poles and residues of the MPA($\q$) representation of Ca.
}
    \label{tab:poles_Ca}
\end{table*} \newpage

\subsection{Ti} \FloatBarrier \begin{figure*}[hbt!]
    \centering
    \includegraphics[width=0.250\textwidth]{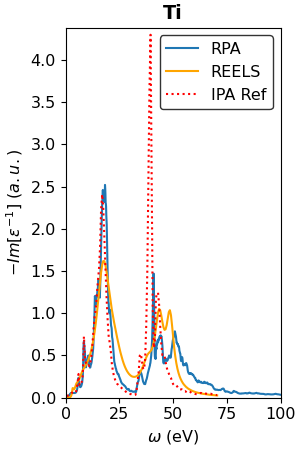}
    \includegraphics[width=0.333\textwidth]{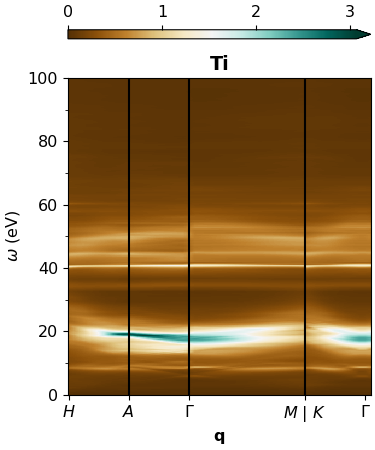}
    \includegraphics[width=0.406\textwidth]{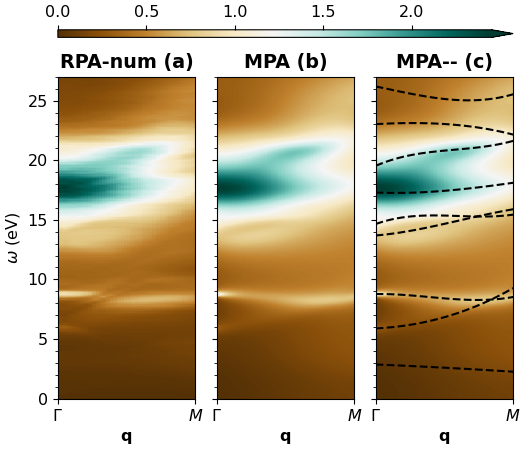}
    \caption{(left panel) Comparison of the computed RPA loss function of Ti in the optical limit with REELS measurements and reference IPA results. (middle panel) Spectral $Y (\omega, \q)$ band structure of Ti computed from the numerical RPA results along the $H A \Gamma M|K \Gamma$ $\q$-path. (right panels) Spectral $Y (\q, \omega)$ band structure of Ti in the $\Gamma M$ direction reconstructed with MPA($\q$) with a number of $n_Y = 9$ poles.}
    \label{fig:Ti}
\end{figure*} 

\begin{table*}[hbt!]
  \begin{ruledtabular}
    \begin{tabular}{cccccccc} 
      \\[-8pt]
       $\Omega_p~$(eV) & $\Omega'_{p}$ & $\Omega''_{p}$ & $\Omega'''_{p}$  & $R_p~$(eV) & $R'_{p}$ & $R''_{p}$ & $R'''_{p}$ \\
      \hline\\[-8pt]     
      $2.86 - 2.87 i$ & $-0.42 + 0.09 i$ & $-0.07 - 0.03 i$ & $0.06 - 0.06 i$ & $0.02 - 0.10 i$ & $1.68 - 2.64 i$ & $0.88 + 0.58 i$ & $0.28 + 1.37 i$ \\
$5.89 - 0.39 i$ & $0.25 - 0.89 i$ & $0.76 + 1.02 i$ & $0.15 - 0.22 i$ & $0.10 - 0.01 i$ & $-2.61 - 7.46 i$ & $-1.49 + 6.02 i$ & $1.59 - 1.19 i$ \\
$8.77 - 0.12 i$ & $0.06 - 0.62 i$ & $-0.76 + 0.89 i$ & $0.65 - 0.38 i$ & $0.22 - 0.06 i$ & $-1.32 + 0.14 i$ & $4.30 + 4.49 i$ & $-2.08 - 4.00 i$ \\
$13.68 - 1.11 i$ & $0.12 + 0.26 i$ & $0.44 - 0.50 i$ & $-0.22 + 0.09 i$ & $0.08 + 0.35 i$ & $4.56 - 1.87 i$ & $-9.18 + 11.98 i$ & $6.65 - 8.84 i$ \\
$14.64 - 1.11 i$ & $0.57 + 0.02 i$ & $-1.08 + 0.55 i$ & $0.63 - 0.70 i$ & $0.68 - 0.01 i$ & $-3.55 + 0.81 i$ & $3.98 - 2.74 i$ & $-2.74 + 0.83 i$ \\
$17.29 - 2.01 i$ & $-0.06 + 0.03 i$ & $0.20 + 0.02 i$ & $-0.05 - 0.01 i$ & $3.95 - 1.78 i$ & $-1.21 + 0.45 i$ & $-1.50 + 0.85 i$ & $1.24 - 0.85 i$ \\
$19.52 - 4.10 i$ & $0.27 + 1.18 i$ & $-0.45 - 1.28 i$ & $0.33 + 0.41 i$ & $0.96 + 0.60 i$ & $1.17 + 1.63 i$ & $-4.13 - 7.01 i$ & $1.46 + 5.57 i$ \\
$23.03 - 3.74 i$ & $0.08 - 0.17 i$ & $-0.17 + 0.58 i$ & $-0.01 - 0.24 i$ & $-1.32 + 2.22 i$ & $-0.63 - 1.78 i$ & $-2.19 + 0.22 i$ & $0.74 + 0.89 i$ \\
$26.19 - 3.33 i$ & $-0.15 - 0.28 i$ & $-0.11 + 0.52 i$ & $0.21 - 0.24 i$ & $1.52 + 0.30 i$ & $2.45 - 2.21 i$ & $-3.52 + 2.19 i$ & $1.40 - 0.39 i$ \\
$31.50 - 5.00 i$ & $-0.08 - 0.05 i$ & $0.14 - 0.16 i$ & $0.11 - 0.02 i$ & $2.56 - 0.25 i$ & $0.84 - 1.31 i$ & $-0.21 + 5.23 i$ & $-1.25 - 0.94 i$ \\
    \end{tabular}
  \end{ruledtabular}
\caption{List of poles and residues of the MPA($\q$) representation of Ti.
}
    \label{tab:poles_Ti}
\end{table*} \newpage

\subsection{V} \FloatBarrier \begin{figure*}[hbt!]
    \centering
    \includegraphics[width=0.266\textwidth]{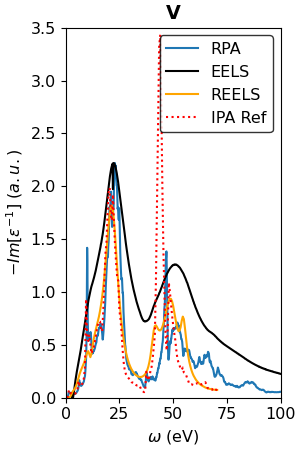}
    \includegraphics[width=0.2\textwidth]{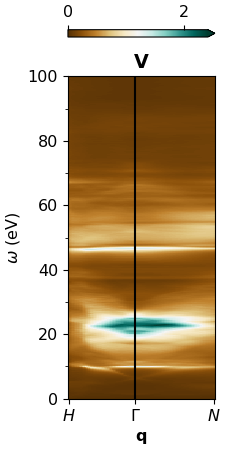}
    \includegraphics[width=0.465\textwidth]{spectral_V_k36_ff_vs_mpa.png}
    \caption{(left panel) Comparison of the computed RPA loss function of V in the optical limit with EELS, REELS measurements, and reference IPA results. (middle panel) Spectral $Y (\omega, \q)$ band structure of V computed from the numerical RPA results along the $H \Gamma N$ $\q$-path. (right panels) Spectral $Y (\q, \omega)$ band structure of V in the $\Gamma N$ direction reconstructed with MPA($\q$) with a number of $n_Y = 13$ poles.}
    \label{fig:V}
\end{figure*} 

\begin{table*}[hbt!]
  \begin{ruledtabular}
    \begin{tabular}{cccccccc} 
      \\[-8pt]
       $\Omega_p~$(eV) & $\Omega'_{p}$ & $\Omega''_{p}$ & $\Omega'''_{p}$  & $R_p~$(eV) & $R'_{p}$ & $R''_{p}$ & $R'''_{p}$ \\
      \hline\\[-8pt]     
      $1.23 - 1.04 i$ & $0.32 + 0.10 i$ & $0.12 - 0.11 i$ & $-0.56 - 0.14 i$ & $0.07 - 0.00 i$ & $-1.04 - 0.04 i$ & $-0.61 + 0.01 i$ & $-0.36 + 0.01 i$ \\
$3.02 - 0.98 i$ & $0.22 - 0.23 i$ & $0.00 + 0.13 i$ & $0.12 + 0.33 i$ & $0.07 - 0.00 i$ & $-1.67 - 0.11 i$ & $-0.57 - 0.13 i$ & $0.75 - 0.10 i$ \\
$5.37 - 1.16 i$ & $0.19 - 0.58 i$ & $-0.03 + 0.01 i$ & $0.13 + 0.37 i$ & $0.07 - 0.00 i$ & $-1.82 - 0.01 i$ & $-0.82 - 0.06 i$ & $1.02 - 0.05 i$ \\
$7.20 - 1.17 i$ & $0.61 + 0.10 i$ & $0.20 - 0.01 i$ & $0.34 + 0.23 i$ & $0.17 - 0.00 i$ & $-1.31 - 0.22 i$ & $-0.19 - 0.08 i$ & $0.07 - 0.02 i$ \\
$10.01 - 0.19 i$ & $0.03 - 0.01 i$ & $-0.11 - 0.25 i$ & $-0.01 + 0.15 i$ & $0.23 - 0.01 i$ & $-0.70 + 0.13 i$ & $-0.36 - 0.05 i$ & $0.55 - 0.06 i$ \\
$11.30 - 1.32 i$ & $0.00 - 0.23 i$ & $-0.05 - 0.08 i$ & $-0.21 + 0.45 i$ & $0.74 - 0.03 i$ & $-1.13 - 0.10 i$ & $-0.51 - 0.04 i$ & $-0.08 - 0.01 i$ \\
$13.83 - 1.36 i$ & $0.40 - 0.57 i$ & $0.30 - 0.28 i$ & $0.39 + 0.26 i$ & $0.35 - 0.00 i$ & $-0.92 - 0.13 i$ & $-0.50 - 0.12 i$ & $0.51 - 0.06 i$ \\
$15.50 - 1.16 i$ & $0.13 - 0.45 i$ & $0.29 - 0.05 i$ & $0.25 + 0.36 i$ & $0.59 - 0.00 i$ & $-0.22 - 0.31 i$ & $-0.42 - 0.17 i$ & $-0.03 - 0.05 i$ \\
$18.86 - 1.35 i$ & $0.29 - 0.87 i$ & $0.61 - 0.45 i$ & $0.23 + 0.27 i$ & $1.23 - 0.13 i$ & $0.34 - 0.76 i$ & $-0.18 - 0.55 i$ & $0.26 - 0.28 i$ \\
$20.86 - 1.19 i$ & $-0.02 - 0.30 i$ & $0.28 - 0.51 i$ & $-0.08 - 0.23 i$ & $1.63 - 0.02 i$ & $0.78 - 0.40 i$ & $-0.27 - 0.40 i$ & $0.21 - 0.22 i$ \\
$22.98 - 0.92 i$ & $-0.08 + 0.05 i$ & $0.07 + 0.12 i$ & $-0.02 - 0.30 i$ & $0.76 - 0.21 i$ & $-0.43 + 0.22 i$ & $-0.85 - 0.32 i$ & $0.55 - 0.34 i$ \\
$23.41 - 1.33 i$ & $-0.42 - 0.77 i$ & $0.07 - 0.45 i$ & $-0.09 - 0.17 i$ & $1.39 - 0.35 i$ & $-0.17 - 0.33 i$ & $0.06 - 0.25 i$ & $0.03 + 0.86 i$ \\
$25.00 - 1.30 i$ & $-0.26 + 0.15 i$ & $0.32 - 0.67 i$ & $-0.36 + 0.18 i$ & $1.06 - 0.13 i$ & $0.06 - 0.26 i$ & $-0.03 + 0.22 i$ & $-0.06 + 0.28 i$ \\
    \end{tabular}
  \end{ruledtabular}
\caption{List of poles and residues of the MPA($\q$) representation of V.
}
    \label{tab:poles_V}
\end{table*} \newpage

\subsection{Cr} \FloatBarrier \begin{figure*}[hbt!]
    \centering
    \includegraphics[width=0.266\textwidth]{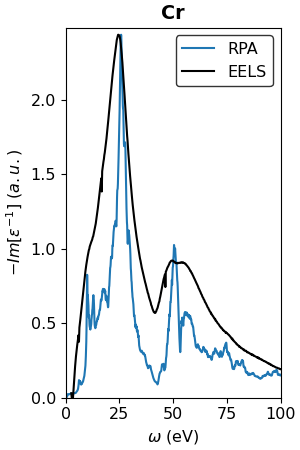}
    \includegraphics[width=0.2\textwidth]{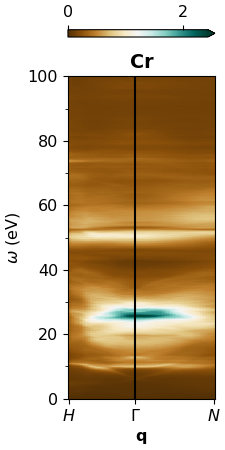}
    \includegraphics[width=0.465\textwidth]{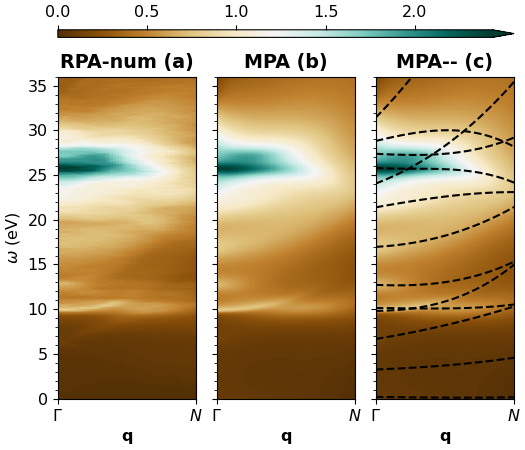}
    \caption{(left panel) Comparison of the computed RPA loss function of Cr in the optical limit with EELS measurements. (middle panel) Spectral $Y (\omega, \q)$ band structure of Cr computed from the numerical RPA results along the $H \Gamma N$ $\q$-path. (right panels) Spectral $Y (\q, \omega)$ band structure of Cr in the $\Gamma N$ direction reconstructed with MPA($\q$) with a number of $n_Y = 12$ poles.}
    \label{fig:Cr}
\end{figure*} 

\begin{table*}[hbt!]
  \begin{ruledtabular}
    \begin{tabular}{cccccccc} 
      \\[-8pt]
       $\Omega_p~$(eV) & $\Omega'_{p}$ & $\Omega''_{p}$ & $\Omega'''_{p}$  & $R_p~$(eV) & $R'_{p}$ & $R''_{p}$ & $R'''_{p}$ \\
      \hline\\[-8pt]     
      $0.19 - 2.01 i$ & $0.69 - 0.29 i$ & $0.04 + 0.07 i$ & $-0.20 + 0.13 i$ & $0.07 - 0.00 i$ & $-1.49 + 0.02 i$ & $-0.95 - 0.25 i$ & $-0.62 + 0.01 i$ \\
$3.25 - 1.86 i$ & $0.90 - 0.92 i$ & $0.49 - 0.31 i$ & $0.17 - 0.06 i$ & $0.07 - 0.00 i$ & $-1.61 - 0.31 i$ & $-0.97 - 0.60 i$ & $-0.52 - 0.38 i$ \\
$6.66 - 1.49 i$ & $0.95 - 0.79 i$ & $0.36 - 0.36 i$ & $0.08 - 0.15 i$ & $0.14 - 0.00 i$ & $-1.47 - 0.71 i$ & $-1.09 - 0.69 i$ & $-0.18 - 0.43 i$ \\
$9.81 - 0.33 i$ & $0.03 - 0.04 i$ & $0.56 - 0.11 i$ & $0.47 - 0.01 i$ & $0.11 - 0.14 i$ & $-1.63 + 0.37 i$ & $-0.41 - 0.19 i$ & $0.04 - 0.19 i$ \\
$10.13 - 1.71 i$ & $-0.06 + 0.24 i$ & $-0.12 + 0.17 i$ & $0.27 - 0.44 i$ & $0.44 - 0.01 i$ & $0.50 - 0.47 i$ & $0.29 - 0.03 i$ & $0.03 - 0.00 i$ \\
$12.72 - 1.00 i$ & $-0.09 - 0.27 i$ & $0.37 - 0.97 i$ & $0.26 - 0.44 i$ & $0.44 - 0.10 i$ & $0.10 - 0.50 i$ & $-0.22 - 0.56 i$ & $-0.14 - 0.35 i$ \\
$16.97 - 2.01 i$ & $0.12 - 0.24 i$ & $0.36 - 0.01 i$ & $0.07 - 0.01 i$ & $1.16 - 0.01 i$ & $-0.32 - 0.59 i$ & $-0.13 - 0.35 i$ & $-0.04 - 0.05 i$ \\
$21.39 - 1.95 i$ & $0.26 + 0.06 i$ & $-0.02 - 0.23 i$ & $-0.04 - 0.23 i$ & $0.91 - 0.04 i$ & $0.05 - 0.07 i$ & $-0.06 - 0.69 i$ & $-0.08 - 0.30 i$ \\
$24.06 - 2.11 i$ & $0.53 - 0.44 i$ & $0.41 - 0.22 i$ & $0.10 - 0.36 i$ & $1.82 - 0.00 i$ & $0.53 - 0.16 i$ & $-0.16 - 0.13 i$ & $-0.26 + 0.08 i$ \\
$25.81 - 0.76 i$ & $-0.07 - 0.07 i$ & $0.26 + 0.22 i$ & $-0.31 - 0.35 i$ & $1.01 - 0.00 i$ & $-0.27 - 0.29 i$ & $-0.25 + 0.13 i$ & $-0.13 + 0.30 i$ \\
$27.39 - 1.53 i$ & $-0.07 + 0.25 i$ & $0.06 - 0.40 i$ & $0.16 - 0.22 i$ & $0.87 - 0.00 i$ & $-0.35 - 0.39 i$ & $0.16 - 0.21 i$ & $0.10 - 0.13 i$ \\
$28.82 - 2.36 i$ & $0.31 - 0.64 i$ & $0.01 - 0.85 i$ & $-0.20 - 0.55 i$ & $1.95 - 0.00 i$ & $1.02 + 0.04 i$ & $-0.13 + 0.43 i$ & $-0.27 + 0.59 i$ \\
$31.44 - 1.79 i$ & $1.05 - 1.22 i$ & $0.62 - 0.45 i$ & $-0.41 - 0.32 i$ & $0.60 - 0.03 i$ & $0.06 - 0.83 i$ & $0.21 - 0.13 i$ & $0.13 + 0.09 i$ \\
    \end{tabular}
  \end{ruledtabular}
\caption{List of poles and residues of the MPA($\q$) representation of Cr.
}
    \label{tab:poles_Cr}
\end{table*} \newpage

\subsection{Fe} \FloatBarrier \begin{figure*}[hbt!]
    \centering
    \includegraphics[width=0.266\textwidth]{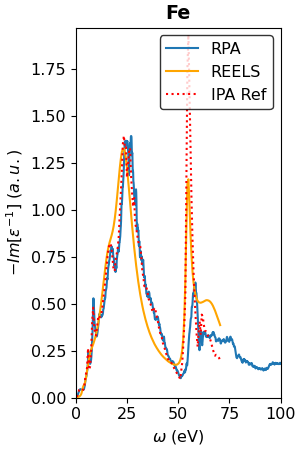}
    \includegraphics[width=0.2\textwidth]{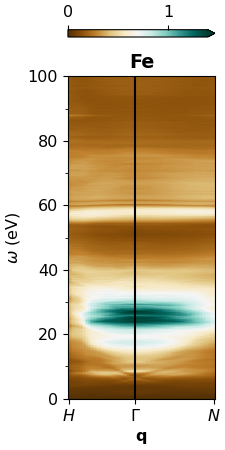}
    \includegraphics[width=0.465\textwidth]{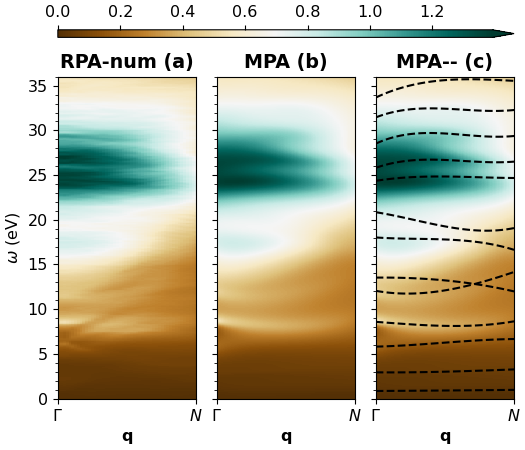}
    \caption{(left panel) Comparison of the computed RPA loss function of Fe in the optical limit with REELS measurements and reference IPA results. (middle panel) Spectral $Y (\omega, \q)$ band structure of Fe computed from the numerical RPA results along the $H \Gamma N$ $\q$-path. (right panels) Spectral $Y (\q, \omega)$ band structure of Fe in the $\Gamma N$ direction reconstructed with MPA($\q$) with a number of $n_Y = 13$ poles.}
    \label{fig:Fe}
\end{figure*} 

\begin{table*}[hbt!]
  \begin{ruledtabular}
    \begin{tabular}{cccccccc} 
      \\[-8pt]
       $\Omega_p~$(eV) & $\Omega'_{p}$ & $\Omega''_{p}$ & $\Omega'''_{p}$  & $R_p~$(eV) & $R'_{p}$ & $R''_{p}$ & $R'''_{p}$ \\
      \hline\\[-8pt]     
      $0.85 - 0.85 i$ & $0.24 - 0.00 i$ & $0.05 + 0.00 i$ & $0.01 - 0.00 i$ & $-0.01 - 0.02 i$ & $1.95 + 1.33 i$ & $-0.71 - 1.03 i$ & $-0.12 + 0.26 i$ \\
$2.93 - 0.33 i$ & $-0.08 - 0.14 i$ & $0.45 + 0.11 i$ & $-0.09 - 0.49 i$ & $0.00 - 0.00 i$ & $0.39 - 7.48 i$ & $3.45 + 5.56 i$ & $1.48 - 1.16 i$ \\
$5.81 - 1.34 i$ & $0.27 - 0.64 i$ & $0.43 + 0.58 i$ & $-0.34 - 0.20 i$ & $0.22 - 0.08 i$ & $-5.85 - 6.09 i$ & $7.26 + 8.90 i$ & $-1.87 - 3.33 i$ \\
$8.58 - 0.50 i$ & $-0.22 - 0.74 i$ & $-0.11 + 1.01 i$ & $0.37 - 0.59 i$ & $0.16 - 0.14 i$ & $-2.71 + 5.78 i$ & $5.41 - 6.22 i$ & $-2.81 + 0.92 i$ \\
$12.06 - 0.86 i$ & $-0.44 - 0.39 i$ & $1.08 + 0.45 i$ & $-0.27 - 0.29 i$ & $0.12 + 0.04 i$ & $-1.00 + 6.14 i$ & $3.18 - 7.89 i$ & $-2.01 + 2.60 i$ \\
$13.53 - 2.44 i$ & $0.11 - 0.51 i$ & $-0.18 - 0.12 i$ & $-0.09 + 0.21 i$ & $0.32 - 0.26 i$ & $2.28 + 0.20 i$ & $-7.72 + 1.58 i$ & $4.44 - 2.40 i$ \\
$18.03 - 2.35 i$ & $-0.12 - 0.02 i$ & $0.29 + 0.20 i$ & $-0.33 - 0.15 i$ & $0.83 + 0.60 i$ & $3.05 + 0.99 i$ & $-9.43 - 1.79 i$ & $4.48 + 0.42 i$ \\
$20.85 - 2.24 i$ & $-0.24 + 0.03 i$ & $-0.34 + 0.01 i$ & $0.42 - 0.10 i$ & $0.39 - 0.01 i$ & $1.06 - 3.48 i$ & $-1.49 + 0.24 i$ & $-1.50 + 1.91 i$ \\
$24.35 - 1.83 i$ & $0.20 + 0.05 i$ & $-0.31 - 0.11 i$ & $0.14 + 0.08 i$ & $0.78 + 0.98 i$ & $4.53 + 0.20 i$ & $-6.33 + 2.26 i$ & $0.63 - 2.08 i$ \\
$25.81 - 2.77 i$ & $0.39 + 0.31 i$ & $-0.75 - 0.33 i$ & $0.40 + 0.10 i$ & $2.11 - 0.77 i$ & $-0.92 + 3.67 i$ & $-0.63 - 6.01 i$ & $-0.21 + 2.69 i$ \\
$28.51 - 3.77 i$ & $0.46 + 0.35 i$ & $-0.94 + 0.01 i$ & $0.54 - 0.41 i$ & $2.12 + 1.50 i$ & $-1.16 - 0.11 i$ & $-2.30 - 1.22 i$ & $2.86 + 1.22 i$ \\
$31.45 - 3.66 i$ & $0.36 + 0.20 i$ & $-0.71 - 0.04 i$ & $0.40 - 0.19 i$ & $1.90 + 0.01 i$ & $2.35 - 0.12 i$ & $-5.48 + 1.44 i$ & $2.07 - 1.21 i$ \\
$33.67 - 3.88 i$ & $0.40 + 0.28 i$ & $-0.45 - 0.29 i$ & $0.16 - 0.01 i$ & $0.86 - 0.56 i$ & $4.13 + 1.19 i$ & $-9.22 + 3.82 i$ & $3.57 - 2.35 i$ \\
$37.40 - 4.16 i$ & $0.17 - 0.06 i$ & $0.16 + 0.19 i$ & $-0.09 - 0.02 i$ & $2.45 - 0.65 i$ & $1.50 - 2.32 i$ & $-4.41 + 9.42 i$ & $0.80 - 6.06 i$ \\
    \end{tabular}
  \end{ruledtabular}
\caption{List of poles and residues of the MPA($\q$) representation of Fe.
}
    \label{tab:poles_Fe}
\end{table*} \newpage

\subsection{Co} \FloatBarrier \begin{figure*}[hbt!]
    \centering
    \includegraphics[width=0.250\textwidth]{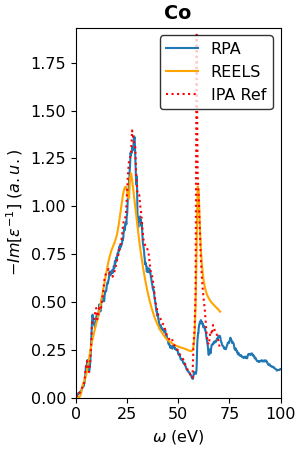}
    \includegraphics[width=0.333\textwidth]{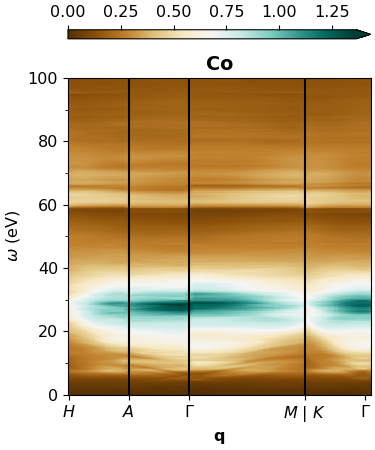}
    \includegraphics[width=0.406\textwidth]{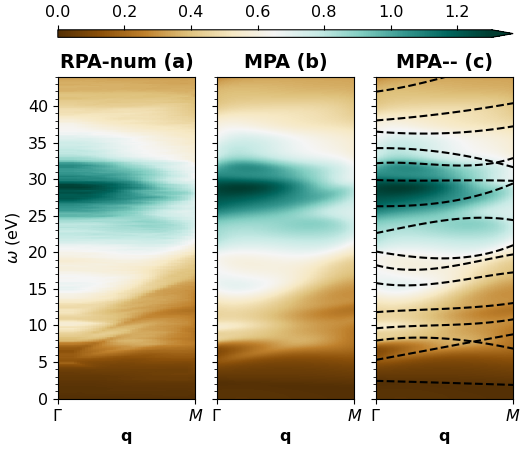}
    \caption{(left panel) Comparison of the computed RPA loss function of Co in the optical limit with REELS measurements and reference IPA results. (middle panel) Spectral $Y (\omega, \q)$ band structure of Co computed from the numerical RPA results along the $H A \Gamma M|K \Gamma$ $\q$-path. (right panels) Spectral $Y (\q, \omega)$ band structure of Co in the $\Gamma M$ direction reconstructed with MPA($\q$) with a number of $n_Y = 16$ poles.}
    \label{fig:Co}
\end{figure*} 

\begin{table*}[hbt!]
  \begin{ruledtabular}
    \begin{tabular}{cccccccc} 
      \\[-8pt]
       $\Omega_p~$(eV) & $\Omega'_{p}$ & $\Omega''_{p}$ & $\Omega'''_{p}$  & $R_p~$(eV) & $R'_{p}$ & $R''_{p}$ & $R'''_{p}$ \\
      \hline\\[-8pt]     
$1.04 - 1.04 i$ & $0.16 + 0.00 i$ & $0.04 - 0.00 i$ & $-0.00 + 0.00 i$ & $0.01 - 0.03 i$ & $0.14 + 0.55 i$ & $0.07 + 0.17 i$ & $0.03 + 0.02 i$ \\
$2.42 - 0.68 i$ & $-0.28 - 0.50 i$ & $-0.03 - 0.05 i$ & $-0.00 + 0.02 i$ & $-0.02 - 0.01 i$ & $1.10 + 0.56 i$ & $0.24 + 0.16 i$ & $0.07 + 0.06 i$ \\
$5.28 - 1.10 i$ & $1.16 + 0.65 i$ & $0.25 - 0.71 i$ & $-0.00 - 0.33 i$ & $0.13 + 0.01 i$ & $0.63 - 0.60 i$ & $0.35 + 0.66 i$ & $0.05 + 0.10 i$ \\
$7.92 - 0.94 i$ & $0.72 - 0.25 i$ & $-1.18 - 0.21 i$ & $0.25 - 0.07 i$ & $0.25 - 0.22 i$ & $0.02 + 1.85 i$ & $0.26 - 0.84 i$ & $0.04 - 0.59 i$ \\
$9.60 - 0.80 i$ & $0.38 - 0.09 i$ & $-0.58 - 0.43 i$ & $0.46 + 0.44 i$ & $0.14 - 0.13 i$ & $-0.98 + 1.10 i$ & $0.17 + 0.20 i$ & $0.40 - 0.34 i$ \\
$11.81 - 2.79 i$ & $0.08 + 0.64 i$ & $-0.19 - 0.38 i$ & $0.25 + 0.06 i$ & $0.11 - 0.01 i$ & $-0.55 - 1.30 i$ & $-0.16 - 0.23 i$ & $0.25 - 0.25 i$ \\
$15.80 - 3.00 i$ & $-0.44 + 0.21 i$ & $1.15 - 0.37 i$ & $-0.53 + 0.17 i$ & $1.25 + 0.37 i$ & $-0.80 - 0.59 i$ & $-0.29 - 0.50 i$ & $0.34 + 0.17 i$ \\
$18.26 - 2.18 i$ & $-0.52 - 0.41 i$ & $1.20 + 0.64 i$ & $-0.52 - 0.21 i$ & $0.57 + 0.03 i$ & $-0.97 - 1.24 i$ & $0.03 - 0.67 i$ & $-0.07 + 0.31 i$ \\
$20.11 - 2.78 i$ & $-0.25 - 0.53 i$ & $-0.02 + 1.06 i$ & $0.35 - 0.53 i$ & $0.01 - 1.07 i$ & $-0.68 - 0.88 i$ & $0.40 - 0.24 i$ & $-0.28 + 0.08 i$ \\
$22.61 - 2.93 i$ & $0.36 - 0.17 i$ & $-0.08 + 1.13 i$ & $-0.13 - 0.86 i$ & $0.19 + 0.47 i$ & $-0.85 + 0.63 i$ & $0.45 - 0.13 i$ & $-0.70 - 0.25 i$ \\
$26.30 - 2.47 i$ & $-0.00 - 0.12 i$ & $0.08 + 0.49 i$ & $0.15 - 0.27 i$ & $0.76 + 0.21 i$ & $-1.32 + 0.28 i$ & $0.07 + 0.38 i$ & $-0.26 - 0.11 i$ \\
$29.89 - 2.33 i$ & $-0.07 + 0.22 i$ & $0.14 - 0.37 i$ & $-0.09 + 0.19 i$ & $0.42 + 1.75 i$ & $-2.20 + 0.46 i$ & $-0.38 - 0.39 i$ & $0.44 + 0.14 i$ \\
$32.17 - 2.36 i$ & $0.08 + 0.16 i$ & $-0.42 - 0.34 i$ & $0.38 + 0.19 i$ & $1.41 + 1.88 i$ & $-1.17 + 0.43 i$ & $-1.19 + 0.13 i$ & $0.38 - 0.33 i$ \\
$34.24 - 2.43 i$ & $0.01 + 0.04 i$ & $-0.14 - 0.40 i$ & $-0.01 + 0.20 i$ & $1.59 + 0.27 i$ & $-0.32 + 0.89 i$ & $-0.23 + 0.83 i$ & $0.31 - 0.19 i$ \\
$36.51 - 3.31 i$ & $-0.10 + 0.16 i$ & $0.13 - 0.21 i$ & $0.00 + 0.04 i$ & $1.75 + 0.84 i$ & $-1.31 + 1.39 i$ & $-0.50 - 0.00 i$ & $0.15 - 0.06 i$ \\
$38.03 - 3.54 i$ & $0.08 - 0.00 i$ & $0.04 + 0.07 i$ & $0.01 - 0.14 i$ & $2.54 - 0.78 i$ & $-0.83 + 1.24 i$ & $-0.15 + 0.65 i$ & $0.09 + 0.18 i$ \\
$41.96 - 4.31 i$ & $0.14 - 0.25 i$ & $0.15 + 0.27 i$ & $-0.05 - 0.06 i$ & $3.06 - 2.62 i$ & $0.00 + 2.68 i$ & $-0.80 + 0.10 i$ & $-0.62 + 0.23 i$ \\
$46.11 - 4.02 i$ & $0.51 - 0.30 i$ & $-0.16 + 0.12 i$ & $-0.05 + 0.09 i$ & $-1.41 - 3.15 i$ & $1.01 + 3.32 i$ & $0.69 + 0.32 i$ & $0.48 + 0.01 i$ \\
    \end{tabular}
  \end{ruledtabular}
\caption{List of poles and residues of the MPA($\q$) representation of Co.
}
    \label{tab:poles_Co}
\end{table*} \newpage

\subsection{Ni} \FloatBarrier \begin{figure*}[hbt!]
    \centering
    \includegraphics[width=0.266\textwidth]{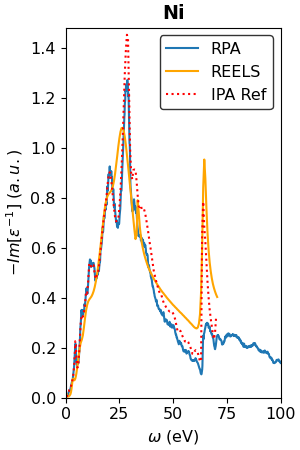}
    \includegraphics[width=0.2\textwidth]{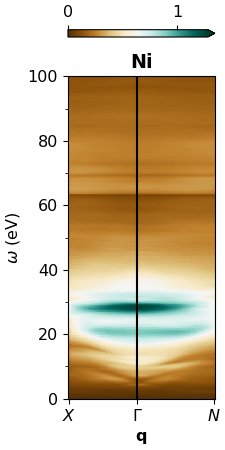}
    \includegraphics[width=0.465\textwidth]{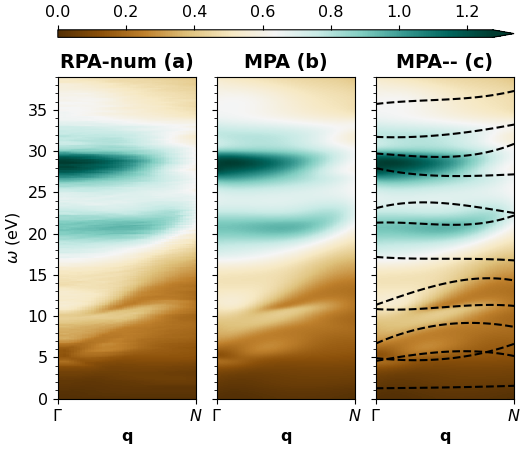}
    \caption{(left panel) Comparison of the computed RPA loss function of Ni in the optical limit with REELS measurements and reference IPA results. (middle panel) Spectral $Y (\omega, \q)$ band structure of Ni computed from the numerical RPA results along the $X \Gamma N$ $\q$-path. (right panels) Spectral $Y (\q, \omega)$ band structure of Ni in the $\Gamma N$ direction reconstructed with MPA($\q$) with a number of $n_Y = 6$ poles.}
    \label{fig:Ni}
\end{figure*} 

\begin{table*}[hbt!]
  \begin{ruledtabular}
    \begin{tabular}{cccccccc} 
      \\[-8pt]
       $\Omega_p~$(eV) & $\Omega'_{p}$ & $\Omega''_{p}$ & $\Omega'''_{p}$  & $R_p~$(eV) & $R'_{p}$ & $R''_{p}$ & $R'''_{p}$ \\
      \hline\\[-8pt]     
$1.25 - 1.25 i$ & $0.17 + 0.03 i$ & $0.21 - 0.02 i$ & $0.09 - 0.01 i$ & $0.00 - 0.02 i$ & $1.39 + 1.40 i$ & $0.50 - 0.25 i$ & $-0.07 - 0.54 i$ \\
$4.55 - 2.09 i$ & $0.39 + 2.04 i$ & $1.21 - 3.38 i$ & $-1.09 + 0.81 i$ & $0.03 - 0.16 i$ & $18.07 + 14.32 i$ & $-27.23 - 29.25 i$ & $11.13 + 14.58 i$ \\
$4.90 - 0.62 i$ & $-0.63 - 0.55 i$ & $1.12 - 0.39 i$ & $0.31 + 0.14 i$ & $0.03 + 0.10 i$ & $-0.21 - 0.77 i$ & $1.00 + 1.50 i$ & $0.65 - 1.61 i$ \\
$6.66 - 2.47 i$ & $1.73 + 1.36 i$ & $-1.54 - 0.52 i$ & $0.28 - 0.48 i$ & $0.28 - 0.42 i$ & $-1.71 - 6.13 i$ & $-4.76 + 13.35 i$ & $4.63 - 6.51 i$ \\
$10.85 - 2.35 i$ & $-0.51 + 1.33 i$ & $1.25 - 1.49 i$ & $-0.69 + 0.22 i$ & $0.20 - 0.50 i$ & $-10.73 - 4.15 i$ & $13.80 + 11.53 i$ & $-4.22 - 6.32 i$ \\
$11.35 - 4.07 i$ & $0.66 + 1.06 i$ & $0.30 - 0.81 i$ & $-0.56 + 0.08 i$ & $0.56 + 0.48 i$ & $-0.11 + 0.45 i$ & $-3.47 - 5.61 i$ & $0.62 + 3.78 i$ \\
$17.17 - 2.08 i$ & $-0.16 + 0.10 i$ & $0.30 - 0.07 i$ & $-0.18 - 0.08 i$ & $0.27 - 0.13 i$ & $-2.46 + 1.61 i$ & $0.48 - 5.19 i$ & $-0.43 + 2.42 i$ \\
$21.32 - 2.64 i$ & $0.07 - 0.04 i$ & $-0.51 + 0.43 i$ & $0.53 - 0.40 i$ & $1.00 + 0.80 i$ & $1.80 + 0.33 i$ & $-4.36 - 3.54 i$ & $0.62 + 2.95 i$ \\
$23.06 - 3.11 i$ & $0.40 - 0.05 i$ & $-0.74 + 0.24 i$ & $0.30 - 0.23 i$ & $0.56 + 0.26 i$ & $-2.15 + 1.85 i$ & $6.04 - 4.15 i$ & $-2.55 + 1.78 i$ \\
$27.94 - 0.83 i$ & $-0.26 - 0.46 i$ & $0.29 + 0.92 i$ & $-0.09 - 0.47 i$ & $0.11 + 0.13 i$ & $0.85 + 0.94 i$ & $-0.50 + 0.95 i$ & $-1.52 - 1.69 i$ \\
$29.70 - 1.87 i$ & $-0.08 + 0.05 i$ & $-0.05 - 0.21 i$ & $0.21 + 0.10 i$ & $0.52 + 2.27 i$ & $-2.34 - 0.51 i$ & $2.58 - 2.94 i$ & $-1.21 + 3.27 i$ \\
$31.72 - 3.09 i$ & $-0.04 - 0.08 i$ & $0.16 + 0.15 i$ & $-0.04 + 0.05 i$ & $3.93 + 0.02 i$ & $0.33 + 2.39 i$ & $-5.19 - 5.05 i$ & $3.06 + 2.67 i$ \\
$35.70 - 3.95 i$ & $0.11 + 0.06 i$ & $-0.16 - 0.22 i$ & $0.14 + 0.20 i$ & $2.54 - 0.62 i$ & $1.45 + 0.91 i$ & $-5.54 - 0.51 i$ & $2.42 + 0.27 i$ \\
    \end{tabular}
  \end{ruledtabular}
\caption{List of poles and residues of the MPA($\q$) representation of Ni.
}
    \label{tab:poles_Ni}
\end{table*} \newpage

\subsection{Cu} \FloatBarrier \begin{figure*}[hbt!]
    \centering
    \includegraphics[width=0.266\textwidth]{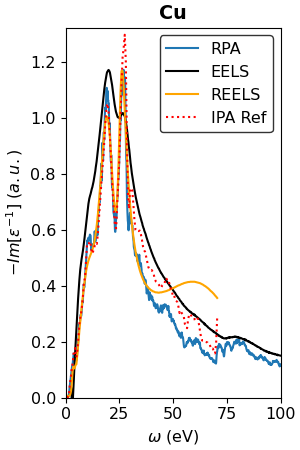}
    \includegraphics[width=0.2\textwidth]{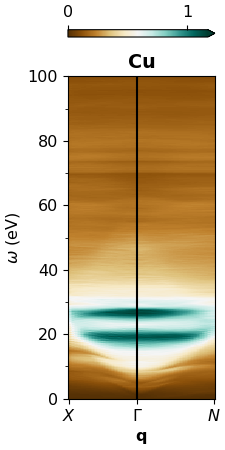}
    \includegraphics[width=0.465\textwidth]{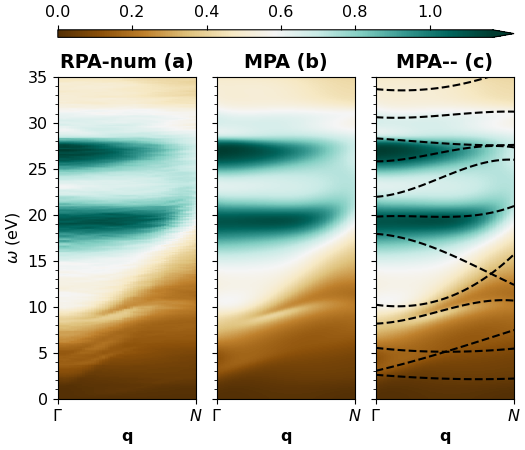}
    \caption{(left panel) Comparison of the computed RPA loss function of Cu in the optical limit with EELS, REELS measurements, and reference IPA results. (middle panel) Spectral $Y (\omega, \q)$ band structure of Cu computed from the numerical RPA results along the $X \Gamma N$ $\q$-path. (right panels) Spectral $Y (\q, \omega)$ band structure of Cu in the $\Gamma N$ direction reconstructed with MPA($\q$) with a number of $n_Y = 12$ poles.}
    \label{fig:Cu}
\end{figure*} 

\begin{table*}[hbt!]
  \begin{ruledtabular}
    \begin{tabular}{cccccccc} 
      \\[-8pt]
       $\Omega_p~$(eV) & $\Omega'_{p}$ & $\Omega''_{p}$ & $\Omega'''_{p}$  & $R_p~$(eV) & $R'_{p}$ & $R''_{p}$ & $R'''_{p}$ \\
      \hline\\[-8pt]     
      $2.58 - 2.59 i$ & $-0.96 + 0.06 i$ & $0.30 + 0.13 i$ & $0.18 - 0.02 i$ & $0.04 - 0.05 i$ & $0.51 + 0.28 i$ & $-0.06 - 0.73 i$ & $0.08 + 0.24 i$ \\
$3.02 - 1.26 i$ & $1.77 + 1.34 i$ & $1.25 - 0.68 i$ & $-0.29 - 0.15 i$ & $0.17 - 0.05 i$ & $-4.42 - 3.70 i$ & $4.29 + 9.95 i$ & $-0.93 - 5.09 i$ \\
$5.52 - 2.53 i$ & $0.03 - 1.39 i$ & $0.37 + 0.55 i$ & $-0.02 - 0.05 i$ & $-0.18 - 0.56 i$ & $-4.54 - 0.87 i$ & $5.13 + 1.71 i$ & $-1.28 - 1.96 i$ \\
$8.15 - 1.56 i$ & $-0.17 + 1.54 i$ & $2.35 - 2.17 i$ & $-1.60 + 0.79 i$ & $0.02 + 0.05 i$ & $0.23 + 0.89 i$ & $7.13 - 1.60 i$ & $-3.42 - 0.36 i$ \\
$10.21 - 1.89 i$ & $-0.40 - 0.09 i$ & $1.30 - 0.29 i$ & $0.44 - 1.05 i$ & $0.40 - 0.11 i$ & $5.52 - 3.89 i$ & $-13.67 - 2.50 i$ & $9.28 + 1.02 i$ \\
$17.91 - 3.14 i$ & $-0.04 - 0.29 i$ & $-0.97 + 0.54 i$ & $0.40 - 0.25 i$ & $0.16 + 0.24 i$ & $3.20 + 3.54 i$ & $-9.27 - 10.23 i$ & $3.38 + 5.96 i$ \\
$19.78 - 2.34 i$ & $0.11 - 0.08 i$ & $-0.47 + 0.33 i$ & $0.48 - 0.28 i$ & $0.90 + 0.77 i$ & $5.18 - 1.77 i$ & $-8.14 - 0.91 i$ & $1.76 + 2.21 i$ \\
$21.94 - 3.11 i$ & $0.12 - 0.31 i$ & $0.87 + 1.06 i$ & $-0.65 - 0.64 i$ & $1.07 + 1.62 i$ & $-2.40 + 0.44 i$ & $-1.33 + 0.42 i$ & $1.81 - 0.29 i$ \\
$25.80 - 2.64 i$ & $-0.00 + 0.02 i$ & $0.50 + 0.27 i$ & $-0.37 - 0.36 i$ & $1.93 - 0.43 i$ & $-0.35 + 1.20 i$ & $-2.85 + 1.04 i$ & $1.75 - 0.42 i$ \\
$28.31 - 1.05 i$ & $-0.08 + 0.07 i$ & $-0.03 - 0.25 i$ & $0.06 + 0.09 i$ & $-0.10 + 0.57 i$ & $-2.68 - 4.25 i$ & $-0.13 + 7.70 i$ & $-0.75 - 5.05 i$ \\
$30.61 - 3.13 i$ & $-0.08 + 0.12 i$ & $0.28 - 0.07 i$ & $-0.18 + 0.11 i$ & $2.08 + 1.77 i$ & $-3.23 + 0.01 i$ & $1.40 - 0.21 i$ & $-0.10 + 0.16 i$ \\
$33.66 - 3.92 i$ & $-0.09 + 0.02 i$ & $0.22 + 0.05 i$ & $0.01 - 0.10 i$ & $2.59 - 0.85 i$ & $-0.60 + 0.11 i$ & $-2.99 + 3.14 i$ & $1.38 - 2.42 i$ \\
$39.06 - 6.01 i$ & $-0.01 - 0.08 i$ & $0.11 + 0.22 i$ & $0.13 - 0.29 i$ & $4.52 - 1.97 i$ & $-0.90 + 1.44 i$ & $-1.73 + 3.81 i$ & $-0.45 - 2.89 i$ \\
$42.73 - 5.29 i$ & $0.13 - 0.28 i$ & $0.18 + 0.21 i$ & $-0.14 - 0.39 i$ & $-1.89 - 3.02 i$ & $1.42 + 1.66 i$ & $0.37 + 7.13 i$ & $-1.39 - 1.77 i$ \\
    \end{tabular}
  \end{ruledtabular}
\caption{List of poles and residues of the MPA($\q$) representation of Cu.
}
    \label{tab:poles_Cu}
\end{table*} \newpage

\subsection{Zn} \FloatBarrier \begin{figure*}[hbt!]
    \centering
    \includegraphics[width=0.250\textwidth]{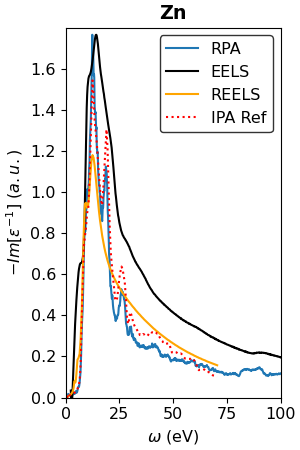}
    \includegraphics[width=0.333\textwidth]{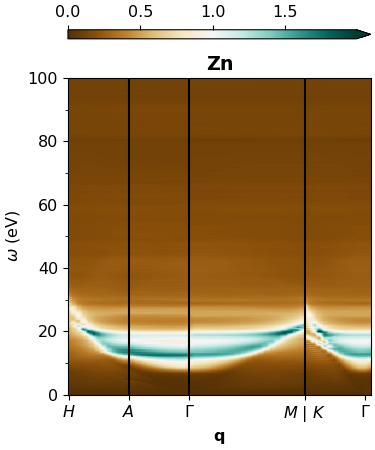}
    \includegraphics[width=0.406\textwidth]{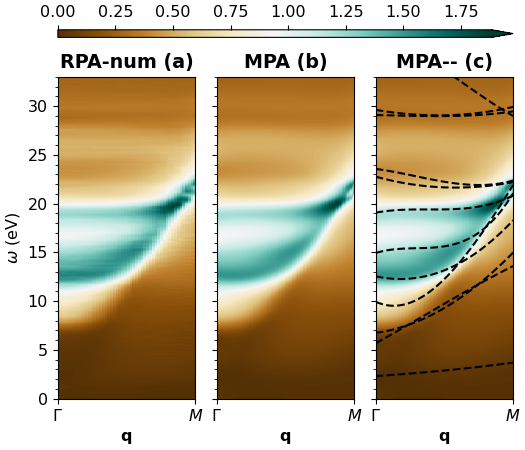}
    \caption{(left panel) Comparison of the computed RPA loss function of Zn in the optical limit with EELS, REELS measurements, and reference IPA results. (middle panel) Spectral $Y (\omega, \q)$ band structure of Zn computed from the numerical RPA results along the $H A \Gamma M|K \Gamma$ $\q$-path. (right panels) Spectral $Y (\q, \omega)$ band structure of Zn in the $\Gamma M$ direction reconstructed with MPA($\q$) with a number of $n_Y = 12$ poles.}
    \label{fig:Zn}
\end{figure*} 

\begin{table*}[hbt!]
  \begin{ruledtabular}
    \begin{tabular}{cccccccc} 
      \\[-8pt]
       $\Omega_p~$(eV) & $\Omega'_{p}$ & $\Omega''_{p}$ & $\Omega'''_{p}$  & $R_p~$(eV) & $R'_{p}$ & $R''_{p}$ & $R'''_{p}$ \\
      \hline\\[-8pt]     
      $2.31 - 2.12 i$ & $0.82 - 0.00 i$ & $0.26 + 0.04 i$ & $0.12 - 0.04 i$ & $-0.16 + 0.08 i$ & $-6.82 + 4.87 i$ & $4.72 - 0.41 i$ & $1.36 - 2.02 i$ \\
$6.32 - 2.74 i$ & $0.70 + 1.62 i$ & $2.09 - 0.96 i$ & $0.02 - 0.44 i$ & $0.11 - 1.50 i$ & $-11.24 - 0.79 i$ & $17.13 + 2.85 i$ & $-7.15 - 2.54 i$ \\
$9.93 - 4.64 i$ & $-1.72 + 0.90 i$ & $5.11 + 0.25 i$ & $-1.73 + 0.45 i$ & $1.24 + 0.28 i$ & $-3.39 - 7.05 i$ & $-2.35 + 12.34 i$ & $3.21 - 5.76 i$ \\
$12.54 - 1.26 i$ & $-0.47 - 0.14 i$ & $1.27 + 0.54 i$ & $0.13 - 0.48 i$ & $0.81 - 0.17 i$ & $3.38 - 6.63 i$ & $-14.46 + 11.40 i$ & $8.28 - 3.50 i$ \\
$14.92 - 1.78 i$ & $0.67 - 0.72 i$ & $-2.09 + 2.02 i$ & $2.26 - 1.49 i$ & $0.83 + 0.48 i$ & $6.88 + 5.18 i$ & $-6.24 - 12.29 i$ & $1.22 + 6.35 i$ \\
$19.06 - 1.61 i$ & $0.27 + 0.09 i$ & $-0.70 - 0.41 i$ & $0.60 + 0.48 i$ & $1.22 + 0.49 i$ & $0.07 + 2.74 i$ & $8.42 - 6.04 i$ & $-10.55 + 3.23 i$ \\
$22.76 - 3.56 i$ & $-0.40 + 0.33 i$ & $0.37 - 0.37 i$ & $-0.05 + 0.22 i$ & $2.52 + 0.47 i$ & $-1.97 - 4.29 i$ & $0.65 + 4.22 i$ & $3.47 - 1.57 i$ \\
$23.55 - 2.13 i$ & $-0.13 - 0.25 i$ & $-0.38 + 0.43 i$ & $0.40 - 0.10 i$ & $-1.12 - 0.60 i$ & $1.36 - 2.77 i$ & $0.41 - 2.79 i$ & $0.95 + 0.71 i$ \\
$29.42 - 2.89 i$ & $-0.07 + 0.06 i$ & $0.07 - 0.08 i$ & $0.03 + 0.03 i$ & $0.63 + 1.39 i$ & $1.18 - 1.18 i$ & $-2.30 + 3.70 i$ & $0.40 - 1.18 i$ \\
$37.46 - 9.32 i$ & $-0.07 - 0.63 i$ & $-0.50 + 0.10 i$ & $0.20 + 0.20 i$ & $3.77 - 2.44 i$ & $1.57 - 0.12 i$ & $-0.66 + 0.13 i$ & $0.27 + 0.08 i$ \\
    \end{tabular}
  \end{ruledtabular}
\caption{List of poles and residues of the MPA($\q$) representation of Zn.
}
    \label{tab:poles_Zn}
\end{table*} \newpage

\subsection{Mo} \FloatBarrier \begin{figure*}[hbt!]
    \centering
    \includegraphics[width=0.266\textwidth]{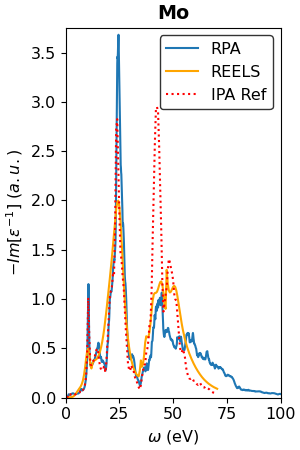}
    \includegraphics[width=0.2\textwidth]{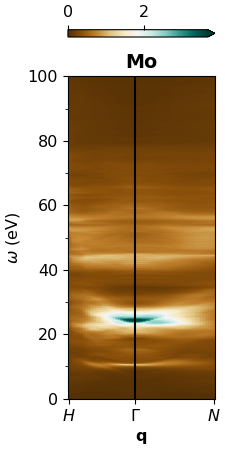}
    \includegraphics[width=0.465\textwidth]{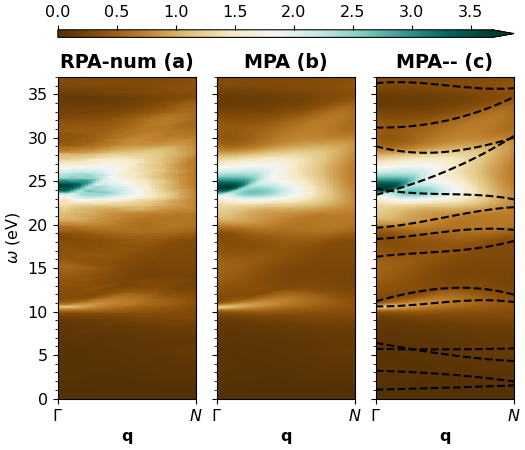}
    \caption{(left panel) Comparison of the computed RPA loss function of Mo in the optical limit with REELS measurements and reference IPA results. (middle panel) Spectral $Y (\omega, \q)$ band structure of Mo computed from the numerical RPA results along the $H \Gamma N$ $\q$-path. (right panels) Spectral $Y (\q, \omega)$ band structure of Mo in the $\Gamma N$ direction reconstructed with MPA($\q$) with a number of $n_Y = 14$ poles.}
    \label{fig:Mo}
\end{figure*} 

\begin{table*}[hbt!]
  \begin{ruledtabular}
    \begin{tabular}{cccccccc} 
      \\[-8pt]
       $\Omega_p~$(eV) & $\Omega'_{p}$ & $\Omega''_{p}$ & $\Omega'''_{p}$  & $R_p~$(eV) & $R'_{p}$ & $R''_{p}$ & $R'''_{p}$ \\
      \hline\\[-8pt]     
      $1.02 - 1.02 i$ & $0.78 - 0.01 i$ & $0.14 + 0.01 i$ & $0.03 - 0.00 i$ & $-0.00 - 0.02 i$ & $0.07 + 1.80 i$ & $-0.03 + 0.64 i$ & $0.34 - 0.06 i$ \\
$3.20 - 1.25 i$ & $-0.24 - 0.17 i$ & $-0.32 - 0.08 i$ & $-0.11 - 0.03 i$ & $0.02 - 0.01 i$ & $-0.13 - 2.86 i$ & $1.07 - 0.52 i$ & $-0.52 + 0.52 i$ \\
$5.70 - 1.71 i$ & $-0.10 + 0.09 i$ & $0.02 + 0.16 i$ & $0.00 + 0.09 i$ & $0.00 + 0.02 i$ & $2.57 - 3.11 i$ & $-0.07 - 0.86 i$ & $-0.80 + 2.12 i$ \\
$6.41 - 1.05 i$ & $-0.61 - 1.65 i$ & $-0.07 + 0.53 i$ & $0.16 + 0.30 i$ & $0.01 - 0.03 i$ & $2.16 - 10.71 i$ & $0.97 + 2.44 i$ & $-1.55 + 2.06 i$ \\
$10.59 - 0.27 i$ & $-0.03 - 0.30 i$ & $0.76 - 0.17 i$ & $-0.63 + 0.11 i$ & $0.29 + 0.07 i$ & $8.46 - 7.34 i$ & $-8.49 + 8.81 i$ & $4.14 - 1.80 i$ \\
$11.17 - 2.46 i$ & $0.41 + 1.85 i$ & $0.02 - 2.09 i$ & $-0.29 + 0.23 i$ & $-0.10 - 0.38 i$ & $-7.82 - 1.88 i$ & $7.41 + 6.49 i$ & $-0.21 - 6.19 i$ \\
$16.30 - 3.07 i$ & $0.17 + 0.38 i$ & $-0.31 - 0.12 i$ & $0.34 - 0.12 i$ & $1.15 + 2.29 i$ & $-6.09 - 1.33 i$ & $5.34 + 1.87 i$ & $-0.59 - 0.08 i$ \\
$18.36 - 2.06 i$ & $0.04 + 0.21 i$ & $0.43 - 0.24 i$ & $-0.36 + 0.04 i$ & $2.39 + 0.92 i$ & $-11.54 - 1.14 i$ & $21.64 + 2.58 i$ & $-11.32 - 2.40 i$ \\
$19.64 - 2.13 i$ & $0.06 + 0.44 i$ & $0.50 - 0.76 i$ & $-0.31 + 0.23 i$ & $-0.23 - 4.00 i$ & $-10.23 + 3.34 i$ & $15.41 - 8.57 i$ & $-5.87 + 5.09 i$ \\
$23.42 - 0.59 i$ & $0.35 + 0.30 i$ & $0.20 - 1.34 i$ & $0.03 + 0.97 i$ & $-0.15 - 0.24 i$ & $12.43 - 6.17 i$ & $-19.67 + 22.03 i$ & $7.59 - 6.43 i$ \\
$24.17 - 1.35 i$ & $-0.27 - 0.04 i$ & $0.47 + 0.05 i$ & $-0.30 - 0.17 i$ & $3.05 - 1.22 i$ & $2.77 - 1.11 i$ & $-8.85 + 0.81 i$ & $4.18 + 1.51 i$ \\
$29.03 - 2.96 i$ & $-0.34 + 0.34 i$ & $0.56 - 0.77 i$ & $-0.16 + 0.52 i$ & $-3.03 + 3.48 i$ & $-5.49 - 2.92 i$ & $7.18 + 4.53 i$ & $-2.93 - 1.94 i$ \\
$31.16 - 3.29 i$ & $-0.00 - 0.15 i$ & $0.18 + 0.35 i$ & $0.04 - 0.12 i$ & $4.65 + 1.06 i$ & $-3.83 + 2.10 i$ & $5.83 - 1.13 i$ & $-3.64 + 0.10 i$ \\
$36.26 - 1.74 i$ & $0.10 - 0.12 i$ & $-0.36 + 0.15 i$ & $0.23 - 0.02 i$ & $0.87 - 0.14 i$ & $0.16 + 0.12 i$ & $0.45 - 0.54 i$ & $-0.58 - 0.97 i$ \\
$42.04 - 2.72 i$ & $0.17 - 0.09 i$ & $0.08 - 0.13 i$ & $-0.14 + 0.01 i$ & $4.43 - 3.21 i$ & $0.35 + 0.48 i$ & $1.97 + 0.48 i$ & $-1.16 - 0.34 i$ \\
    \end{tabular}
  \end{ruledtabular}
\caption{List of poles and residues of the MPA($\q$) representation of Mo.
}
    \label{tab:poles_Mo}
\end{table*} \newpage

\subsection{Pd} \FloatBarrier \begin{figure*}[hbt!]
    \centering
    \includegraphics[width=0.266\textwidth]{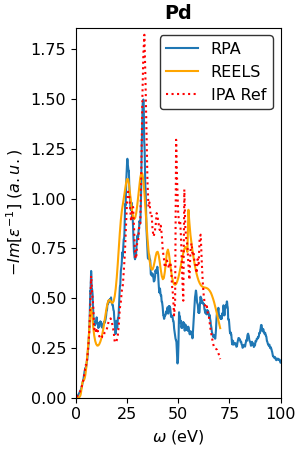}
    \includegraphics[width=0.2\textwidth]{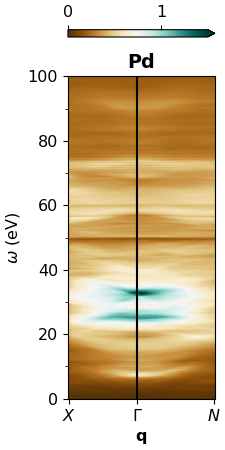}
    \includegraphics[width=0.465\textwidth]{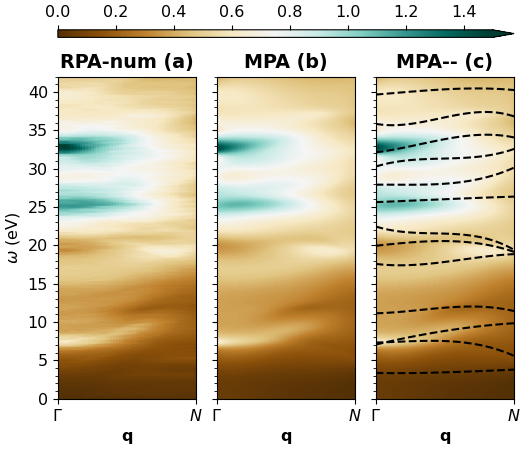}
    \caption{(left panel) Comparison of the computed RPA loss function of Pd in the optical limit with REELS measurements and reference IPA results. (middle panel) Spectral $Y (\omega, \q)$ band structure of Pd computed from the numerical RPA results along the $X \Gamma N$ $\q$-path. (right panels) Spectral $Y (\q, \omega)$ band structure of Pd in the $\Gamma N$ direction reconstructed with MPA($\q$) with a number of $n_Y = 13$ poles.}
    \label{fig:Pd}
\end{figure*} 

\begin{table*}[hbt!]
  \begin{ruledtabular}
    \begin{tabular}{cccccccc} 
      \\[-8pt]
       $\Omega_p~$(eV) & $\Omega'_{p}$ & $\Omega''_{p}$ & $\Omega'''_{p}$  & $R_p~$(eV) & $R'_{p}$ & $R''_{p}$ & $R'''_{p}$ \\
      \hline\\[-8pt]     
      $3.34 - 3.35 i$ & $-0.24 + 0.04 i$ & $0.87 - 0.13 i$ & $-0.39 + 0.11 i$ & $-0.48 - 0.16 i$ & $-3.89 - 0.11 i$ & $5.32 + 4.51 i$ & $-1.29 - 3.64 i$ \\
$7.04 - 3.33 i$ & $0.55 + 1.45 i$ & $-0.45 + 0.06 i$ & $0.53 - 1.16 i$ & $0.26 + 0.24 i$ & $2.34 - 10.01 i$ & $-9.14 + 14.80 i$ & $5.48 - 4.72 i$ \\
$7.31 - 0.64 i$ & $0.15 - 0.33 i$ & $0.46 + 0.94 i$ & $-1.08 - 0.71 i$ & $0.26 - 0.12 i$ & $-4.61 + 0.96 i$ & $1.25 - 2.43 i$ & $1.48 + 1.81 i$ \\
$11.13 - 3.52 i$ & $-0.32 + 1.27 i$ & $0.84 - 0.23 i$ & $-0.54 - 0.78 i$ & $0.47 + 0.23 i$ & $-7.82 - 2.17 i$ & $8.01 + 4.92 i$ & $-2.05 - 2.35 i$ \\
$17.58 - 3.15 i$ & $-0.33 + 0.37 i$ & $0.92 - 0.31 i$ & $-0.49 + 0.15 i$ & $0.76 + 0.52 i$ & $-3.19 + 1.62 i$ & $1.54 - 2.39 i$ & $-1.37 + 0.57 i$ \\
$19.94 - 2.01 i$ & $0.19 + 0.02 i$ & $-0.05 + 0.37 i$ & $-0.23 - 0.35 i$ & $-0.52 + 0.28 i$ & $-0.61 + 2.15 i$ & $-0.40 - 13.58 i$ & $0.31 + 9.61 i$ \\
$22.47 - 2.30 i$ & $-0.50 - 0.08 i$ & $1.08 + 0.37 i$ & $-0.86 - 0.29 i$ & $0.61 + 0.45 i$ & $4.44 - 5.29 i$ & $-15.82 + 7.74 i$ & $11.41 - 4.17 i$ \\
$25.64 - 1.28 i$ & $0.06 - 0.02 i$ & $0.01 + 0.08 i$ & $-0.00 - 0.18 i$ & $0.24 + 0.64 i$ & $3.70 - 7.18 i$ & $-6.38 + 21.29 i$ & $3.00 - 12.29 i$ \\
$27.92 - 2.56 i$ & $0.00 + 0.12 i$ & $-0.11 - 0.11 i$ & $0.28 - 0.13 i$ & $2.32 + 3.85 i$ & $-6.52 - 1.19 i$ & $10.25 + 0.12 i$ & $-4.15 + 0.65 i$ \\
$30.28 - 2.52 i$ & $0.37 - 0.03 i$ & $-0.70 - 0.03 i$ & $0.48 + 0.10 i$ & $5.84 - 1.02 i$ & $-4.92 + 2.21 i$ & $6.07 - 6.16 i$ & $-2.86 + 3.98 i$ \\
$32.14 - 1.20 i$ & $0.14 - 0.13 i$ & $0.32 + 0.30 i$ & $-0.34 - 0.20 i$ & $0.58 - 2.30 i$ & $-5.87 + 4.40 i$ & $4.59 - 9.71 i$ & $-0.46 + 5.47 i$ \\
$35.87 - 0.56 i$ & $-0.17 - 0.43 i$ & $0.85 + 1.01 i$ & $-0.62 - 0.60 i$ & $0.08 - 0.19 i$ & $-1.98 + 10.22 i$ & $-8.06 - 17.34 i$ & $9.00 + 7.15 i$ \\
$39.68 - 4.03 i$ & $0.10 - 0.24 i$ & $0.00 + 0.24 i$ & $-0.06 - 0.06 i$ & $4.11 - 2.04 i$ & $0.81 + 4.31 i$ & $-8.87 - 0.38 i$ & $6.38 - 2.97 i$ \\
$43.72 - 3.21 i$ & $0.37 - 0.52 i$ & $-0.28 + 0.41 i$ & $-0.04 - 0.26 i$ & $-1.35 - 3.01 i$ & $3.50 + 4.16 i$ & $-3.66 + 11.11 i$ & $1.00 - 10.05 i$ \\
    \end{tabular}
  \end{ruledtabular}
\caption{List of poles and residues of the MPA($\q$) representation of Pd.
}
    \label{tab:poles_Pd}
\end{table*} \newpage

\subsection{Ag} \FloatBarrier \begin{figure*}[hbt!]
    \centering
    \includegraphics[width=0.266\textwidth]{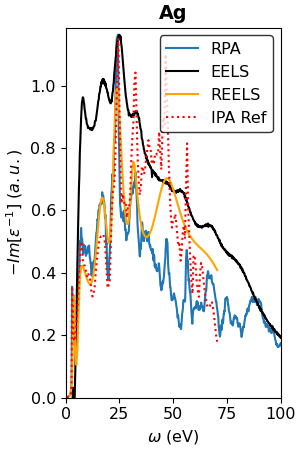}
    \includegraphics[width=0.2\textwidth]{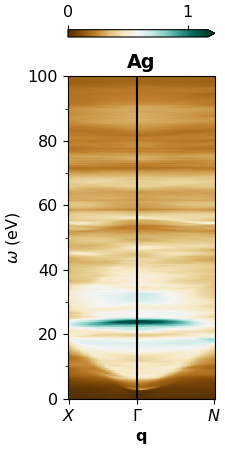}
    \includegraphics[width=0.465\textwidth]{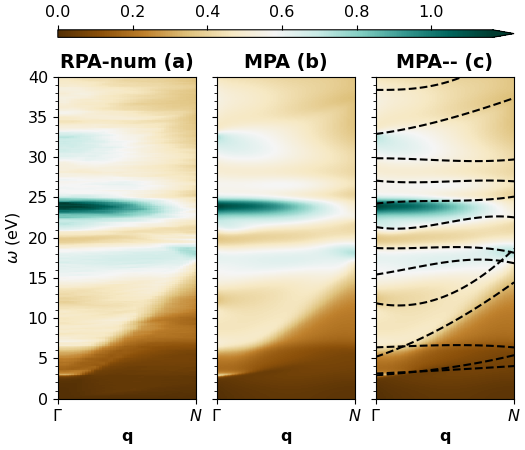}
    \caption{(left panel) Comparison of the computed RPA loss function of Ag in the optical limit with EELS, REELS  measurements, and reference IPA results. (middle panel) Spectral $Y (\omega, \q)$ band structure of Ag computed from the numerical RPA results along the $X \Gamma N$ $\q$-path. (right panels) Spectral $Y (\q, \omega)$ band structure of Ag in the $\Gamma N$ direction reconstructed with MPA($\q$) with a number of $n_Y = 13$ poles.}
    \label{fig:Ag}
\end{figure*} 

\begin{table*}[hbt!]
  \begin{ruledtabular}
    \begin{tabular}{cccccccc} 
      \\[-8pt]
       $\Omega_p~$(eV) & $\Omega'_{p}$ & $\Omega''_{p}$ & $\Omega'''_{p}$  & $R_p~$(eV) & $R'_{p}$ & $R''_{p}$ & $R'''_{p}$ \\
      \hline\\[-8pt]     
      $2.88 - 0.20 i$ & $0.89 + 0.23 i$ & $0.63 - 1.52 i$ & $0.32 - 0.15 i$ & $0.07 - 0.04 i$ & $-9.33 - 3.12 i$ & $10.70 + 6.05 i$ & $-4.95 - 3.44 i$ \\
$3.11 - 3.11 i$ & $0.49 - 0.01 i$ & $0.09 + 0.02 i$ & $0.01 - 0.01 i$ & $-0.07 - 0.33 i$ & $-0.84 + 3.08 i$ & $2.84 - 4.03 i$ & $-1.53 + 2.46 i$ \\
$5.18 - 2.94 i$ & $0.74 + 2.60 i$ & $1.87 - 0.65 i$ & $0.32 - 0.81 i$ & $0.11 - 0.01 i$ & $14.86 - 13.29 i$ & $-36.56 + 17.63 i$ & $20.80 - 8.32 i$ \\
$6.36 - 1.14 i$ & $0.19 + 0.01 i$ & $0.11 - 0.02 i$ & $-0.28 - 0.10 i$ & $0.20 - 0.25 i$ & $-5.07 - 3.16 i$ & $3.94 + 5.57 i$ & $-0.60 - 3.12 i$ \\
$11.85 - 1.06 i$ & $-0.45 - 1.35 i$ & $1.63 + 1.10 i$ & $-0.05 + 0.36 i$ & $-0.10 + 0.24 i$ & $2.63 - 9.03 i$ & $-8.95 + 5.97 i$ & $4.81 + 3.15 i$ \\
$15.41 - 3.04 i$ & $0.23 + 0.58 i$ & $0.52 - 0.74 i$ & $-0.61 + 0.40 i$ & $0.63 + 0.74 i$ & $-7.92 + 0.14 i$ & $16.14 - 2.71 i$ & $-10.35 + 3.22 i$ \\
$18.72 - 2.11 i$ & $-0.07 - 0.15 i$ & $0.33 + 0.66 i$ & $-0.33 - 0.48 i$ & $0.77 + 1.12 i$ & $-0.90 + 0.62 i$ & $-4.70 - 0.29 i$ & $4.62 - 1.70 i$ \\
$21.33 - 1.47 i$ & $-0.32 + 0.48 i$ & $1.21 - 1.33 i$ & $-0.78 + 0.93 i$ & $0.55 - 0.21 i$ & $-10.06 - 0.13 i$ & $18.74 - 0.03 i$ & $-10.46 + 0.08 i$ \\
$24.24 - 1.08 i$ & $0.16 - 0.05 i$ & $-0.32 + 0.13 i$ & $0.23 - 0.10 i$ & $0.87 + 0.56 i$ & $-0.37 + 6.58 i$ & $-3.94 - 11.69 i$ & $2.32 + 5.77 i$ \\
$27.10 - 2.56 i$ & $-0.13 + 0.01 i$ & $0.31 + 0.20 i$ & $-0.19 - 0.17 i$ & $1.30 + 1.02 i$ & $1.88 + 2.39 i$ & $-9.14 - 3.23 i$ & $5.88 + 1.02 i$ \\
$29.87 - 4.14 i$ & $0.01 - 0.04 i$ & $-0.14 - 0.06 i$ & $0.12 + 0.12 i$ & $2.46 - 1.68 i$ & $3.39 + 4.23 i$ & $-9.78 - 4.66 i$ & $4.97 + 0.99 i$ \\
$32.89 - 0.52 i$ & $0.15 - 0.43 i$ & $0.17 + 0.63 i$ & $-0.04 - 0.21 i$ & $0.09 + 0.14 i$ & $5.29 + 4.50 i$ & $-12.54 - 2.30 i$ & $5.17 - 2.01 i$ \\
$38.38 - 5.92 i$ & $-0.02 + 0.04 i$ & $0.11 - 0.12 i$ & $0.21 + 0.06 i$ & $5.21 + 0.09 i$ & $1.90 + 3.11 i$ & $-7.19 + 3.90 i$ & $1.84 - 4.95 i$ \\
$42.88 - 4.62 i$ & $-0.10 - 0.65 i$ & $-0.01 + 0.93 i$ & $0.37 - 0.60 i$ & $-0.36 - 3.49 i$ & $10.09 - 1.00 i$ & $-5.46 + 12.71 i$ & $-3.11 - 4.03 i$ \\
    \end{tabular}
  \end{ruledtabular}
\caption{List of poles and residues of the MPA($\q$) representation of Ag.
}
    \label{tab:poles_Ag}
\end{table*} \newpage

\subsection{Sn} \FloatBarrier \begin{figure*}[hbt!]
    \centering
    \includegraphics[width=0.266\textwidth]{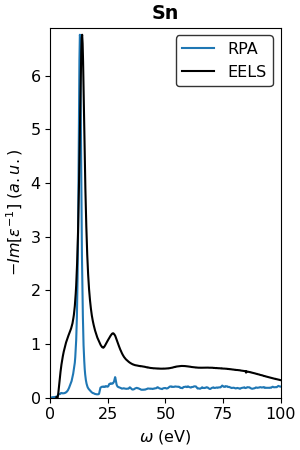}
    \includegraphics[width=0.2\textwidth]{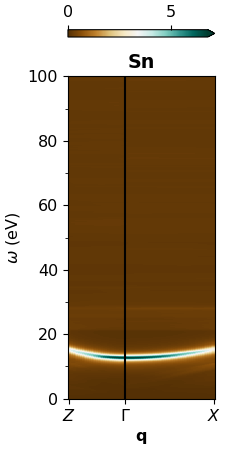}
    \includegraphics[width=0.465\textwidth]{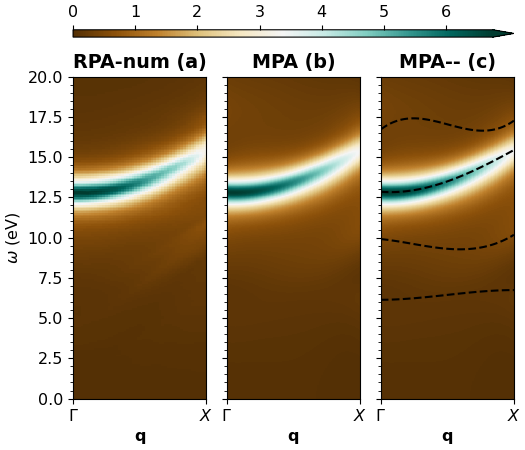}
    \caption{(left panel) Comparison of the computed RPA loss function of Sn in the optical limit with EELS measurements. (middle panel) Spectral $Y (\omega, \q)$ band structure of Sn computed from the numerical RPA results along the $Z \Gamma N$ $\q$-path. (right panels) Spectral $Y (\q, \omega)$ band structure of Sn in the $\Gamma X$ direction reconstructed with MPA($\q$) with a number of $n_Y = 4$ poles.}
    \label{fig:Sn}
\end{figure*} 

\begin{table*}[hbt!]
  \begin{ruledtabular}
    \begin{tabular}{cccccccc} 
      \\[-8pt]
       $\Omega_p~$(eV) & $\Omega'_{p}$ & $\Omega''_{p}$ & $\Omega'''_{p}$  & $R_p~$(eV) & $R'_{p}$ & $R''_{p}$ & $R'''_{p}$ \\
      \hline\\[-8pt]     
      $6.13 - 1.75 i$ & $-0.12 + 0.44 i$ & $0.56 + 0.06 i$ & $-0.27 - 0.40 i$ & $-0.06 + 0.22 i$ & $-6.56 - 1.72 i$ & $5.19 + 10.68 i$ & $1.32 - 11.33 i$ \\
$9.91 - 3.98 i$ & $-0.15 - 0.05 i$ & $-0.96 + 1.08 i$ & $1.03 - 0.70 i$ & $1.46 + 0.35 i$ & $-3.28 + 2.04 i$ & $1.61 - 2.08 i$ & $0.95 - 0.95 i$ \\
$12.85 - 0.74 i$ & $-0.10 + 0.00 i$ & $0.70 - 0.03 i$ & $-0.20 + 0.01 i$ & $5.16 + 0.07 i$ & $-0.26 - 1.01 i$ & $0.20 + 4.11 i$ & $-0.67 - 2.96 i$ \\
$16.71 - 2.57 i$ & $0.76 - 0.08 i$ & $-1.92 - 0.61 i$ & $1.20 + 0.80 i$ & $0.58 - 1.33 i$ & $-5.55 + 4.02 i$ & $16.17 - 6.86 i$ & $-12.23 + 3.10 i$ \\
    \end{tabular}
  \end{ruledtabular}
\caption{List of poles and residues of the MPA($\q$) representation of Sn.
}
    \label{tab:poles_Sn}
\end{table*} \newpage

\subsection{Ta} \FloatBarrier \begin{figure*}[hbt!]
    \centering
    \includegraphics[width=0.266\textwidth]{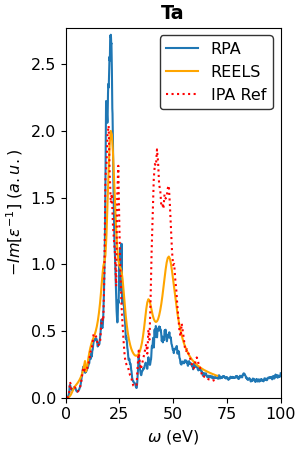}
    \includegraphics[width=0.2\textwidth]{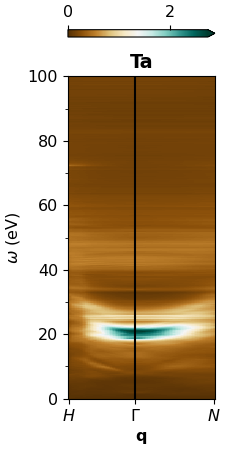}
    \includegraphics[width=0.465\textwidth]{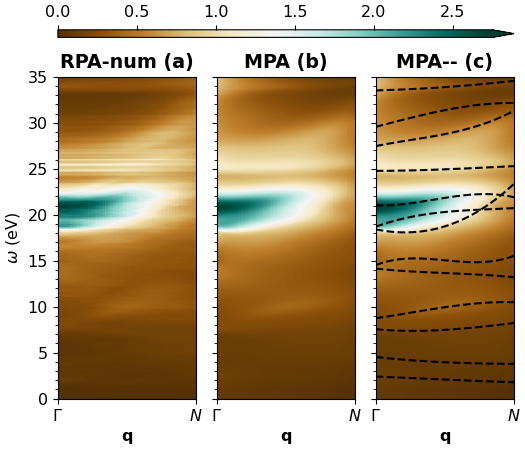}
    \caption{(left panel) Comparison of the computed RPA loss function of Ta in the optical limit with REELS measurements and reference IPA results. (middle panel) Spectral $Y (\omega, \q)$ band structure of Ta computed from the numerical RPA results along the $H \Gamma N$ $\q$-path. (right panels) Spectral $Y (\q, \omega)$ band structure of Ta in the $\Gamma N$ direction reconstructed with MPA($\q$) with a number of $n_Y = 13$ poles.}
    \label{fig:Ta}
\end{figure*} 

\begin{table*}[hbt!]
  \begin{ruledtabular}
    \begin{tabular}{cccccccc} 
      \\[-8pt]
       $\Omega_p~$(eV) & $\Omega'_{p}$ & $\Omega''_{p}$ & $\Omega'''_{p}$  & $R_p~$(eV) & $R'_{p}$ & $R''_{p}$ & $R'''_{p}$ \\
      \hline\\[-8pt]     
      $2.40 - 1.19 i$ & $-0.22 - 0.64 i$ & $-0.07 + 0.01 i$ & $0.06 + 0.03 i$ & $0.08 - 0.01 i$ & $-2.16 - 3.33 i$ & $-0.85 + 0.94 i$ & $2.76 + 0.31 i$ \\
$4.52 - 1.38 i$ & $-0.68 - 0.20 i$ & $0.48 + 0.14 i$ & $-0.10 - 0.05 i$ & $0.01 - 0.01 i$ & $-0.89 - 17.86 i$ & $10.00 + 3.95 i$ & $-1.13 + 2.53 i$ \\
$7.54 - 0.43 i$ & $-0.35 - 0.50 i$ & $0.80 + 0.12 i$ & $-0.26 + 0.07 i$ & $0.04 - 0.05 i$ & $-1.86 - 3.58 i$ & $-0.66 + 4.98 i$ & $1.53 - 2.39 i$ \\
$8.73 - 1.58 i$ & $0.38 + 0.68 i$ & $0.37 - 0.42 i$ & $-0.34 - 0.25 i$ & $-0.01 - 0.04 i$ & $-15.06 + 26.77 i$ & $19.53 - 20.28 i$ & $-6.55 - 2.41 i$ \\
$14.14 - 1.24 i$ & $-0.21 - 0.21 i$ & $0.31 - 0.42 i$ & $-0.23 + 0.56 i$ & $0.11 + 0.27 i$ & $-2.02 + 0.38 i$ & $1.67 + 11.10 i$ & $-0.88 - 9.60 i$ \\
$14.51 - 0.79 i$ & $0.89 - 1.30 i$ & $-2.10 + 1.76 i$ & $1.37 - 0.72 i$ & $0.05 + 0.10 i$ & $6.22 - 29.65 i$ & $-2.01 + 32.60 i$ & $5.94 - 6.07 i$ \\
$18.41 - 0.53 i$ & $-0.33 - 0.08 i$ & $0.74 + 0.26 i$ & $0.14 - 0.38 i$ & $-0.36 - 0.51 i$ & $-9.69 - 1.80 i$ & $17.19 + 3.32 i$ & $-8.73 - 0.29 i$ \\
$18.69 - 1.16 i$ & $0.59 + 0.01 i$ & $-0.62 - 0.03 i$ & $0.25 - 0.07 i$ & $1.14 + 0.82 i$ & $-2.88 + 1.77 i$ & $4.66 - 1.16 i$ & $-2.12 + 1.40 i$ \\
$21.02 - 1.63 i$ & $-0.06 - 0.02 i$ & $0.70 + 0.02 i$ & $-0.56 - 0.06 i$ & $4.35 + 0.48 i$ & $0.52 - 1.22 i$ & $-2.65 + 2.95 i$ & $1.30 - 2.46 i$ \\
$24.77 - 0.59 i$ & $0.00 - 0.09 i$ & $0.06 + 0.07 i$ & $-0.02 - 0.03 i$ & $0.06 - 0.29 i$ & $2.35 + 3.62 i$ & $5.85 - 1.82 i$ & $-7.40 - 1.23 i$ \\
$27.46 - 2.33 i$ & $0.24 - 0.19 i$ & $-0.22 + 0.30 i$ & $0.26 - 0.03 i$ & $0.06 + 1.63 i$ & $2.78 - 1.36 i$ & $-6.19 - 4.46 i$ & $2.30 + 6.53 i$ \\
$29.56 - 3.65 i$ & $0.27 + 0.01 i$ & $-0.02 + 0.20 i$ & $-0.09 - 0.03 i$ & $3.21 + 0.78 i$ & $-1.80 - 2.44 i$ & $-3.12 + 6.35 i$ & $3.43 - 3.83 i$ \\
$33.52 - 1.09 i$ & $0.02 + 0.17 i$ & $0.05 - 0.30 i$ & $-0.00 + 0.16 i$ & $0.11 - 0.08 i$ & $-0.06 - 6.63 i$ & $0.25 + 17.50 i$ & $-1.22 - 9.36 i$ \\
$35.04 - 4.08 i$ & $0.17 + 0.07 i$ & $0.03 + 0.09 i$ & $-0.03 - 0.09 i$ & $4.15 - 1.07 i$ & $-5.64 + 0.59 i$ & $7.04 + 0.27 i$ & $-3.52 - 0.11 i$ \\
    \end{tabular}
  \end{ruledtabular}
\caption{List of poles and residues of the MPA($\q$) representation of Ta.
}
    \label{tab:poles_Ta}
\end{table*} \newpage

\subsection{W} \FloatBarrier \begin{figure*}[hbt!]
    \centering
    \includegraphics[width=0.266\textwidth]{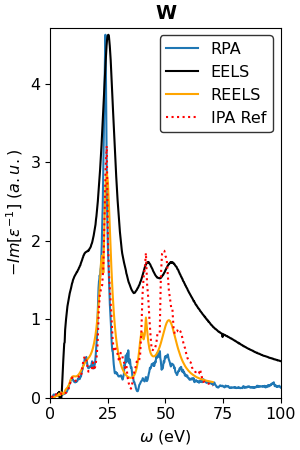}
    \includegraphics[width=0.2\textwidth]{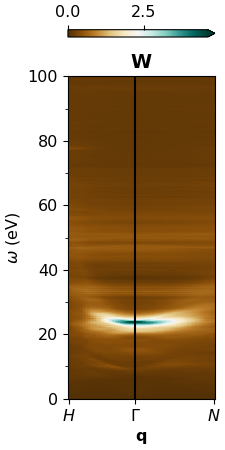}
    \includegraphics[width=0.465\textwidth]{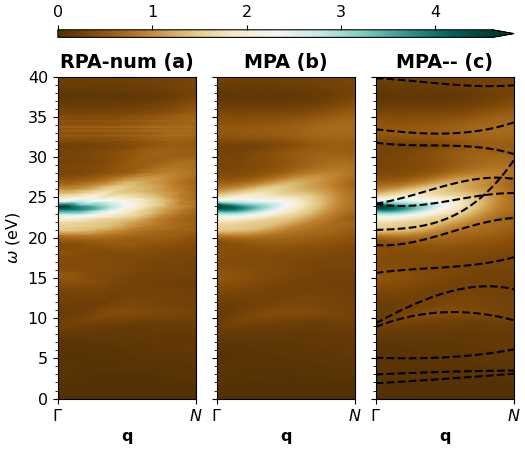}
    \caption{(left panel) Comparison of the computed RPA loss function of W in the optical limit with EELS, REELS measurements, and reference IPA results. (middle panel) Spectral $Y (\omega, \q)$ band structure of W computed from the numerical RPA results along the $H \Gamma N$ $\q$-path. (right panels) Spectral $Y (\q, \omega)$ band structure of W in the $\Gamma N$ direction reconstructed with MPA($\q$) with a number of $n_Y = 13$ poles.}
    \label{fig:W}
\end{figure*} 

\begin{table*}[hbt!]
  \begin{ruledtabular}
    \begin{tabular}{cccccccc} 
      \\[-8pt]
       $\Omega_p~$(eV) & $\Omega'_{p}$ & $\Omega''_{p}$ & $\Omega'''_{p}$  & $R_p~$(eV) & $R'_{p}$ & $R''_{p}$ & $R'''_{p}$ \\
      \hline\\[-8pt]     
      $1.90 - 1.90 i$ & $1.08 + 0.00 i$ & $0.15 + 0.00 i$ & $0.00 + 0.00 i$ & $0.19 - 0.11 i$ & $0.15 + 1.32 i$ & $0.19 + 0.20 i$ & $0.05 - 0.04 i$ \\
$3.00 - 1.95 i$ & $0.77 - 0.38 i$ & $-0.05 - 0.13 i$ & $-0.03 - 0.05 i$ & $-0.15 - 0.24 i$ & $1.06 - 0.69 i$ & $0.41 + 0.09 i$ & $0.12 - 0.02 i$ \\
$5.07 - 3.12 i$ & $0.22 - 0.72 i$ & $0.44 - 0.02 i$ & $0.17 + 0.07 i$ & $-0.45 + 0.37 i$ & $-0.28 - 0.12 i$ & $-0.57 + 0.14 i$ & $-0.07 + 0.08 i$ \\
$8.90 - 1.48 i$ & $1.41 + 0.68 i$ & $-1.35 - 0.60 i$ & $0.11 + 0.02 i$ & $0.20 + 0.04 i$ & $-0.58 - 1.01 i$ & $-0.72 + 0.55 i$ & $-0.37 - 0.20 i$ \\
$9.36 - 4.07 i$ & $1.19 + 1.78 i$ & $-0.06 - 0.68 i$ & $-0.47 - 0.53 i$ & $-0.04 + 0.21 i$ & $1.06 + 0.68 i$ & $0.01 + 0.26 i$ & $-0.27 - 0.01 i$ \\
$15.61 - 1.72 i$ & $0.33 - 0.14 i$ & $-0.44 - 0.24 i$ & $0.36 + 0.39 i$ & $0.41 + 0.53 i$ & $-0.39 + 1.01 i$ & $-0.70 + 0.07 i$ & $-0.70 - 0.51 i$ \\
$19.08 - 1.30 i$ & $-0.10 - 0.42 i$ & $1.16 + 0.69 i$ & $-0.72 - 0.14 i$ & $-0.05 + 0.72 i$ & $-2.15 + 1.10 i$ & $0.36 - 0.25 i$ & $-0.15 - 0.74 i$ \\
$20.96 - 1.00 i$ & $0.06 - 0.03 i$ & $-0.03 + 0.37 i$ & $0.83 - 0.94 i$ & $0.84 - 0.10 i$ & $-1.48 + 2.63 i$ & $-1.27 + 0.65 i$ & $-0.88 - 0.69 i$ \\
$24.22 - 0.63 i$ & $-0.27 - 0.17 i$ & $0.89 + 0.31 i$ & $-0.49 - 0.56 i$ & $1.25 + 1.29 i$ & $5.97 - 3.01 i$ & $-3.18 + 1.04 i$ & $0.29 + 0.89 i$ \\
$24.22 - 2.10 i$ & $0.20 + 0.22 i$ & $0.57 - 0.19 i$ & $-0.53 + 0.09 i$ & $5.08 - 0.47 i$ & $-2.94 + 0.22 i$ & $0.37 - 0.96 i$ & $0.54 + 0.66 i$ \\
$31.82 - 1.24 i$ & $-0.15 + 0.14 i$ & $0.35 - 0.25 i$ & $-0.29 + 0.06 i$ & $-0.27 - 0.12 i$ & $-0.58 - 2.63 i$ & $-1.42 - 1.18 i$ & $-0.15 - 0.07 i$ \\
$33.48 - 4.75 i$ & $-0.08 - 0.22 i$ & $-0.03 + 0.51 i$ & $0.18 - 0.37 i$ & $5.34 + 0.19 i$ & $2.19 - 0.74 i$ & $-1.25 + 0.74 i$ & $0.81 - 0.17 i$ \\
$39.83 - 2.71 i$ & $-0.03 - 0.11 i$ & $-0.14 + 0.25 i$ & $0.13 - 0.12 i$ & $2.04 - 1.84 i$ & $-0.71 - 1.11 i$ & $-1.11 + 1.49 i$ & $0.45 - 0.74 i$ \\
$43.91 - 1.91 i$ & $0.14 - 0.23 i$ & $-0.38 + 0.12 i$ & $0.26 - 0.06 i$ & $-0.49 - 3.88 i$ & $1.02 + 0.58 i$ & $0.73 - 0.32 i$ & $-0.23 - 0.03 i$ \\
    \end{tabular}
  \end{ruledtabular}
\caption{List of poles and residues of the MPA($\q$) representation of W.
}
    \label{tab:poles_W}
\end{table*} \newpage

\subsection{Os} \FloatBarrier \begin{figure*}[hbt!]
    \centering
    \includegraphics[width=0.250\textwidth]{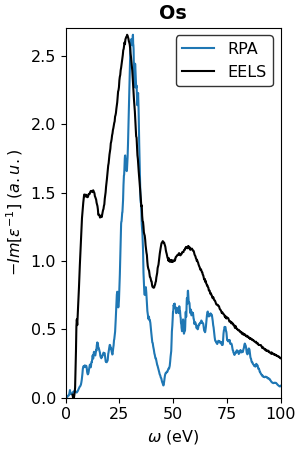}
    \includegraphics[width=0.333\textwidth]{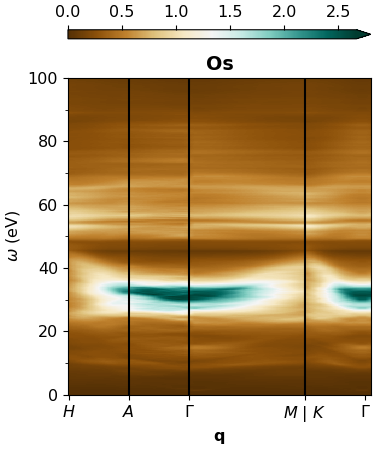}
    \includegraphics[width=0.406\textwidth]{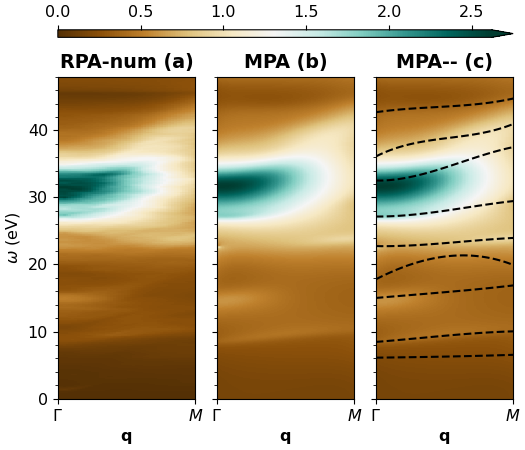}
    \caption{(left panel) Comparison of the computed RPA loss function of Os in the optical limit with EELS measurements. (middle panel) Spectral $Y (\omega, \q)$ band structure of Os computed from the numerical RPA results along the $H A \Gamma M|K\Gamma$ $\q$-path. (right panels) Spectral $Y (\q, \omega)$ band structure of Os in the $\Gamma M$ direction reconstructed with MPA($\q$) with a number of $n_Y = 9$ poles.}
    \label{fig:Os}
\end{figure*} 

\begin{table*}[hbt!]
  \begin{ruledtabular}
    \begin{tabular}{cccccccc} 
      \\[-8pt]
       $\Omega_p~$(eV) & $\Omega'_{p}$ & $\Omega''_{p}$ & $\Omega'''_{p}$  & $R_p~$(eV) & $R'_{p}$ & $R''_{p}$ & $R'''_{p}$ \\
      \hline\\[-8pt]     
      $6.10 - 4.66 i$ & $0.50 - 0.55 i$ & $-0.05 + 0.12 i$ & $-0.06 + 0.11 i$ & $-0.11 - 0.28 i$ & $2.68 + 4.13 i$ & $2.11 + 1.19 i$ & $-0.73 - 1.60 i$ \\
$8.45 - 0.68 i$ & $0.39 - 0.17 i$ & $0.21 + 0.22 i$ & $-0.22 - 0.12 i$ & $0.07 - 0.02 i$ & $7.67 - 5.48 i$ & $-10.71 + 1.59 i$ & $3.39 + 1.81 i$ \\
$15.00 - 1.99 i$ & $0.21 + 0.21 i$ & $0.05 - 0.75 i$ & $0.10 - 0.29 i$ & $0.66 + 0.38 i$ & $-0.93 - 1.09 i$ & $-4.13 + 4.30 i$ & $3.91 - 2.78 i$ \\
$17.77 - 5.29 i$ & $1.50 - 1.06 i$ & $-0.98 + 0.83 i$ & $-0.21 + 0.02 i$ & $-0.21 - 0.72 i$ & $-0.78 - 15.74 i$ & $25.88 + 17.15 i$ & $-19.81 - 5.10 i$ \\
$22.74 + 0.06 i$ & $-0.03 - 0.90 i$ & $0.29 + 1.68 i$ & $-0.15 - 0.87 i$ & $0.03 + 0.05 i$ & $-5.42 - 9.26 i$ & $16.54 + 2.31 i$ & $-9.22 + 1.23 i$ \\
$27.16 - 0.77 i$ & $-0.01 + 0.09 i$ & $0.37 - 0.49 i$ & $-0.19 + 0.34 i$ & $0.37 - 0.21 i$ & $-2.32 + 2.64 i$ & $-0.66 - 4.48 i$ & $0.13 + 0.51 i$ \\
$32.52 - 3.36 i$ & $-0.02 - 0.06 i$ & $0.72 + 0.00 i$ & $-0.35 - 0.29 i$ & $7.44 + 4.32 i$ & $1.94 - 1.15 i$ & $-3.12 + 6.07 i$ & $1.06 - 2.63 i$ \\
$36.10 - 5.71 i$ & $0.47 + 0.71 i$ & $-0.65 - 1.00 i$ & $0.42 + 0.48 i$ & $3.64 - 0.76 i$ & $-7.18 - 0.18 i$ & $14.13 + 0.25 i$ & $-8.14 + 0.25 i$ \\
$42.71 - 2.88 i$ & $0.13 + 0.24 i$ & $-0.18 - 0.38 i$ & $0.15 + 0.18 i$ & $0.61 + 0.30 i$ & $-4.40 + 4.79 i$ & $5.14 - 9.72 i$ & $-2.52 + 6.12 i$ \\
$50.22 - 6.73 i$ & $0.38 - 0.30 i$ & $-0.41 - 0.11 i$ & $0.08 + 0.08 i$ & $3.70 - 5.76 i$ & $2.33 + 3.83 i$ & $-0.85 - 4.10 i$ & $1.83 + 1.97 i$ \\
    \end{tabular}
  \end{ruledtabular}
\caption{List of poles and residues of the MPA($\q$) representation of Os.
}
    \label{tab:poles_Os}
\end{table*} \newpage

\subsection{Pt} \FloatBarrier \begin{figure*}[hbt!]
    \centering
    \includegraphics[width=0.266\textwidth]{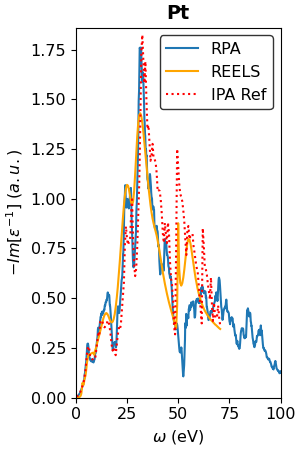}
    \includegraphics[width=0.2\textwidth]{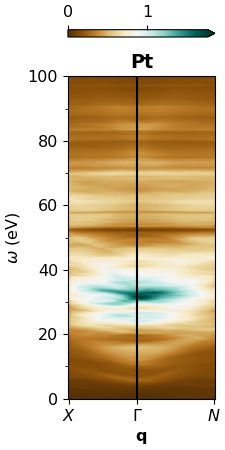}
    \includegraphics[width=0.465\textwidth]{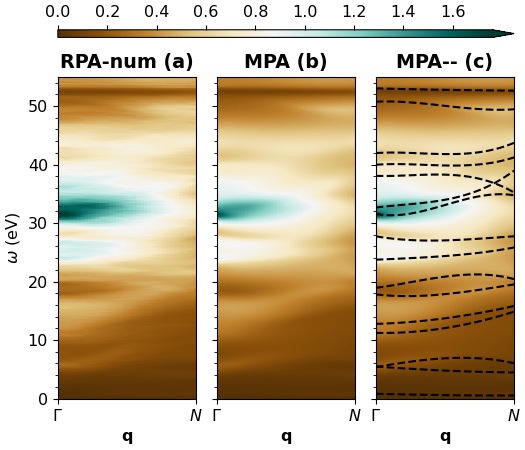}
    \caption{(left panel) Comparison of the computed RPA loss function of Pt in the optical limit with REELS measurements and reference IPA results. (middle panel) Spectral $Y (\omega, \q)$ band structure of Pt computed from the numerical RPA results along the $X \Gamma N$ $\q$-path. (right panels) Spectral $Y (\q, \omega)$ band structure of Pt in the $\Gamma N$ direction reconstructed with MPA($\q$) with a number of $n_Y = 15$ poles.}
    \label{fig:Pt}
\end{figure*} 

\begin{table*}[hbt!]
  \begin{ruledtabular}
    \begin{tabular}{cccccccc} 
      \\[-8pt]
       $\Omega_p~$(eV) & $\Omega'_{p}$ & $\Omega''_{p}$ & $\Omega'''_{p}$  & $R_p~$(eV) & $R'_{p}$ & $R''_{p}$ & $R'''_{p}$ \\
      \hline\\[-8pt]     
      $0.82 - 0.82 i$ & $-1.41 - 0.00 i$ & $0.45 + 0.00 i$ & $0.24 - 0.00 i$ & $-0.00 - 0.01 i$ & $0.56 + 2.63 i$ & $-0.05 + 0.25 i$ & $-0.06 - 0.01 i$ \\
$5.38 - 1.31 i$ & $1.40 + 0.39 i$ & $0.00 - 1.10 i$ & $-1.09 + 0.42 i$ & $0.19 + 0.00 i$ & $-0.14 - 3.07 i$ & $-1.10 + 0.41 i$ & $-0.01 + 0.15 i$ \\
$5.48 - 1.52 i$ & $-0.50 - 0.51 i$ & $0.11 + 0.60 i$ & $-0.01 - 0.01 i$ & $0.09 - 0.10 i$ & $-2.38 + 2.20 i$ & $0.40 - 0.56 i$ & $0.25 - 0.40 i$ \\
$11.23 - 2.48 i$ & $-0.05 - 0.17 i$ & $0.89 - 0.50 i$ & $-0.10 + 0.24 i$ & $0.55 - 0.10 i$ & $1.44 + 0.93 i$ & $0.61 - 1.05 i$ & $0.42 - 0.41 i$ \\
$12.76 - 4.34 i$ & $-0.34 + 1.32 i$ & $0.83 - 1.30 i$ & $-0.17 + 0.42 i$ & $-0.03 - 0.84 i$ & $-3.01 - 0.57 i$ & $0.60 + 0.16 i$ & $0.92 + 0.03 i$ \\
$17.81 - 2.31 i$ & $-0.31 + 0.05 i$ & $0.73 + 0.25 i$ & $-0.24 - 0.17 i$ & $-0.29 + 1.55 i$ & $-4.69 - 0.52 i$ & $1.76 - 0.22 i$ & $1.16 + 0.57 i$ \\
$18.93 - 3.25 i$ & $0.48 - 0.92 i$ & $0.26 + 1.38 i$ & $-0.60 - 0.32 i$ & $0.90 - 1.50 i$ & $0.59 + 0.56 i$ & $-1.30 - 0.43 i$ & $-0.37 - 0.31 i$ \\
$23.73 - 0.66 i$ & $0.10 - 0.35 i$ & $-0.05 + 0.60 i$ & $0.12 - 0.28 i$ & $0.32 - 0.17 i$ & $-0.77 - 2.04 i$ & $-0.87 + 1.98 i$ & $-0.04 + 0.43 i$ \\
$27.74 - 1.22 i$ & $-0.29 + 0.01 i$ & $0.46 + 0.28 i$ & $-0.16 - 0.46 i$ & $0.08 + 0.76 i$ & $-4.30 - 0.62 i$ & $1.35 + 0.69 i$ & $-0.15 + 0.70 i$ \\
$31.58 - 0.70 i$ & $-0.28 - 0.15 i$ & $1.41 + 0.19 i$ & $-0.93 - 0.12 i$ & $0.49 + 0.11 i$ & $-5.71 + 3.56 i$ & $-0.60 - 1.15 i$ & $0.87 - 0.66 i$ \\
$32.65 - 2.36 i$ & $0.18 + 0.07 i$ & $-0.28 - 0.07 i$ & $0.49 - 0.07 i$ & $2.39 + 0.96 i$ & $-0.25 + 1.28 i$ & $-1.37 - 0.37 i$ & $-0.18 - 0.80 i$ \\
$38.07 - 3.61 i$ & $-0.04 - 0.04 i$ & $0.37 - 0.11 i$ & $-0.47 + 0.03 i$ & $3.58 + 4.52 i$ & $-0.30 - 0.96 i$ & $-0.80 + 0.60 i$ & $0.17 - 0.67 i$ \\
$39.97 - 1.74 i$ & $0.08 + 0.04 i$ & $-0.34 - 0.39 i$ & $0.32 + 0.39 i$ & $1.33 + 0.17 i$ & $-1.80 - 1.11 i$ & $-0.43 + 0.97 i$ & $0.27 + 0.11 i$ \\
$41.96 - 3.65 i$ & $0.07 - 0.01 i$ & $-0.37 - 0.01 i$ & $0.37 + 0.10 i$ & $2.22 - 5.08 i$ & $-1.65 - 0.25 i$ & $-0.12 + 0.14 i$ & $-0.01 + 0.29 i$ \\
$50.74 - 2.17 i$ & $0.04 + 0.06 i$ & $-0.31 + 0.08 i$ & $0.21 - 0.11 i$ & $-0.02 - 0.12 i$ & $-1.73 + 1.97 i$ & $-0.49 + 1.33 i$ & $0.02 + 1.33 i$ \\
$53.03 - 0.67 i$ & $-0.03 - 0.03 i$ & $0.02 + 0.06 i$ & $-0.01 - 0.03 i$ & $-0.09 - 0.17 i$ & $2.28 - 3.65 i$ & $-2.80 + 1.75 i$ & $0.65 - 0.08 i$ \\
$62.54 - 5.92 i$ & $0.44 - 0.11 i$ & $-0.70 - 0.29 i$ & $0.40 + 0.05 i$ & $4.15 - 1.29 i$ & $1.11 + 1.69 i$ & $0.98 - 0.65 i$ & $-0.28 + 0.27 i$ \\
    \end{tabular}
  \end{ruledtabular}
\caption{List of poles and residues of the MPA($\q$) representation of Pt.
}
    \label{tab:poles_Pt}
\end{table*} \newpage

\subsection{Au} \FloatBarrier \begin{figure*}[hbt!]
    \centering
    \includegraphics[width=0.266\textwidth]{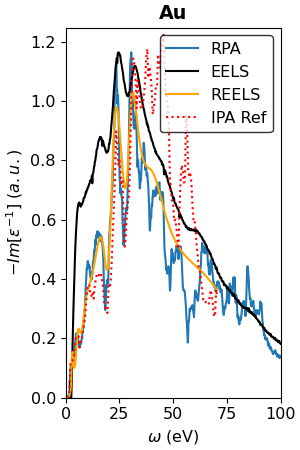}
    \includegraphics[width=0.2\textwidth]{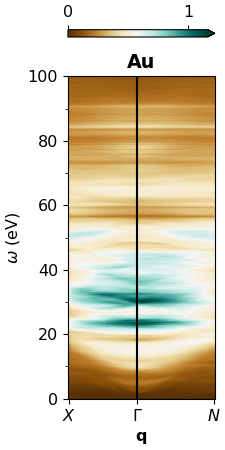}
    \includegraphics[width=0.465\textwidth]{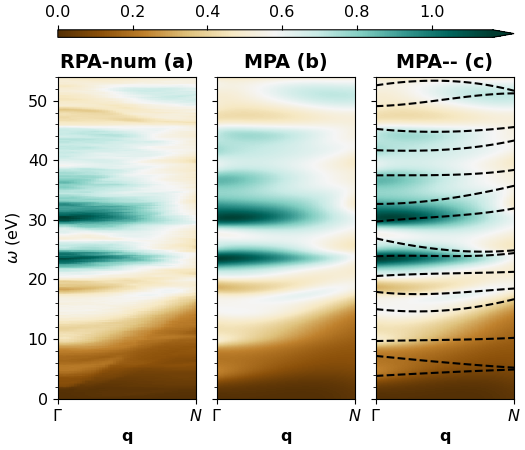}
    \caption{(left panel) Comparison of the computed RPA loss function of Au in the optical limit with EELS, REELS measurements,and reference IPAresults. (middle panel) Spectral $Y (\omega, \q)$ band structure of Au computed from the numerical RPA results along the $X \Gamma N$ $\q$-path. (right panels) Spectral $Y (\q, \omega)$ band structure of Au in the $\Gamma N$ direction reconstructed with MPA($\q$) with a number of $n_Y = 15$ poles.}
    \label{fig:Au}
\end{figure*} 

\begin{table*}[hbt!]
  \begin{ruledtabular}
    \begin{tabular}{cccccccc} 
      \\[-8pt]
       $\Omega_p~$(eV) & $\Omega'_{p}$ & $\Omega''_{p}$ & $\Omega'''_{p}$  & $R_p~$(eV) & $R'_{p}$ & $R''_{p}$ & $R'''_{p}$ \\
      \hline\\[-8pt]     
      $3.80 - 2.25 i$ & $0.85 - 0.28 i$ & $0.03 - 0.20 i$ & $-0.01 - 0.03 i$ & $0.32 - 0.41 i$ & $-1.46 - 1.35 i$ & $0.55 + 1.30 i$ & $0.34 + 0.36 i$ \\
$7.16 - 2.44 i$ & $-0.37 - 0.84 i$ & $0.04 + 0.06 i$ & $0.06 - 0.08 i$ & $-0.15 - 0.10 i$ & $0.89 - 1.31 i$ & $-0.77 + 0.41 i$ & $-0.14 + 0.17 i$ \\
$9.65 - 1.10 i$ & $0.16 - 0.35 i$ & $-0.14 - 0.02 i$ & $0.14 - 0.04 i$ & $0.22 - 0.13 i$ & $-2.18 - 1.74 i$ & $0.20 + 0.90 i$ & $0.55 + 0.39 i$ \\
$15.02 - 1.78 i$ & $-0.35 + 0.06 i$ & $0.59 - 0.12 i$ & $-0.03 + 0.12 i$ & $0.23 + 0.15 i$ & $-0.89 - 0.81 i$ & $-0.93 + 0.22 i$ & $0.01 + 0.60 i$ \\
$17.92 - 1.64 i$ & $-0.30 - 0.13 i$ & $0.60 + 0.42 i$ & $-0.23 - 0.37 i$ & $0.06 + 0.63 i$ & $-0.62 - 2.06 i$ & $-1.03 - 0.27 i$ & $-0.54 + 0.45 i$ \\
$20.56 - 1.21 i$ & $0.15 + 0.00 i$ & $-0.16 - 0.32 i$ & $0.09 + 0.10 i$ & $0.20 + 0.03 i$ & $0.43 + 1.43 i$ & $-0.08 + 1.08 i$ & $0.08 + 0.36 i$ \\
$23.87 - 1.86 i$ & $0.09 - 0.01 i$ & $-0.28 + 0.06 i$ & $0.24 - 0.01 i$ & $1.58 + 0.85 i$ & $-1.31 + 0.19 i$ & $-0.59 + 0.49 i$ & $0.20 - 0.54 i$ \\
$26.89 - 1.96 i$ & $-0.39 - 0.07 i$ & $0.17 + 0.28 i$ & $0.09 - 0.48 i$ & $0.60 + 0.44 i$ & $-3.13 - 3.46 i$ & $0.54 + 0.90 i$ & $0.36 + 0.11 i$ \\
$29.69 - 1.76 i$ & $0.16 + 0.02 i$ & $-0.14 - 0.22 i$ & $0.14 + 0.15 i$ & $1.03 - 0.54 i$ & $-0.74 + 2.59 i$ & $-2.36 - 1.23 i$ & $0.12 - 0.22 i$ \\
$32.67 - 3.20 i$ & $0.02 - 0.25 i$ & $0.19 + 0.42 i$ & $-0.02 - 0.20 i$ & $1.44 + 0.99 i$ & $1.03 + 0.74 i$ & $-0.79 - 0.18 i$ & $0.08 + 0.04 i$ \\
$37.46 - 2.72 i$ & $0.01 - 0.04 i$ & $-0.05 - 0.02 i$ & $0.08 + 0.09 i$ & $1.77 + 1.26 i$ & $-1.92 - 0.75 i$ & $-0.28 + 0.67 i$ & $0.49 + 0.01 i$ \\
$41.69 - 3.03 i$ & $-0.02 - 0.19 i$ & $0.04 + 0.37 i$ & $0.06 - 0.17 i$ & $2.54 + 0.20 i$ & $-1.24 + 1.82 i$ & $-0.34 - 0.34 i$ & $-0.13 - 0.38 i$ \\
$45.27 - 2.40 i$ & $-0.11 + 0.01 i$ & $0.14 - 0.08 i$ & $-0.02 + 0.04 i$ & $1.39 + 0.51 i$ & $0.17 - 0.52 i$ & $0.76 - 0.03 i$ & $-0.76 + 0.84 i$ \\
$49.08 - 3.87 i$ & $0.04 - 0.21 i$ & $0.19 + 0.19 i$ & $-0.14 - 0.04 i$ & $3.87 - 1.05 i$ & $-0.04 + 0.75 i$ & $0.54 - 0.75 i$ & $0.48 + 0.62 i$ \\
$52.56 - 3.73 i$ & $0.14 - 0.11 i$ & $-0.14 - 0.02 i$ & $-0.02 - 0.03 i$ & $1.94 - 4.04 i$ & $0.53 + 3.63 i$ & $-0.69 + 0.05 i$ & $-0.29 - 0.26 i$ \\
$55.57 - 3.22 i$ & $0.10 - 0.11 i$ & $-0.12 - 0.03 i$ & $-0.02 + 0.02 i$ & $-2.52 - 1.08 i$ & $3.40 + 0.97 i$ & $0.78 + 0.81 i$ & $0.25 + 0.11 i$ \\
    \end{tabular}
  \end{ruledtabular}
\caption{List of poles and residues of the MPA($\q$) representation of Au.
}
    \label{tab:poles_Au}
\end{table*} \newpage

\subsection{Tl} \FloatBarrier \begin{figure*}[hbt!]
    \centering
    \includegraphics[width=0.250\textwidth]{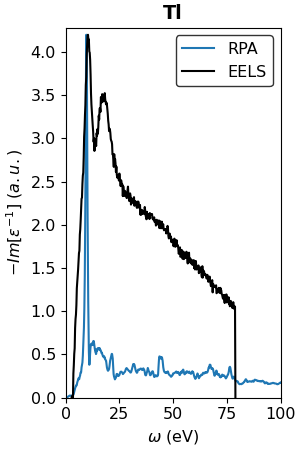}
    \includegraphics[width=0.333\textwidth]{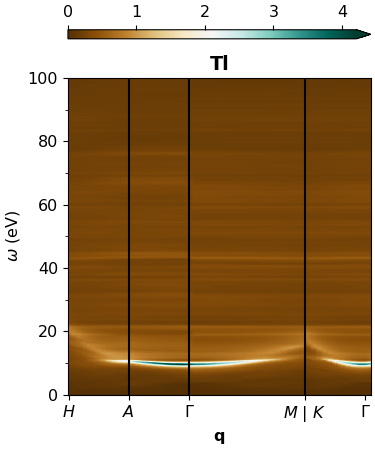}
    \includegraphics[width=0.406\textwidth]{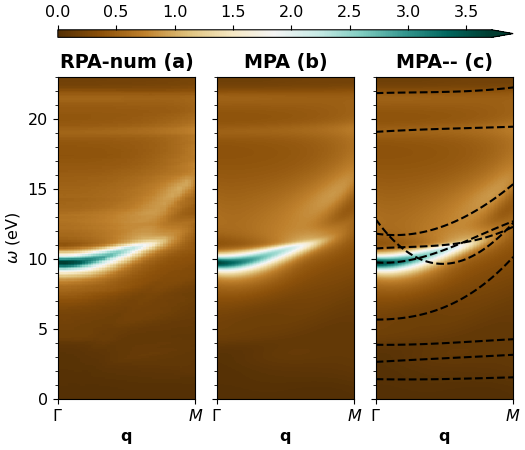}
    \caption{(left panel) Comparison of the computed RPA loss function of Tl in the optical limit with EELS measurements. (middle panel) Spectral $Y (\omega, \q)$ band structure of Tl computed from the numerical RPA results along the $H A \Gamma M|K \Gamma$ $\q$-path. (right panels) Spectral $Y (\q, \omega)$ band structure of Tl in the $\Gamma M$ direction reconstructed with MPA($\q$) with a number of $n_Y = 10$ poles.}
    \label{fig:Tl}
\end{figure*} 

\begin{table*}[hbt!]
  \begin{ruledtabular}
    \begin{tabular}{cccccccc} 
      \\[-8pt]
       $\Omega_p~$(eV) & $\Omega'_{p}$ & $\Omega''_{p}$ & $\Omega'''_{p}$  & $R_p~$(eV) & $R'_{p}$ & $R''_{p}$ & $R'''_{p}$ \\
      \hline\\[-8pt]     
$1.39 - 0.86 i$ & $0.36 - 1.07 i$ & $0.18 + 0.98 i$ & $0.02 - 0.50 i$ & $0.02 + 0.01 i$ & $1.00 - 4.66 i$ & $-2.77 - 2.29 i$ & $-1.22 - 0.17 i$ \\
$2.62 - 1.76 i$ & $-0.03 + 0.70 i$ & $0.38 - 0.74 i$ & $0.24 - 0.26 i$ & $0.16 - 0.22 i$ & $0.81 - 2.70 i$ & $-0.46 + 0.53 i$ & $-0.11 + 0.08 i$ \\
$3.85 - 1.24 i$ & $-0.26 + 0.59 i$ & $0.77 - 0.84 i$ & $-0.08 - 0.42 i$ & $-0.11 - 0.29 i$ & $-2.72 - 2.99 i$ & $-1.52 + 0.44 i$ & $0.08 + 0.09 i$ \\
$5.64 - 2.42 i$ & $-0.16 + 0.38 i$ & $0.92 + 0.04 i$ & $0.72 - 0.16 i$ & $-0.69 - 0.09 i$ & $-0.79 - 2.65 i$ & $-2.37 + 2.28 i$ & $-0.08 + 0.23 i$ \\
$9.74 - 0.53 i$ & $-0.16 - 0.17 i$ & $1.36 + 0.27 i$ & $-0.60 - 0.10 i$ & $2.10 - 0.16 i$ & $0.02 - 1.68 i$ & $-3.96 + 3.55 i$ & $1.99 - 1.99 i$ \\
$10.73 - 1.16 i$ & $0.08 + 0.43 i$ & $-0.29 - 0.08 i$ & $0.49 - 0.30 i$ & $-1.28 - 0.27 i$ & $-5.44 - 1.86 i$ & $4.12 + 1.23 i$ & $-1.20 + 0.49 i$ \\
$11.79 - 7.16 i$ & $-0.53 + 0.45 i$ & $0.16 + 1.27 i$ & $0.39 - 0.77 i$ & $5.62 - 0.26 i$ & $-1.71 + 0.04 i$ & $0.45 - 1.08 i$ & $-0.46 + 0.83 i$ \\
$12.82 - 3.93 i$ & $-2.12 - 0.04 i$ & $2.07 + 1.25 i$ & $-0.03 - 1.09 i$ & $-0.37 + 0.74 i$ & $-0.38 + 1.09 i$ & $0.89 - 2.29 i$ & $-1.98 + 0.24 i$ \\
$19.06 - 0.54 i$ & $0.07 - 0.02 i$ & $-0.05 + 0.01 i$ & $0.03 + 0.00 i$ & $0.10 - 0.01 i$ & $0.74 + 2.91 i$ & $-0.52 - 2.33 i$ & $0.08 + 0.98 i$ \\
$21.83 - 0.73 i$ & $0.03 - 0.01 i$ & $-0.05 + 0.02 i$ & $0.06 - 0.02 i$ & $0.17 + 0.22 i$ & $0.42 + 0.77 i$ & $-1.74 + 0.02 i$ & $0.34 + 0.68 i$ \\
$26.30 - 4.19 i$ & $0.48 - 0.45 i$ & $-0.32 - 0.31 i$ & $-0.23 - 0.05 i$ & $1.80 - 1.87 i$ & $3.11 + 1.16 i$ & $-1.05 + 1.24 i$ & $0.56 - 0.77 i$ \\
    \end{tabular}
  \end{ruledtabular}
\caption{List of poles and residues of the MPA($\q$) representation of Tl.
}
    \label{tab:poles_Tl}
\end{table*} \newpage

\subsection{Pb} \FloatBarrier \begin{figure*}[hbt!]
    \centering
    \includegraphics[width=0.266\textwidth]{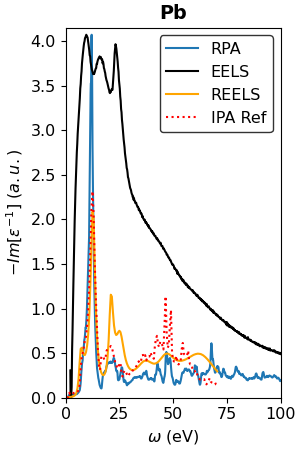}
    \includegraphics[width=0.2\textwidth]{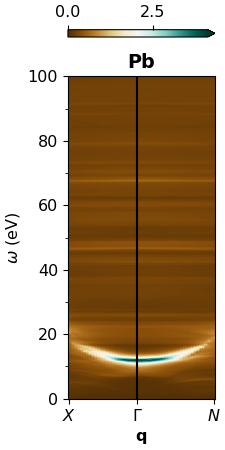}
    \includegraphics[width=0.465\textwidth]{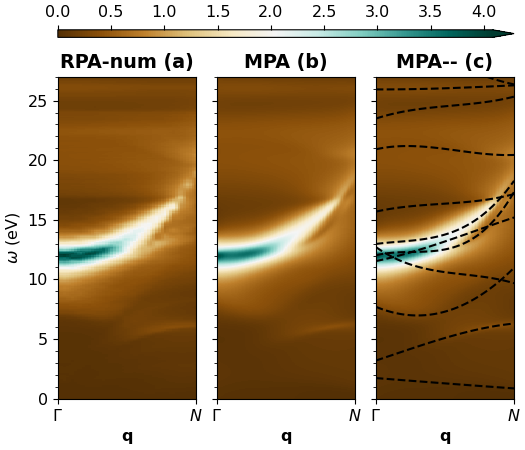}
    \caption{(left panel) Comparison of the computed RPA loss function of Pb in the optical limit with optical measurements. (middle panel) Spectral $Y (\omega, \q)$ band structure of Pb computed from the numerical RPA results along the $X \Gamma N$ $\q$-path. (right panels) Spectral $Y (\q, \omega)$ band structure of Pb in the $\Gamma N$ direction reconstructed with MPA($\q$) with a number of $n_Y = 12$ poles.}
    \label{fig:Pb}
\end{figure*} 

\begin{table*}[hbt!]
  \begin{ruledtabular}
    \begin{tabular}{cccccccc} 
      \\[-8pt]
       $\Omega_p~$(eV) & $\Omega'_{p}$ & $\Omega''_{p}$ & $\Omega'''_{p}$  & $R_p~$(eV) & $R'_{p}$ & $R''_{p}$ & $R'''_{p}$ \\
      \hline\\[-8pt]     
      $1.72 - 1.64 i$ & $-0.95 - 0.10 i$ & $-0.02 + 0.02 i$ & $0.04 - 0.00 i$ & $0.11 - 0.06 i$ & $-5.48 - 2.53 i$ & $2.57 + 0.53 i$ & $1.33 + 0.73 i$ \\
$3.18 - 3.05 i$ & $-0.69 + 3.26 i$ & $1.01 - 0.75 i$ & $-0.13 - 0.67 i$ & $0.01 - 0.07 i$ & $1.03 + 0.31 i$ & $-2.24 + 1.58 i$ & $-0.86 + 0.57 i$ \\
$7.71 - 1.89 i$ & $-1.37 + 0.46 i$ & $1.99 + 0.31 i$ & $0.33 - 1.19 i$ & $0.01 - 0.65 i$ & $-2.52 + 0.42 i$ & $-3.21 - 0.51 i$ & $1.43 - 1.49 i$ \\
$11.53 - 0.88 i$ & $0.41 + 0.03 i$ & $0.39 + 0.23 i$ & $-0.11 - 0.92 i$ & $0.66 + 0.65 i$ & $7.76 - 11.98 i$ & $-11.28 + 11.03 i$ & $9.96 - 4.13 i$ \\
$12.02 - 0.54 i$ & $0.48 - 0.36 i$ & $-1.67 + 1.11 i$ & $2.08 - 0.78 i$ & $1.88 + 0.26 i$ & $-2.77 - 1.17 i$ & $-4.89 + 5.57 i$ & $6.13 - 4.91 i$ \\
$12.73 - 7.96 i$ & $-2.20 + 1.44 i$ & $1.79 - 0.14 i$ & $-0.32 - 0.90 i$ & $-1.43 + 0.77 i$ & $-3.60 - 0.14 i$ & $-0.45 - 1.77 i$ & $2.06 + 2.41 i$ \\
$12.97 - 0.90 i$ & $0.32 - 1.20 i$ & $-0.77 + 3.14 i$ & $1.26 - 1.90 i$ & $1.21 + 0.02 i$ & $-2.08 + 2.88 i$ & $3.38 - 9.94 i$ & $-2.97 + 6.05 i$ \\
$15.68 - 3.04 i$ & $0.31 + 0.14 i$ & $-0.71 + 1.05 i$ & $0.53 - 0.87 i$ & $-0.31 - 0.40 i$ & $-0.69 - 3.07 i$ & $0.16 + 1.92 i$ & $-2.44 + 0.17 i$ \\
$20.91 - 1.58 i$ & $0.25 - 0.04 i$ & $-0.71 + 0.45 i$ & $0.41 - 0.33 i$ & $-0.31 + 0.10 i$ & $-1.62 - 0.40 i$ & $-1.98 + 0.30 i$ & $0.47 + 0.55 i$ \\
$23.49 - 4.13 i$ & $0.29 + 0.21 i$ & $-0.37 - 0.38 i$ & $0.21 + 0.31 i$ & $3.88 + 4.57 i$ & $0.06 + 2.54 i$ & $-0.11 - 0.39 i$ & $-0.97 - 0.00 i$ \\
$25.94 - 1.83 i$ & $-0.01 + 0.13 i$ & $0.06 - 0.24 i$ & $-0.02 + 0.14 i$ & $1.22 - 0.86 i$ & $-0.95 + 0.80 i$ & $0.25 + 0.55 i$ & $-0.10 + 0.58 i$ \\
$31.24 - 4.37 i$ & $-0.30 - 0.39 i$ & $-0.12 + 0.31 i$ & $0.10 + 0.07 i$ & $2.20 - 5.77 i$ & $2.29 - 0.50 i$ & $-0.53 + 1.16 i$ & $-0.42 - 0.73 i$ \\
    \end{tabular}
  \end{ruledtabular}
\caption{List of poles and residues of the MPA($\q$) representation of Pb.
}
    \label{tab:poles_Pb}
\end{table*}

%
%
\bibliography{biblio}